\documentclass[12pt,chapterheads]{ucsd}
\usepackage{amsmath, amscd, amssymb, amsthm}
\usepackage{graphicx}
\usepackage{tikz}

\begin{document}


\newcommand{\comb}[2]{{\begin{pmatrix} #1 \\ #2 \end{pmatrix}}}
\newcommand{\braket}[2]{{\left\langle #1 \middle| #2 \right\rangle}}
\newcommand{\bra}[1]{{\left\langle #1 \right|}}
\newcommand{\ket}[1]{{\left| #1 \right\rangle}}
\newcommand{\ketbra}[2]{{\left| #1 \middle\rangle \middle \langle #2 \right|}}


\title{Nonlinear Quantum Search}

\author{Thomas Giechaung Wong}
\degreeyear{2014}

\degreetitle{Doctor of Philosophy} 

\field{Physics}
\chair{Professor David Meyer}
\othermembers{
Professor Daniel Arovas \\
Professor Michael Holst \\
Professor Jeffrey Rabin \\
Professor Lu Sham \\
}
\numberofmembers{5} 

\begin{frontmatter}

\makefrontmatter 

\begin{dedication} 
	\normalsize To my biological, spiritual, and academic families. I love you all dearly.
\end{dedication}

\tableofcontents
\listoffigures  

\begin{acknowledgements} 
	I am thankful for my research advisor, David Meyer, for his unselfish and genuine care for my success. His advice, guidance, and mentoring has modeled for me the qualities of an academic advisor, and my future students and I are indebted to him for the role model he's been in my life. I am also thankful for my fellow research group members whose help and friendship is cherished.

	My appreciation also goes to Origins, my spiritual family in San Diego, which has been a grace-filled community for me to grow and mature. There are too many people to name, but I thank each one for reflecting a unique part of the goodness of God and for calling out the greatness within me. I excitedly look forward to where each of us will go in life.

	I give my deepest gratitude to my parents, who have supported me throughout my life. Wherever life takes me, I can always turn to them for love, refreshment, guidance, and acceptance. I am also grateful for my brother, whose lifelong friendship makes him closer than a brother, and to my extended family.

	Finally, I praise the Lord, whose love for me began before I did a thing. I give thanks that this dissertation could be done from a place of love and acceptance, and not as a means to gain love and acceptance. I look forward to partnering with Him to see His goodness and love, wisdom and revelation impact every area of society, including the sciences.

	\bigskip
	Chapter 2, nearly in full, is a reprint of the material as it appears in ``Nonlinear Quantum Search Using the Gross-Pitaevskii Equation'' in New Journal of Physics 15, 063014 (2013). D.~A.~Meyer and T.~G.~Wong both contributed significantly to the work.

	Chapter 3, nearly in full, is a reprint of the material as it appears in ``Quantum Search with General Nonlinearities'' in Physical Review A 89, 012312 (2014). D.~A.~Meyer and T.~G.~Wong both contributed significantly to the work.

	Chapter 4 is based on a paper, ``Global Symmetry is Unnecessary for Fast Quantum Search,'' published in Physical Review Letters 112, 210502 (2014). J.~Janmark, D.~A.~Meyer and T.~G.~Wong all contributed significantly to the work.

	Chapter 5 is preliminary work for a paper to be published. D.~A.~Meyer and T.~G.~Wong both contributed significantly to the work.
\end{acknowledgements}

\begin{vitapage}
	\begin{vita}
      		\item[2008] B.S.~in Physics, Computer Science, and Mathematics \emph{summa cum laude}, Santa Clara University
      		\item[2009] Intern Single Subject Teaching Credential, Santa Clara University
      		\item[2011] M.S.~in Physics, University of California, San Diego 
      		\item[2014] Ph.D.~in Physics, University of California, San Diego 
	\end{vita}
	\begin{publications}
		\item D. A. Meyer and T. G. Wong, ``Nonlinear Quantum Search on Sufficiently Complete Graphs,'' preparing for publication.

		\item J. Janmark, D. A. Meyer and T. G. Wong, ``Global Symmetry is Unnecessary for Fast Quantum Search,'' Physical Review Letters 112, 210502 (2014).

		\item D. A. Meyer and T. G. Wong, ``Quantum Search with General Nonlinearities,'' Physical Review A 89, 012312 (2014).

		\item D. A. Meyer and T. G. Wong, ``Nonlinear Quantum Search Using the Gross-Pitaevskii Equation,'' New Journal of Physics 15, 063014 (2013).

		\item D. N. Ostrov and T. G. Wong, ``Optimal Asset Allocation for Passive Investing with Capital Loss Harvesting,'' Applied Mathematical Finance 18, 291 (2011).

		\item T. G. Wong, M. Foster, J. Colgan, and D. H. Madison, ``Treatment of ion-atom collisions using a partial-wave expansion of the projectile wavefunction,'' European Journal of Physics 30, 447 (2009).
	\end{publications}
\end{vitapage}

\begin{abstract}
	Although quantum mechanics is linear, there are nevertheless quantum systems with multiple interacting particles in which the effective evolution of a single particle is governed by a nonlinear equation. This includes Bose-Einstein condensates, which are governed by the Gross-Pitaevskii equation, which is a cubic nonlinear Schr\"odinger equation with a term proportional to $|\psi|^2\psi$. Evolution by this equation solves the unstructured search problem in constant time, but at the novel expense of increasing the time-measurement precision. Jointly optimizing these resources results in an overall scaling of $N^{1/4}$, which is a significant, but not unreasonable, improvement over the $N^{1/2}$ scaling of Grover's algorithm. Since the Gross-Pitaevskii equation effectively approximates the multi-particle Schr\"odinger equation, for which Grover's algorithm is optimal, our result leads to a quantum information-theoretic bound on the number of particles needed for this approximation to hold, asymptotically. The Gross-Pitaevskii equation is not the only nonlinearity of the form $f(|\psi|^2)\psi$ that arises in effective equations for the evolution of real quantum physical systems, however:\ The cubic-quintic nonlinear Schr\"odinger equation describes light propagation in nonlinear Kerr media with defocusing corrections, and the logarithmic nonlinear Schr\"odinger equation describes Bose liquids under certain conditions. Analysis of computation with such systems yields some surprising results; for example, when time-measurement precision is included in the resource accounting, searching a ``database'' when there is a single correct answer may be easier than searching when there are multiple correct answers. These results lead to quantum information-theoretic bounds on the physical resources required for these effective nonlinear theories to hold, asymptotically. Furthermore, strongly regular graphs, which have no global symmetry, are sufficiently complete for quantum search on them to asymptotically behave like unstructured search. Certain sufficiently complete graphs retain the improved runtime and resource scalings for some nonlinearities, so our scheme for nonlinear, analog quantum computation retains its benefits even when some structure is introduced.
\end{abstract}

\end{frontmatter}


\chapter{Introduction}

\section{Linear Quantum Search}

	Imagine having a shuffled deck of playing cards where we are searching for the Ace of Spades. Since there is no ordering or structure to the cards, one must check each card, one by one, until the Ace of Spades is found. It might be the first card, or it might be the last card; on average, one must search half of them. If there are $N$ cards, then one checks $O(N/2) = O(N)$ of them on average. This is the best that a classical computer can do.

	A quantum computer, on the other hand, can solve this problem in $O(\sqrt{N})$ steps using Grover's algorithm \cite{Grover1996}. Rather than explaining it in the digital, or discrete-time, paradigm in which it was originally proposed, we will focus on its equivalent analog, or continuous-time, analogue. This was first given by Farhi and Gutmann \cite{FG1998}, but we use Childs and Goldstone's notation and interpretation \cite{CG2004}.

	\begin{figure}
		\begin{center}
			\includegraphics[width=2.0in]{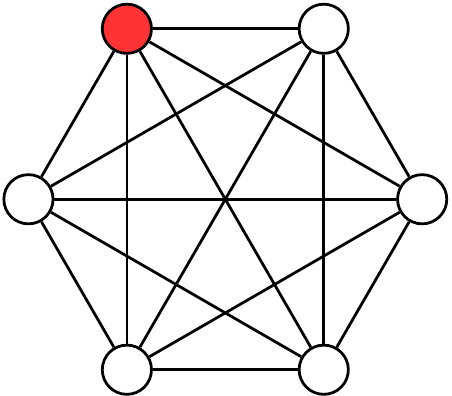}
			\caption{\label{fig:complete}The complete graph with $N = 6$ vertices and a single marked vertex (colored red). The non-marked vertices are colored white, and the state components at them evolve identically by symmetry.}
		\end{center}
	\end{figure}

	The system evolves in a $N$-dimensional Hilbert space with computational basis $\{ | 0 \rangle, \dots, | N-1 \rangle\}$. The initial state $| \psi(0) \rangle$ is an equal superposition $| s \rangle$ of all these basis states:
	\[ | \psi(0) \rangle = | s \rangle = \frac{1}{\sqrt{N}} \sum_{i=0}^{N-1} | i \rangle. \]
	The goal is to find a particular ``marked'' basis state, which we label $| w \rangle$. We do this by evolving by Schr\"odinger's equation
	\[ i \frac{\partial}{\partial t} \ket{\psi} = H_0 \ket{\psi} \]
	with Hamiltonian
	\[ H_0 = -\gamma N | s \rangle \langle s | - | w \rangle \langle w |, \]
	where the first term effects a quantum random walk on the complete graph, so $\gamma$ is a parameter that's inversely proportional to mass, and the second term is a potential well at the marked vertex, causing amplitude to build up there. Since the probability amplitudes of finding the randomly walking quantum particle at the non-marked vertices evolve identically by symmetry, as shown in figure \ref{fig:complete}, the system evolves in a two-dimensional subspace spanned by $\{\ket{w}, \ket{r}\}$, where
	\[ \ket{r} = \frac{1}{\sqrt{N-1}} \sum_{i \ne w} \ket{i} \]
	is the equal superposition of the non-marked vertices.

	\begin{figure}
		\begin{center}
			\includegraphics{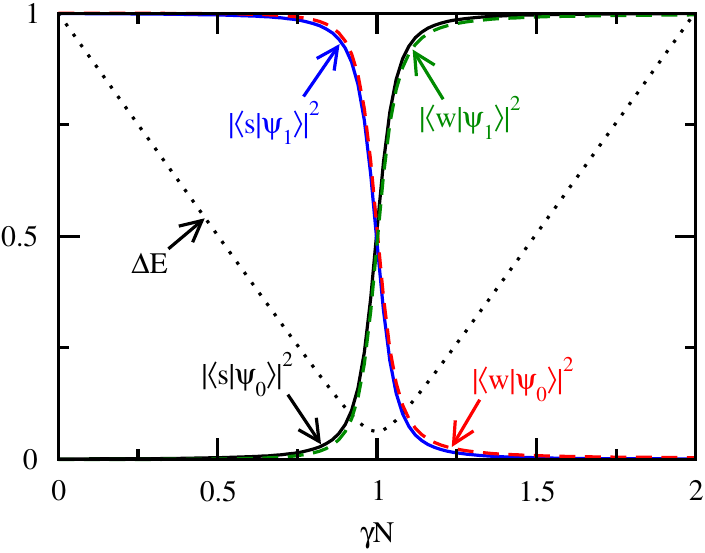}
			\caption{\label{fig:overlap_linear}Eigenvalue gap and eigenstate overlaps of $H_0$ with $N = 1024$.}
		\end{center}
	\end{figure}
	
\begin{figure}
		\begin{center}
			\includegraphics{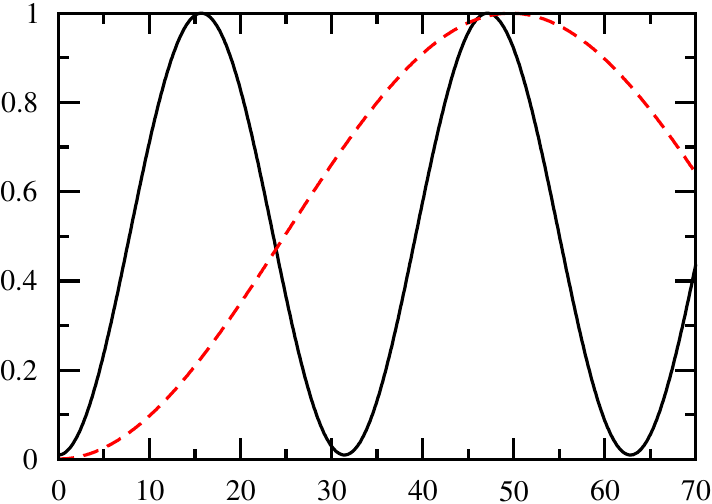}
			\caption{\label{fig:prob_time_linear}Success probability as a function of time for linear search with $\gamma = 1/N$. The solid line is $N = 100$ and the dashed line is $N = 1000$.}
		\end{center}
	\end{figure}

\begin{figure}
		\begin{center}
			\includegraphics[width=2.5in]{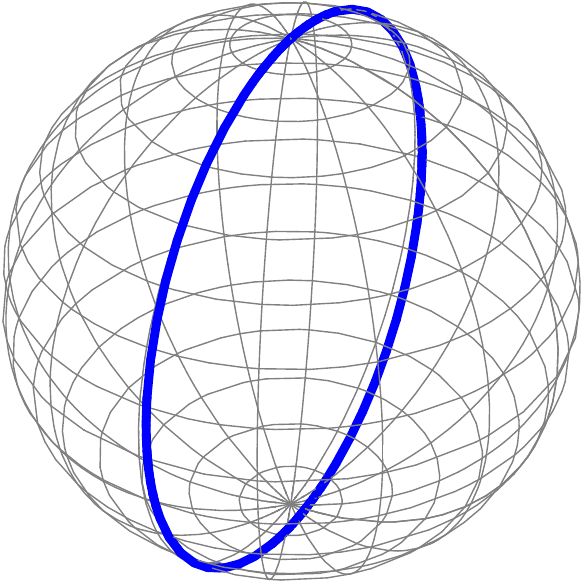}
			\caption{\label{fig:blochsphere_linear}The evolution of $\ket{\psi(t)}$ for linear search with $N = 1024$ and $\gamma = 1/N$, depicted on a Bloch sphere with $\ket{w}$ at the north pole and $\ket{r}$ at the south pole.}
		\end{center}
	\end{figure}

	One might (correctly) reason that the success of the algorithm in finding the marked vertex with probability $1$ depends on the value of $\gamma$. This can be seen in figure \ref{fig:overlap_linear}, which shows the difference in eigenvalues of $H_0$ and the overlaps of its eigenvectors with $| s \rangle$ and $| w \rangle$. When $\gamma$ takes a critical value of $\gamma_{\rm c} = 1/N$, the Hamiltonian becomes
	\[ H_0 = - \ketbra{s}{s} - \ketbra{w}{w}, \]
	and its eigenstates are
	\[ \ket{\psi_{0,1}} = \frac{1}{\sqrt{2}} \sqrt{\frac{\sqrt{N}}{\sqrt{N}+1}} \left( \ket{s} \pm \ket{w} \right) \]
	with corresponding eigenvalues
	\[ E_{0,1} = -1 \mp \frac{1}{\sqrt{N}}. \]
	So the energy gap is $\Delta E = 2/\sqrt{N}$. Then the evolution of the system can be directly calculated. We begin in the state
	\[ \ket{\psi(0)} = \ket{s} = \frac{1}{\sqrt{2}} \sqrt{\frac{\sqrt{N}+1}{\sqrt{N}}} \left( \ket{\psi_0} + \ket{\psi_1} \right). \]
	This evolves to
	\[ \ket{\psi(t)} = e^{-iHt}\ket{s} = \frac{1}{\sqrt{2}} \sqrt{\frac{\sqrt{N}+1}{\sqrt{N}}} \left( e^{-iE_0t} \ket{\psi_0} + e^{-iE_1t} \ket{\psi_1} \right). \]
	Plugging in for $\ket{\psi_0}$ and $\ket{\psi_1}$,
	\begin{align*}
		\ket{\psi(t)} 
			&= \frac{1}{2} \left[ e^{-iE_0t}\left( \ket{s} + \ket{w} \right) + e^{-iE_1t} \left( \ket{s} - \ket{w} \right) \right] \\
			&= \frac{1}{2} \left[ \left( e^{-iE_0t} + e^{-iE_1t} \right) \ket{s} + \left( e^{-iE_0t} - e^{-iE_1t} \right) \ket{w} \right] \\
			&= \frac{1}{2} e^{-i(E_0+E_1)t/2} \left[ \left(e^{i \Delta E t/2} + e^{-i \Delta E t/2} \right) \ket{s} + \left(e^{i \Delta E t/2} - e^{-i \Delta E t/2} \right) \ket{w} \right] \\
			&= e^{-i(E_0+E_1)t/2} \left[ \cos \left( \frac{\Delta E}{2} t \right) \ket{s} + i \sin \left( \frac{\Delta E}{2} t \right) \ket{w} \right],
	\end{align*}
	The amplitude of measuring the randomly walking quantum particle in the vertex corresponding to $\ket{w}$ is
	\[ \braket{w}{\psi(t)} = e^{-i(E_0+E_1)t/2} \left[ \frac{1}{\sqrt{N}} \cos \left( \frac{\Delta E}{2} t \right) + i \sin \left( \frac{\Delta E}{2} t \right) \right]. \]
	So the success probability is
	\[ \left| \braket{w}{\psi(t)} \right|^2 = \frac{1}{N} \cos^2 \left( \frac{\Delta E}{2} t \right) + \sin^2 \left( \frac{\Delta E}{2} t \right), \]
	which equals $1$ when $t = \pi / \Delta E$. So the Schr\"odinger evolution rotates the state from $| s \rangle$ to $| w \rangle$ in time $\pi \sqrt{N} / 2$, as shown in figure \ref{fig:prob_time_linear}. We can also visualize this on the Bloch sphere, as shown in figure \ref{fig:blochsphere_linear}, with $\ket{w}$ at the north pole and $\ket{r}$ at the south pole; the state starts at $\ket{s}$ near the south pole, moves directly to the north pole, loops around the other side, and repeats the motion.

	So the critical $\gamma$ is the value of $\gamma$ that causes the eigenstates of $H_0$ to be proportional to $\ket{s} \pm \ket{w}$, which causes the system to evolve to the marked basis state $\ket{w}$ in $O(\sqrt{N})$ time, thus yielding a successful search. But how do we find $\gamma_c = 1/N$ in the first place? Here, we show two methods of finding it.

	The first method of finding the critical $\gamma$ is by explicitly finding the eigenvectors of $H_0$ and choosing $\gamma$ such that they have the desired form proportional to $\ket{s} \pm \ket{w}$. Recall the system evolves in the two-dimensional subspace spanned by the marked vertex $\ket{w}$ and the equal superposition of non-marked vertices $\ket{r}$. In this basis, the Hamiltonian is
	\[ H_0 = \begin{pmatrix}
		-(\gamma + 1) & -\gamma \sqrt{N-1} \\
		-\gamma \sqrt{N-1} & -\gamma(N-1) \\
	\end{pmatrix}. \]
	Let's find the eigenvalues of $H_0$. The characteristic polynomial is
	\[ \det(H_0 - \lambda \mathbb{I}) = \lambda^2 + (\gamma N + 1)\lambda + \gamma(N-1) \]
	Setting this equal to zero and using the quadratic formula, we get
	\[ \lambda = \frac{-(\gamma N + 1) \pm \sqrt{(\gamma N - 1)^2 + 4\gamma}}{2}, \]
	which has a gap of
	\[ \Delta\lambda = \sqrt{(\gamma N - 1)^2 + 4\gamma}. \]
	Now, let's find the eigenvectors of $H_0$:
	\[ | \psi_\pm \rangle = \begin{pmatrix} u \\ v \end{pmatrix}. \]
	Writing our coefficient matrix in a more general form, our eigenvalue equation is
	\[ \begin{pmatrix} a & c \\ c & b \end{pmatrix} \begin{pmatrix} u \\ v \end{pmatrix}= \lambda\begin{pmatrix} u \\ v \end{pmatrix} , \]
	which results in (from the second equation)
	\[ u = \frac{\lambda_\pm - b}{c} v  = \frac{-\gamma N + 2 \gamma + 1 \mp \sqrt{(\gamma N - 1)^2 + 4\gamma}}{2 \gamma \sqrt{N-1}} v, \]
	or
	\begin{align*}
	       	| \psi_\pm \rangle 
			&= \begin{pmatrix} \frac{\lambda_\pm - b}{c}v \\ v \end{pmatrix} \\
			&\propto \begin{pmatrix} \frac{\lambda_\pm - b}{c} \\ 1 \end{pmatrix} \\
			&= \frac{\lambda_\pm - b}{c} | w \rangle + | r \rangle \\
			&= \frac{\lambda_\pm - b}{c} | w \rangle + \sqrt{\frac{N}{N-1}} \left( | s \rangle - \frac{1}{\sqrt{N}} | w \rangle \right) \\
			&= \sqrt{\frac{N}{N-1}} \left[ \frac{-\gamma N + 1 \mp \sqrt{(\gamma N - 1)^2 + 4\gamma}}{2 \gamma \sqrt{N}} | w \rangle + | s \rangle \right] \\
			&\propto \frac{-\gamma N + 1 \mp \sqrt{(\gamma N - 1)^2
+ 4\gamma}}{2 \gamma \sqrt{N}} | w \rangle + | s \rangle.
	\end{align*}
	We want the term in front of $| w \rangle$ to equal $\mp 1$. That way, the eigenstates are proportional to $\ket{s} \mp \ket{w}$. This happens when $\gamma = \gamma_c = 1/N$:
	\[ \left. \frac{-\gamma N + 1 \mp \sqrt{(\gamma N - 1)^2 + 4\gamma}}{2 \gamma \sqrt{N}} \right|_{\gamma_c} = \mp 1 \quad \Rightarrow \quad \gamma_c = \frac{1}{N}. \]

	Another way to determine the critical $\gamma$ and runtime is using degenerate perturbation theory \cite{Sakurai1994}. We again start with $H_0$ in the $\{ \ket{w}, \ket{r} \}$ basis:
	\[ H_0 = \begin{pmatrix}
		-(\gamma + 1) & -\gamma \sqrt{N-1} \\
		-\gamma \sqrt{N-1} & -\gamma(N-1) \\
	\end{pmatrix}. \]
	Assuming $N$ is large so that $N-1 \approx N$, we separate the Hamiltonian into leading order and higher order terms:
	\[ H_0 = \underbrace{\begin{pmatrix}
		-1 & 0 \\
		0 & -\gamma N \\
	\end{pmatrix}}_{H_0^{(0)}} + \underbrace{\begin{pmatrix}
		0 & -\gamma \sqrt{N} \\
		-\gamma \sqrt{N} & 0 \\
	\end{pmatrix}}_{H_0^{(1)}} + \underbrace{\begin{pmatrix}
		-\gamma & 0 \\
		0 & 0 \\
	\end{pmatrix}}_{H_0^{(2)}}. \]
	In lowest order, the eigenstates of the Hamiltonian are $\ket{w}$ and $\ket{r}$ with corresponding eigenvalues $-1$ and $-\gamma N$. If the eigenvalues are nondegenerate, then since the initial superposition state $\ket{s}$ is approximately $\ket{r}$ for large $N$, the system will stay near its initial state, never having large projection on $\ket{w}$. For the eigenstates to be different, namely a superposition of $\ket{r}$ and $\ket{w}$, we need the eigenvalues to be degenerate. That is, when $\gamma = \gamma_c = 1/N$, the first-order perturbation $H_0^{(1)}$ causes the eigenstates to have the form
	\[ \ket{\psi_\pm} = \alpha_w \ket{w} + \alpha_r \ket{r}, \]
	and the coefficients $\alpha_{w,r}$ and eigenvectors $E_\pm$ can be found by solving the eigenvalue problem
	\[ \begin{pmatrix} H_{ww} & H_{wr} \\ H_{rw} & H_{rr} \end{pmatrix} \begin{pmatrix} \alpha_w \\ \alpha_r \end{pmatrix} = E_\pm \begin{pmatrix} \alpha_w \\ \alpha_r \end{pmatrix}, \]
	where $H_{wr} = \langle w | H^{(0)} + H^{(1)} | r \rangle$, etc. These terms are easy to calculate. We get
	\[ \begin{pmatrix} -1 & \frac{-1}{\sqrt{N}} \\ \frac{-1}{\sqrt{N}} & -1 \end{pmatrix} \begin{pmatrix} \alpha_w \\ \alpha_r \end{pmatrix} = E_\pm \begin{pmatrix} \alpha_w \\ \alpha_r \end{pmatrix}. \]
	Solving this eigenvalue problem, we get eigenvectors
	\[ \frac{1}{\sqrt{2}} \begin{pmatrix} 1 \\ -1 \end{pmatrix} \text{ with eigenvalue } E_+ = -1 + \frac{1}{\sqrt{N}} \]
	\[ \frac{1}{\sqrt{2}} \begin{pmatrix} 1 \\ 1 \end{pmatrix} \text{ with eigenvalue } E_- = -1 - \frac{1}{\sqrt{N}} \]
	Then the approximate eigenstates of $H_0$ are
	\[ \ket{\psi_\pm} = \frac{1}{\sqrt{2}} \left( \ket{w} \mp \ket{r} \right) \]
	with eigenvalues
	\[ E_\pm = -1 \pm \frac{1}{\sqrt{N}}. \]
	Note that the energy gap is $\Delta E = \frac{2}{\sqrt{N}}$. Since $\ket{r} \approx \ket{s}$, we approximately have the eigenstates from before, so the system evolves from $\ket{s}$ to $\ket{w}$ in time $t_* = \pi / \Delta E = \pi \sqrt{N} / 2$.

\section{Nonlinear Quantum Search}

	It is proved that Grover's algorithm is optimal \cite{Zalka1999}, meaning $O(\sqrt{N})$ is the fastest runtime in which quantum mechanics can solve the unstructured search problem. To search faster, one must go beyond standard quantum theory, such as nonlinear extensions. Abrams and Lloyd \cite{AL1998} gave two examples of nonlinear algorithms with fundamental nonlinearities that resulted in unreasonable computational advantages, solving NP-complete and \#P problems in polynomial time. Both of their algorithms can be implemented by a nonlinear Schr\"odinger-type evolution in which the time derivatives of the state components depend upon their hyperbolic tangents \cite{C1998a, C1998b}. The derivative of $\tanh x$ at $x=0$ is $1$, so this is a strongly nonlinear system in which $0$ is an unstable fixed point. The strength of the nonlinearity provides a large computational advantage, but it also makes the system highly susceptible to noise \cite{AL1998,C1998a,C1998b}.

	An obvious question is whether a more modest, physically motivated nonlinearity can still produce a computational advantage. While extensive experimental work has shown that, at least in the familiar regimes of atomic and optical physics, the effect of any fundamental nonlinear generalization of quantum mechanics must be tiny \cite{Weinberg1989, Bollinger1989, Sinha2010}, there are nevertheless quantum mechanical systems with multiple interacting particles in which the effective evolution of a single particle is governed by a nonlinear equation. These include Bose-Einstein condensates (BECs) \cite{Bose1924, Einstein1924, Einstein1925}, whose evolution is described by the celebrated Gross-Pitaevskii equation \cite{G1961, P1961}:
	\begin{equation}
		\label{eq:gpwave}
		i \hbar \frac{\partial}{\partial t} \psi(\mathbf{r},t) = \left[ - \frac{\hbar^2}{2m} \nabla^2 + V_\text{ext}(\mathbf{r}) + \frac{4\pi\hbar^2a}{m} N_0 \left| \psi(\mathbf{r},t) \right|^2 \right] \psi(\mathbf{r},t).
	\end{equation}
	This nonlinear Schr\"odinger equation has a cubic nonlinearity, which has zero derivative at zero, making it softer than those considered by Abrams and Lloyd. In this thesis, we explore the consequences of solving the quantum search problem with such a cubic nonlinearity, and we later generalize it to arbitrary nonlinearities of the form $f(|\psi|^2)\psi$, where $f$ is a real-valued function.
	
	Of course, the Gross-Pitaevskii equation is only an effective approximation of the linear, multi-particle dynamics, for which Grover's algorithm is optimal. So any speedup must be at the expense of increasing the ``space'' resource such that the product of the space requirements and the square of the time requirements is lower bounded by $N$ \cite{Zalka1999}. This will yield a lower bound on the number of condensate atoms needed for the Gross-Pitaevskii equation to be valid.

	To elucidate the source of the cubic nonlinearity in the Gross-Pitaevskii equation, let's explicitly derive it \cite{Pitaevskii1999}. The many-body Hamiltonian describing multiple interacting particles trapped in an external potential $V_\text{ext}(\mathbf{r})$ with two-body interaction potential $V(\mathbf{r} - \mathbf{r}')$ is
	\[ \hat H = \! \int \! d\mathbf{r} \hat\Psi^\dagger(\mathbf{r}) \left[ - \frac{\hbar^2}{2m} \nabla^2 + V_\text{ext}(\mathbf{r}) \right] \hat\Psi(\mathbf{r}) + \frac{1}{2} \! \int \! d\mathbf{r} d\mathbf{r}' \hat\Psi^\dagger(\mathbf{r}) \hat\Psi^\dagger(\mathbf{r}) V(\mathbf{r} - \mathbf{r}') \hat\Psi(\mathbf{r}') \hat\Psi(\mathbf{r}), \]
	where we've quantized the classical fields by promoting them to creation and annihilation operators, $\hat\Psi^\dagger(\mathbf{r})$ and $\hat\Psi(\mathbf{r})$, respectively (\textit{i.e.}, second quantization). In the Heisenberg interpretation, the state vectors remain fixed while the operators evolve according to
	\[ i \hbar \frac{\partial}{\partial t} \hat\Psi(\mathbf{r},t) = [ \hat\Psi, \hat H ]. \]
	Plugging in $\hat H$, this becomes
\[ i \hbar \frac{\partial}{\partial t} \hat\Psi(\mathbf{r},t) = \left[ - \frac{\hbar^2}{2m} \nabla^2 + V_\text{ext}(\mathbf{r}) + \int d\mathbf{r}' \hat\Psi^\dagger(\mathbf{r},t) V(\mathbf{r} - \mathbf{r}') \hat\Psi(\mathbf{r}',t) \right] \hat\Psi(\mathbf{r},t). \]
	We express the operator using mean field theory as an order parameter plus a perturbation:
	\[ \hat\Psi(\mathbf{r},t) = \langle \hat \Psi(\mathbf{r},t) \rangle + \hat\Psi'(\mathbf{r},t). \]
	The order parameter, or number density, can be normalized and interpreted as the wave function of the condensate, so we write it as $\psi(\mathbf{r},t) = \langle \hat \Psi(\mathbf{r},t) \rangle / \sqrt{N_0}$, where $N_0$ is the number of condensate atoms. Assuming that the perturbation is negligible, so the temperature of the condensate is near $0$, we get $\hat\Psi(\mathbf{r},t) \to \psi(\mathbf{r},t)\sqrt{N_0} $. When the Bose gas is dilute, meaning the $s$-wave scattering length $a$ is much less than the interparticle spacing, then the effective interaction is (see section 5.2.1 of \cite{PethickBook})
	\[ V(\mathbf{r} - \mathbf{r}') = \frac{4\pi\hbar^2a}{m} \delta(\mathbf{r} - \mathbf{r}'). \]
	Using this, the evolution of the condensate wave function becomes
	\[ i \hbar \frac{\partial}{\partial t} \psi(\mathbf{r},t) = \left[ - \frac{\hbar^2}{2m} \nabla^2 + V_\text{ext}(\mathbf{r}) + \frac{4\pi\hbar^2a}{m} N_0 \left| \psi(\mathbf{r},t) \right|^2 \right] \psi(\mathbf{r},t), \]
	which is the Gross-Pitaevskii equation in Eq.~\ref{eq:gpwave}.

	Physically, we can exert some control over the strength of the cubic nonlinearity in the Gross-Pitaevskii equation by varying the scattering length $a$ via Feshbach resonance \cite{Timmermans1999}. In this process, two condensate atoms interact via the hyperfine interaction (\textit{i.e.}, an interaction between the electronic and nuclear spins of the atoms), forming a quasi-bound state. The energy of this unstable intermediate state is higher than when the atoms are separate by the binding energy $E_b$, and it can be further offset with a ``detuning'' $\epsilon$. When $\epsilon = 0$, it is said that the collision is ``on resonance.'' In an optically trapped BEC, the detuning corresponds to an external magnetic field $B$ \cite{Ketterle1998}, and the effective scattering length near the resonance $B_0$ is
	\[ a_\text{eff} = \tilde{a} \left( 1 - \frac{\Delta B}{B - B_0} \right), \]
	where $\tilde{a}$ is the scattering length away from the resonance and $\Delta B$ is related to the width of the resonance \cite{Timmermans1999}. Thus there is no theoretical limit as to how much the scattering length can be varied using Feshbach resonance. Experimentally, it depends on the precision in which the external magnetic field can be controlled, and the first group to experimentally observe Feshbach resonance in BECs was able to vary the scattering length by a factor of 10 \cite{Ketterle1998}. 

	The first BEC to be experimentally produced was made by a team led by Eric Cornell and Carl Wieman of JILA by trapping and cooling $2 \times 10^4$ rubidium-87 atoms in a magnetically-confined trap \cite{Cornell1995}. About four months later, this was improved by Wolfgang Ketterle's team at MIT, who trapped $5 \times 10^5$ sodium atoms with the addition of an optical plug \cite{Ketterle1995}. Both rubidium-87 and sodium atoms have positive scattering lengths, meaning the bosons are repulsive. While BECs with negative scattering lengths cannot exist as a homogeneous gas since condensation is preempted by a first-order phase transition \cite{Stoof1994}, they are stable against collapse in the non-homogeneous environment of a trap for small numbers of atoms less than a critical value given by
	\[ N_\text{cr} = \frac{ka_\text{ho}}{|a|}, \]
	where $k$ is the dimensionless ``stability coefficient'' depending on the ratio of magnetic trap frequencies, and $a_\text{ho}$ is the harmonic oscillator length \cite{Ruprecht1995, Donley2001}. This was experimentally verified when a condensate of roughly $10^3$ lithium-7 atoms was produced by Randy Hulet's team at Rice University, just one month after the JILA collaboration's discovery \cite{Hulet1995, HuletEratta}. Condensation of bosons with attractive interactions can also be demonstrated using Feshbach resonance, using the detuning to turn the interaction from repulsive to attractive \cite{Cornell2001}.

	So the Gross-Pitaevskii equation is rooted in established physics, making its cubic nonlinearity a physically reasonable term to include in computation. In the next chapter, we quantify the computational advantage that the cubic nonlinearity provides for the unstructured search problem compared to standard quantum computation. This requires introducing a novel physical resource: time-measurement precision. Since this advantage cannot persist when the Gross-Pitaevskii equation is recognized as an approximation to an underlying multi-particle Schr\"odinger equation, for which Grover's algorithm is optimal, we arrive at a quantum information-theoretic lower bound on the number of condensate atoms needed for this approximation to hold, asymptotically.

	In Chapter 3, we generalize nonlinear search on the complete graph to arbitrary Schr\"odinger-type nonlinearities of the form $f(|\psi|^2| \psi)$, where $f$ is a real-valued function. This includes the cubic nonlinearity in the Gross-Pitaevskii equation, as well as other physical systems including the cubic-quintic and logarithmic Schr\"odinger equations. This yields some surprising results; for example, when time-measurement precision is included in the resource accounting, searching a ``database'' when there is a single correct answer may be easier than searching when there are multiple correct answers.

	As previously explained, search on the complete graph evolves in a two-dimensional subspace spanned by the marked vertex $\ket{w}$ and the superposition of non-marked vertices $\ket{r}$, which we colored red and white in figure \ref{fig:complete}. The next level of difficulty is search in a three-dimensional subspace, which strongly regular graphs support. In Chapter 4, we use degenerate perturbation theory in a novel way to solve the quantum search problem on strongly regular graphs, showing that search also achieves the $O(\sqrt{N})$ speedup. This is similar to the hypercube, which evolves in a larger space than the complete graph, but still searches in $O(\sqrt{N})$ time \cite{CG2004}. Search on these ``sufficiently complete'' graphs is sped up by the nonlinearities in the same way that that it is for the complete graph in Chapters 2 and 3, and we work through this in Chapter 5.

	Finally, we conclude with a summary and give some future directions in Chapter 6.
	

\chapter{Nonlinear Quantum Search on the Complete Graph}

\section{Setup}

To review the search problem, the system evolves in a $N$-dimensional Hilbert space with computational basis $\{ | 0 \rangle, \dots, | N-1 \rangle\}$. The initial state $| \psi(0) \rangle$ is an equal superposition $| s \rangle$ of all these basis states:
\[ | \psi(0) \rangle = | s \rangle = \frac{1}{\sqrt{N}} \sum_{i=0}^{N-1} | i \rangle. \]
The goal is to find a particular ``marked'' basis state, which we label $| w \rangle$.

In the nonlinear regime, we include an additional nonlinear ``self-potential'' $V(t)$ so that the system evolves according to the Gross-Pitaevskii equation Eq.~\ref{eq:gpwave}:
\[ i \frac{\partial}{\partial t} \psi(\mathbf{r},t) = \big[ H_0 - \underbrace{g |\psi(\mathbf{r},t)|^2}_{V(t)} \big] \psi(\mathbf{r},t), \]
where $g > 0$. This corresponds to a BEC with attractive interactions, and thus a negative scattering length \cite{Stoof1994, Hulet1995}. Heuristically, as probability accumulates at the marked state due to the $| w \rangle \langle w |$ term in $H_0$, the self-potential attracts more probability, speeding up the search. Thus we expect larger $g$ to result in a faster algorithm.

In the computational basis, the self-potential is
\[ V(t) = g \sum_{i=0}^{N-1} \left| \langle i | \psi \rangle \right|^2 | i \rangle \langle i |. \]
Even with this nonlinearity, the system remains in the subspace spanned by $\{ | w \rangle, \\ | s \rangle \}$ throughout its evolution. We define a vector
\[ | r \rangle = \frac{1}{\sqrt{N-1}} \sum_{i \ne w} | i \rangle, \]
which is orthonormal to $| w \rangle$. Then the state of the system $| \psi(t) \rangle$ can be written as
\[ | \psi(t) \rangle = \alpha(t) | w \rangle + \beta(t) | r \rangle. \]
Writing the Gross-Pitaevskii equation in this $\{ | w \rangle, | r \rangle\}$ basis, we get
\begin{align}
	\frac{d}{dt} \begin{pmatrix} \alpha \\ \beta \end{pmatrix}
		&= -i \left( H_0 - V \right) \begin{pmatrix} \alpha \\ \beta \end{pmatrix} \label{eq::gp}  \nonumber \\
		&= i \underbrace{ \begin{pmatrix} 
			\gamma + 1 + g|\alpha|^2 & \gamma \sqrt{N-1} \\
			\gamma \sqrt{N-1} & \gamma (N-1) + \frac{g}{N-1} |\beta|^2
		  \end{pmatrix}}_A
		\begin{pmatrix} \alpha \\ \beta \end{pmatrix},
\end{align}
where we've defined $A = -(H_0 - V)$.

\section{Critical Gamma}

\begin{figure}
	\begin{center}
		\includegraphics{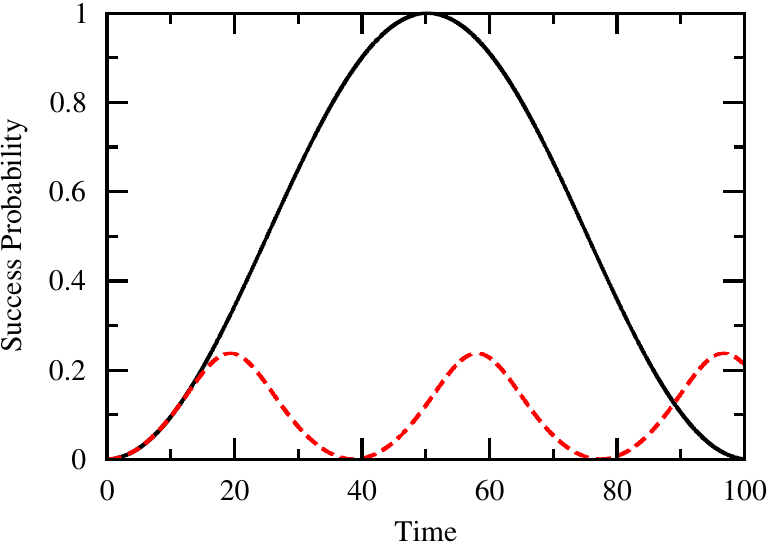}
		\caption{\label{fig:prob_time}Success probability as a function of time for $N = 1024$ and $\gamma = 1/N$ constant. The solid curve is the linear ($g = 0$) case, and the dashed curve is the nonlinear $g = 1$ case.}
	\end{center}
\end{figure}

\begin{figure}
	\begin{center}
		\includegraphics[width=2.5in]{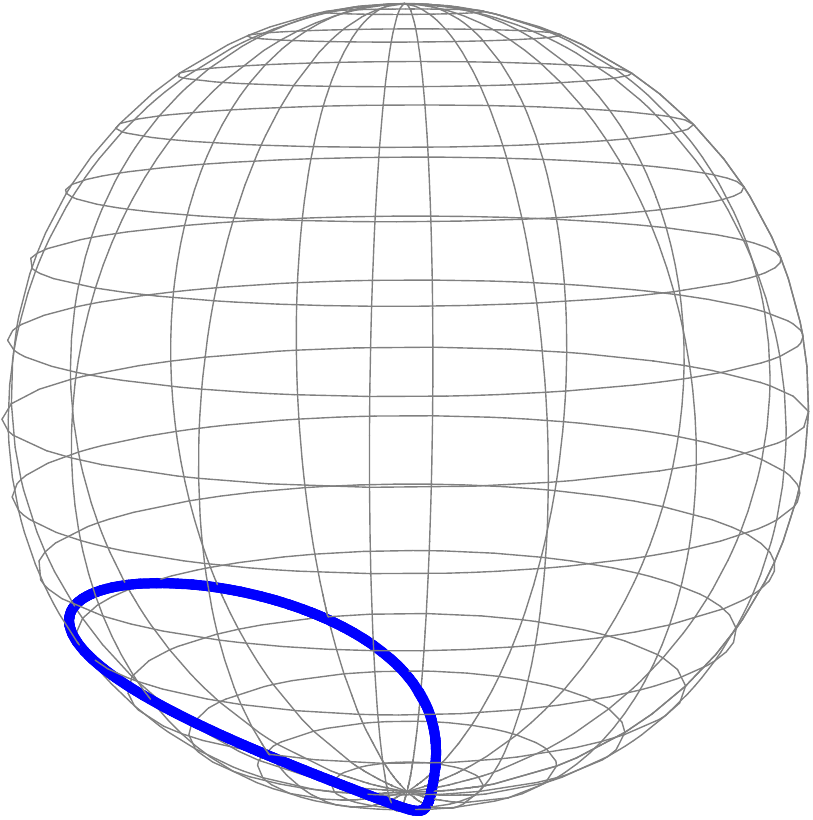}
		\caption[The evolution of $\ket{\psi(t)}$ for nonlinear search with $N = 1024$, $\gamma = 1/N$ constant, and $g = 1$, depicted on a Bloch sphere with $\ket{w}$ at the north pole and $\ket{r}$ at the south pole.]{\label{fig:blochsphere_nonlinear_gammaconst}The evolution of $\ket{\psi(t)}$ for nonlinear search with $N = 1024$, $\gamma = 1/N$ constant, and $g = 1$, depicted on a Bloch sphere with $\ket{w}$ at the north pole and $\ket{r}$ at the south pole. The system fails to reach the north pole, and so it fails to reach a success probability of $1$.}
	\end{center}
\end{figure}

\begin{figure}
	\begin{center}
		\includegraphics{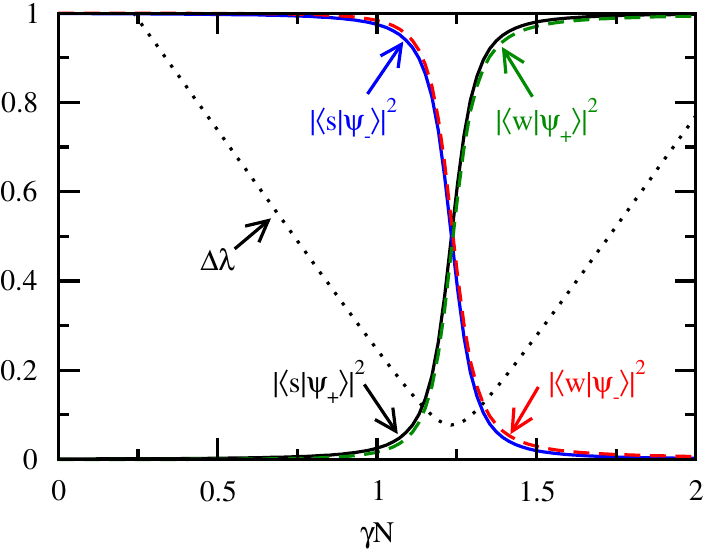}
		\caption{\label{fig:overlap_nonlinear}Eigenvalue gap and eigenstate overlaps of $A$ at $t = 20$ for nonlinear search with $N = 1024$, $g = 1$, and $\gamma = 1/N$ constant.}
	\end{center}
\end{figure}

Before proceeding with further analytical calculations, we build some intuition by examining two plots. For constant $\gamma$ and $g$, the success probability as a function of time, $|\alpha(t)|^2$, is plotted in figure \ref{fig:prob_time} along with the linear result. The nonlinear algorithm underperforms the linear one in this case. As shown on the Bloch sphere in figure \ref{fig:blochsphere_nonlinear_gammaconst}, with $\ket{w}$ at the north pole and $\ket{r}$ at the south pole, the system starts near the south pole and begins moving towards the north pole. But then it veers to the side, looping near the bottom of the sphere, and returning to its initial position. So the system never has high success probability. This is true in general for constant $\gamma$ and $g$, and it can be understood by examining the time-dependence of the critical value of $\gamma$, which is the value of $\gamma$ that ensures that the eigenstates of $A$ are in the form $\pm | w \rangle + | s \rangle$. Initially, $\gamma_{\rm c} = 1/N$. Then, as shown in figure \ref{fig:overlap_nonlinear}, it shifts to a larger value. If $\gamma$ is constant, it will not follow this shift, we will no longer have the desired eigenstates, and the algorithm will perform poorly.

To determine how $\gamma_{\rm c}$ varies with time, we find the eigenvectors of $A$ and choose $\gamma$ so that they have the desired form $\pm | w \rangle + | s \rangle$. To eliminate fractions in the subsequent algebra, we rescale the nonlinearity coefficient $g$ by defining
\[ G = \frac{g}{N-1}. \]
Solving the characteristic equation gives the eigenvalues of $A$:
\[ \lambda_\pm = \frac{1}{2} \left( \gamma N + 1 + G \sigma \right) \pm \frac{1}{2} \Delta\lambda, \]
where the gap between them is
\[ \Delta\lambda = \sqrt{(\gamma N - 1)^2 + 4\gamma + G^2 \delta^2 + 2G\delta \left[ 1 - \gamma(N-2) \right]}, \]
and we've defined
\[ \sigma = (N-1)|\alpha|^2 + |\beta|^2 \quad \mathrm{and} \quad \delta = (N-1)|\alpha|^2 - |\beta|^2. \]
The corresponding eigenvectors of $A$ are
\[ | \psi_\pm \rangle = \sqrt{\frac{N}{N-1}} \left[ \frac{-\gamma N + 1 + \delta G \pm \Delta\lambda}{2 \gamma \sqrt{N}} | w \rangle + | s \rangle \right]. \]
The critical value of $\gamma$ ensures that these eigenvectors have the form $\pm | w \rangle + | s \rangle$. That is,
\[ \left. \frac{-\gamma N + 1 + \delta G \pm \Delta\lambda}{2 \gamma \sqrt{N}} \right|_{\gamma_{\rm c}} = 1. \]
Solving this yields:
\begin{equation}
	\label{eq::criticalgamma}
	\gamma_{\rm c} = \frac{1 + G \delta}{N}.
\end{equation}
Note that in the linear limit ($G=0$), this reduces to $\gamma_{\rm c} = 1/N$, as expected and calculated in Chapter 1. Since $\delta$ varies with time, Eq.~\ref{eq::criticalgamma} implies $\gamma_{\rm c}$ also varies with time, in agreement with our previous discussion about figures \ref{fig:prob_time} and \ref{fig:overlap_nonlinear}. Furthermore, it can be precomputed without needing to know the location of the marked vertex, so there is no issue of having to measure the system during the computation.

\section{Runtime}

\begin{figure}
	\begin{center}
		\includegraphics{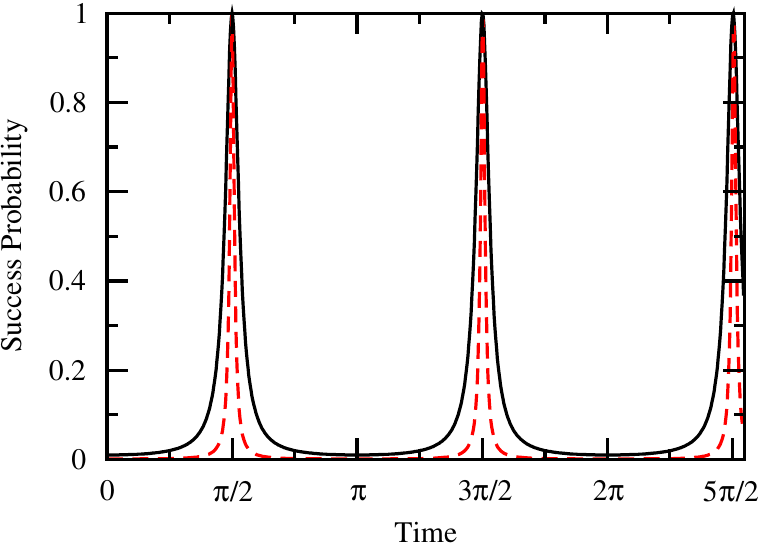}
		\caption{\label{fig:prob_time_critical}Success probability as a function of time for nonlinear search with $G = 1$ and $\gamma = \gamma_{\rm c}$ as defined in Eq.~\ref{eq::criticalgamma}. The solid line is $N = 100$ and the dashed line is $N = 1000$.}
	\end{center}
\end{figure}

For the remainder of the chapter, we choose time-varying $\gamma = \gamma_{\rm c}$ according to Eq.~\ref{eq::criticalgamma}. Before analytically determining the consequences of this, let's again consider a plot. Figure \ref{fig:prob_time_critical} shows the success probability as a function of time. There are several observations. First, the success probability reaches $1$, which occurs because we constructed the eigenstates to make this happen. Second, as $N$ increases, the runtime remains constant. Third, the success probability is periodic. Finally, the peak in success probability becomes increasingly narrow for large $N$. Let's now analytically prove the second, third, and fourth observations.

To begin, we explicitly write out Eq.~\ref{eq::gp} to get two coupled, first-order ordinary differential equations for $\alpha(t)$ and $\beta(t)$:
\begin{align}
	\label{eq::dadt} \frac{d\alpha}{dt} &= i \left\{ \left[ \gamma_{\rm c} + 1 + G(N-1) |\alpha|^2 \right] \alpha + \gamma_{\rm c} \sqrt{N-1} \beta \right\} \\
	\label{eq::dbdt} \frac{d\beta}{dt} &= i \left\{ \gamma_{\rm c} \sqrt{N-1} \alpha + \left[ \gamma_{\rm c} (N-1) + G |\beta|^2 \right] \beta \right\}.
\end{align}
We decouple these equations by defining three real variables $x(t)$, $y(t)$, and $z(t)$ such that
\begin{align}
	\label{eq::x} x &= |\alpha|^2 \\
	\label{eq::yz} y + i z &= \alpha \beta^*.
\end{align}
Note that $x(t)$ defined by Eq.~\ref{eq::x} is the success probability. Differentiating it and utilizing Eq.~\ref{eq::dadt}, we find that
\[ \frac{dx}{dt} = \frac{d|\alpha|^2}{dt} = \alpha \frac{d\alpha^*}{dt} + \frac{d\alpha}{dt} \alpha^* = 2 \gamma_{\rm c} \sqrt{N-1} z. \]
Solving this for $z$, we get
\begin{equation}
	\label{eq::z}
	z = \frac{1}{2 \gamma_{\rm c} \sqrt{N-1}} \frac{dx}{dt}.
\end{equation}
Noting that $d\gamma_{\rm c}/dt = G \, dx/dt$, we differentiate Eq.~\ref{eq::z} to get
\begin{equation}
	\label{eq::dzdt1}
	\frac{dz}{dt} = \frac{1}{2 \sqrt{N-1}} \left[ \frac{-1}{\gamma_{\rm c}^2} G \left( \frac{dx}{dt} \right)^2 + \frac{1}{\gamma_{\rm c}} \frac{d^2x}{dt^2} \right].
\end{equation}
Now we want to find another expression for $dz/dt$, which we can then set equal to Eq.~\ref{eq::dzdt1}. We do this by differentiating Eq.~\ref{eq::yz}, utilizing Eq.~\ref{eq::dadt} and Eq.~\ref{eq::dbdt}, and equating the real and imaginary parts, which yields
\begin{align}
	\label{eq::dydt} \frac{dy}{dt} &= - 2\gamma_{\rm c} z \\
	\label{eq::dzdt2} \frac{dz}{dt} &= 2\gamma_{\rm c} y + \gamma_{\rm c} \sqrt{N-1} (1-2x).
\end{align}
Substituting Eq.~\ref{eq::z} for $z$ into Eq.~\ref{eq::dydt}, we get
\[ \frac{dy}{dt} = \frac{-1}{\sqrt{N-1}} \frac{dx}{dt}, \]
which integrates to
\[ y = \frac{1}{\sqrt{N-1}} (1-x), \]
where the constant of integration was found using $x(0) = 1/N$ and $y(0) = \\ \sqrt{N-1}/N$. Now we can plug this into Eq.~\ref{eq::dzdt2} to get
\begin{align*}
	\frac{dz}{dt} 
		&= 2 \gamma_{\rm c} \frac{1}{\sqrt{N-1}} (1-x) + \gamma_{\rm c} \sqrt{N-1} (1-2x) \\
		&= \frac{\gamma_{\rm c}}{\sqrt{N-1}} \left( 1 + N - 2Nx \right).
\end{align*}
Equating this to Eq.~\ref{eq::dzdt1} and simplifying yields
\[ \frac{d^2x}{dt^2} = \frac{G}{\gamma_{\rm c}} \left( \frac{dx}{dt} \right)^2 + 2 \gamma_{\rm c}^2 \left( 1 + N - 2Nx \right). \]
Plugging in for $\gamma_{\rm c}$ as defined in Eq.~\ref{eq::criticalgamma}, this becomes
\begin{equation}
	\label{eq::uncoupled}
	\frac{d^2x}{dt^2} = \frac{NG}{1-G+NGx} \left( \frac{dx}{dt} \right)^2 + \frac{2}{N^2} \left( 1-G+NGx \right)^2 \left( 1+N-2Nx \right)
\end{equation}
Now let $f(x) = (dx/dt)^2$ so that $df/dx = 2 d^2x/dt^2$. Then Eq.~\ref{eq::uncoupled} becomes
\[
	\frac{1}{2} \frac{df}{dx} = \frac{NG}{1-G+NGx} f + \frac{2}{N^2} \left( 1-G+NGx \right)^2 \left( 1+N-2Nx \right).
\]
Solving this first-order ODE and using the initial condition $f(x=1/N) = 0$, we get
\[ f(x) = \frac{4(Nx-1)(1-x)\left[1+G(Nx-1)\right]^2}{N^2}. \]
Taking the square root and noting that $dx/dt = \pm \sqrt{f(x)}$,
\begin{equation}
	\label{eq::firstderiv}
	\frac{dx}{dt} = \pm \sqrt{ \frac{4(Nx-1)(1-x)\left[1+G(Nx-1)\right]^2}{N^2} }.
\end{equation}
To solve this uncoupled equation, we use separation of variables and integrate from $t = 0$ to $t$ and $x = 1/N$ to $x$, which yields
\begin{equation}
	\label{eq::timeprob}
	t = -\sqrt{\frac{N}{1+G(N-1)}} \left\{ \tan^{-1}\left[\frac{\sqrt{N} \sqrt{1-x}}{\sqrt{1+G(N-1)} \sqrt{Nx-1}}\right] - \frac{\pi}{2} \right\}.
\end{equation}
Solving for $x$, the success probability as a function of time is
\begin{equation}
	\label{eq::probtime}
	x(t) = \frac{N + \left[ 1+G(N-1) \right] \tan^2 \left[ \frac{\pi}{2} - \sqrt{\frac{1+G(N-1)}{N}} t \right]}{N + N  \left[ 1+G(N-1) \right] \tan^2 \left[ \frac{\pi}{2} - \sqrt{\frac{1+G(N-1)}{N}} t \right]} .
\end{equation}
From this, the success probability reaches $1$ when the tangent term is zero, which first occurs at time
\[ t_* = \frac{1}{\sqrt{1+G (N-1)}} \frac{\pi \sqrt{N}}{2}. \]
This runtime is exactly constant for $G = 1$. Also, when $G = \Theta(1)$, the runtime for large $N$ is $\pi/2\sqrt{G}$, and thus asymptotically constant (and arbitrarily small!). From Eq.~\ref{eq::probtime}, we also see that the success probability is periodic with a period of $2t_*$.

Now let's prove that the peak in success probability is narrow by finding its width, thus proving all our observations about figure \ref{fig:prob_time_critical}. Using Eq.~\ref{eq::timeprob}, the difference in time at which the success probability reaches a height of $1 - \epsilon$ is
\[ \Delta t = 2 \sqrt{\frac{N}{1+G(N-1) }} \tan^{-1}\left[\frac{\sqrt{N} \sqrt{\epsilon}}{\sqrt{1+G(N-1)} \sqrt{N(1-\epsilon)-1}}\right]. \]
The $\tan^{-1}$ makes it difficult to determine the scaling with $N$, so we Taylor expand it:
\[ \Delta t = \frac{2N}{1+G(N-1)} \sqrt{\frac{\epsilon}{N-1}} + O(\epsilon^{3/2}). \]
When $G = N^\kappa$, the first term scales as $\Theta(N^{1/2})$ when $\kappa \le -1$ and $\Theta(N^{-1/2 - \kappa})$ when $\kappa > -1$, for large $N$. To determine whether keeping this first term alone is sufficient, we use Taylor's remainder theorem to bound the error
\[ R_1(\epsilon) \le \frac{N^2 (1+3G(N(1-\epsilon)-1))}{\left(N (1-\epsilon)-1\right)^{3/2} \left(1+G\left(N \left(1-\epsilon\right)-1\right)\right)^2} \epsilon^{3/2}, \]
which has the same scaling for large $N$ as the first term in the Taylor series for $\Delta t$. Thus it suffices to keep only the first term.

For constant $G$, the width in success probability is $\Theta(1/\sqrt{N})$, which agrees with our observation from figure \ref{fig:prob_time_critical} that the peak in success probability is increasingly narrow as $N$ increases. Thus we must measure the system with increasing time precision. This behavior is opposite the linear case. That is, when $G = 0$ the width is $\Theta(\sqrt{N})$, so the time at which we measure the result can be increasingly imprecise as $N$ increases.

\section{Time-Measurement Precision}

This time-measurement precision requirement of the nonlinear algorithm requires additional resources. In particular, time and frequency standards are currently defined by atomic clocks, such as NIST-F1 in the United States \cite{NIST-F1}. An atomic clock with $n_\mathrm{clock}$ ions used as atomic oscillators has a time-measurement precision of $1/\sqrt{n_\mathrm{clock}}$ when the ions are acted upon independently. This can be improved using quantum entanglement, reducing the time-measurement precision to $1/n_\mathrm{clock}$ \cite{BIWH1996, GLM2004}. Even with this improvement, our constant-time nonlinear search algorithm would require $O(\sqrt{N})$ ions in an atomic clock to have sufficiently high time-measurement precision to measure the peak in success probability. So, although our nonlinear algorithm runs in constant time, the total resource requirement is still $O(\sqrt{N})$, the same as the linear algorithm. This raises the possibility that nonlinear quantum mechanics may not provide efficient solutions to NP-complete and \#P problems when all the resource requirements are taken into consideration \cite{AL1998}.

\begin{figure}
	\begin{center}
		\includegraphics{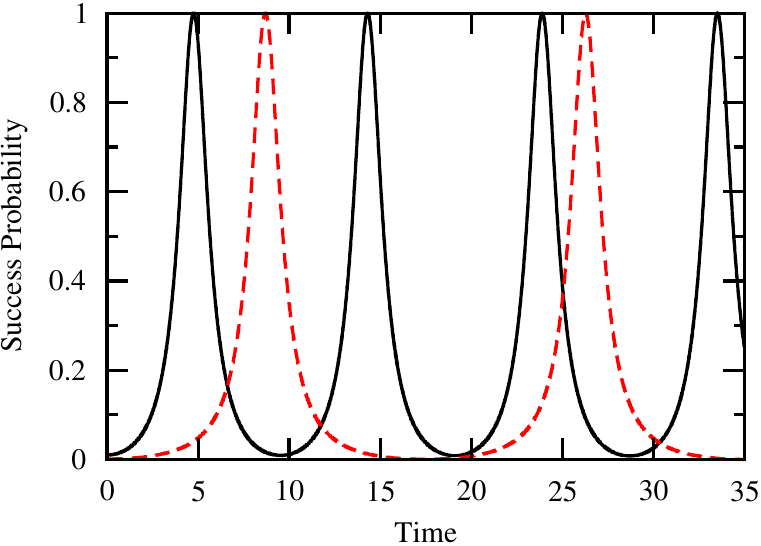}
		\caption{\label{fig:prob_time_jointoptimized}Success probability as a function of time for nonlinear search with $G = N^{-1/2}$ and $\gamma = \gamma_{\rm c}$ as defined in Eq.~\ref{eq::criticalgamma}. The solid line is $N = 100$ and the dashed line is $N = 1000$. The peaks have same width, independent of $N$.}
	\end{center}
\end{figure}

In our case, however, we can settle for a smaller improvement in runtime and reduce the time-measurement precision and total resource requirement. If we let $G$ decrease as $N^{\kappa}$ for $\kappa\le0$, then the runtime is $t_* = \Theta(N^{-\kappa/2})$, and the time-measurement precision is $\Delta t = \Theta(N^{-1/2-\kappa})$, where we've assumed for both that $\kappa > -1$, since for $\kappa\le -1$, $\Delta t = \Theta(N^{1/2})$, independently of $G$. This time-measurement precision requires $O(N^{1/2+\kappa})$ ions in an atomic clock that utilizes entanglement. We assume, as in the setup for Grover's algorithm, that $\log N$ qubits can be used to encode the $N$-dimensional Hilbert space; these should also be included in the required ``space'' resources. Multiplying the time and ``space'' requirements together, which preserves the time-space tradeoff inherent in na\"ive parallelization, the resulting total resource requirement takes a minimum value of $O(N^{1/4} \log N)$ when $\kappa = -1/2$ (so that the runtime is $N^{1/4}$ and the time-measurement precision is constant). The success probability as a function of time at this jointly optimized value of $G$ is plotted in figure \ref{fig:prob_time_jointoptimized}; note that the peak width is independent of $N$.

This significant---but not unreasonable---improvement over the \\ $\Theta(\sqrt{N} \log N)$ time-space resource requirements of the linear quantum search algorithm is consistent with our expectation that a modest nonlinearity should result in a modest speedup.

\section{Repulsive Interactions}

Our nonlinear search algorithm was based on the intuition that attractive interactions speed up the accumulation of success probability. By the same intuition, repulsive interactions, where $G < 0$, should yield a worse runtime. Our derivation of Eq.~\ref{eq::criticalgamma} for the critical value of $\gamma$ is unchanged if we flip the sign of $G$, so Eq.~\ref{eq::uncoupled} and Eq.~\ref{eq::firstderiv} are still valid for repulsive interactions. These equations yield critical points $x_* = 1/N$, $1$, and $(G-1)/NG$, corresponding to minima, maxima, and stationary points, respectively.

\begin{figure}
	\begin{center}
		\includegraphics{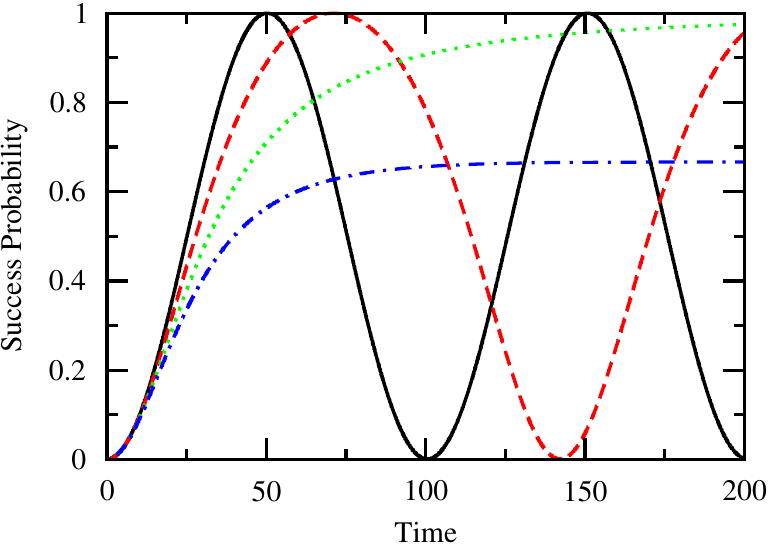}
		\caption[Success probability as a function of time for $N = 1024$ and $\gamma = \gamma_{\rm c}$ as defined in Eq.~\ref{eq::criticalgamma}.]{\label{fig:prob_time_repulsive}Success probability as a function of time for $N = 1024$ and $\gamma = \gamma_{\rm c}$ as defined in Eq.~\ref{eq::criticalgamma}. The solid line is the linear ($g=0$) case, the dashed line is the nonlinear $g = -0.5$ case, the dotted line is the nonlinear $g = -1$ case, and the dot-dashed line is the nonlinear $g = -1.5$ case.}
	\end{center}
\end{figure}

When $G > -1/(N-1)$, the success probability is unhindered by the stationary point and reaches a maximum value of $1$, as shown in the dashed curve of figure \ref{fig:prob_time_repulsive}. When $G < -1/(N-1)$, however, reaching this maximum is precluded by the presence of a stationary point, as shown in the dashed and dot-dashed curves of figure \ref{fig:prob_time_repulsive}.

We can explicitly prove that repulsive interactions ($G < 0$) will underperform the linear ($G=0$) algorithm. From Eq.~\ref{eq::firstderiv},
\[ \frac{dx}{dt} = \pm \frac{2}{N} \sqrt{(Nx-1)(1-x)} \left[ 1 + G(Nx-1) \right]. \]
So when $G < 0$, the magnitude of $dx/dt$ at a particular value of $x$ is less than when $G = 0$. Then success probability will increase more slowly for repulsive interactions (except initially, where they increase at the same rate). Thus it will underperform the linear algorithm.

\section{Validity of the Gross-Pitaevskii Equation}

Of course, the cubic nonlinearity we've exploited is not fundamental, but rather occurs in an effective description of an interacting multi-particle quantum system (\textit{e.g.}, a BEC). So we must include the number of particles $N_0$ in our resource accounting. Each particle interacts with the potential at the marked site, so in the framework of Zalka's optimality proof for Grover's algorithm \cite{Zalka1999} (generalized to continuous time \cite{Cleve2009}), there are $N_0$ oracles, each responding to a $\log N$ bit query.  Zalka showed that the product of the space requirements and the square of the time requirements is lower bounded by $N$, \textit{i.e.}, $(N_0 \log N)(N^{1/4})^2 = \Omega(N)$. Solving for the number of particles, $N_0 = \Omega(N^{1/2}/\log N)$. This is a quantum information-theoretic lower bound on the number of particles necessary for the Gross-Pitaevskii equation to be the correct asymptotic description of the multi-particle (linear) quantum dynamics.

Notice that once we account for the scaling of $N_0$ in the space requirements, the product of the time and space requirements is $O(N^{3/4})$, worse than the $O(N^{1/2}\log N)$ of Grover's algorithm.  In fact, if we calculate for the general case $G = N^{\kappa}$, where $\kappa$ need not be chosen to optimize the product of the time and space (ignoring $N_0$) resources, Zalka's bound implies $N_0 = \Omega(\max\{1,N^{1+\kappa}/\log N\})$, so the total time-space requirements are $O(N^{1+\kappa/2})$ for $\kappa> -1$, and $O(N^{1/2}\log N)$ when $\kappa = -1$.  This is optimized for $\kappa = -1$, {\it i.e.}, by Grover's algorithm.  On the other hand, Zalka's bound is strongest when $\kappa = 0$, in which case it implies that $N_0 = \Omega(N/\log N)$. That is, the existence of the constant time nonlinear algorithm we found in section 4 implies this stronger lower bound on $N_0$, despite the $O(N^{1/2})$ number of clock ions required.  To our knowledge, this is the first lower bound derived on the scaling of $N_0$ required for the Gross-Pitaevskii equation be a good asymptotic approximation.

This bound also is significantly stronger than the bound implied by the physically plausible requirement that the volume of the multi-particle condensate, and thus $N_0$, be of at least the order of the volume of space in which the $N$ possible discrete locations are defined.  Were we working in any fixed, finite dimension, {\it e.g.}, on a cubic lattice, the volume would be proportional to $N$, implying $N_0 = \Omega(N)$.  But we are not; the complete graph with equal pairwise transition rates is realized by the vertices and edges of an equilateral $(N-1)$-dimensional simplex.  With edges of length 1, this has volume $\sqrt{N/2^{N-1}}/(N-1)!$, which is much smaller than $N$, and also much smaller than our bound of $N/\log N$.

\section{Critical Gamma is a Continuous Rescaling of Time}

We previously derived the critical value of $\gamma$ so that the eigenstates of the Hamiltonian are proportional to $\pm | w \rangle + | s \rangle$. Now we examine what the critical value of $\gamma$ does from another perspective. Recall the ``Hamiltonian'' we've been using is
\[ H = - \gamma N | s \rangle \langle s | - | w \rangle \langle w | - g \sum_i |\psi_i|^2 | i \rangle \langle i |, \]
where $\psi_i = \langle i | \psi \rangle$. Explicitly writing the nonlinear term as marked and unmarked terms, we get
\begin{align*}
	H
		&= - \gamma N | s \rangle \langle s | - | w \rangle \langle w | - g |\alpha|^2 | w \rangle \langle w | - g \frac{|\beta|^2}{N-1} \sum_{x \ne w} | x \rangle \langle x | \\
		&= - \gamma N | s \rangle \langle s | - | w \rangle \langle w | - G(N-1) |\alpha|^2 | w \rangle \langle w | - G |\beta|^2 \sum_{x \ne w} | x \rangle \langle x | \\
		&= - \gamma N | s \rangle \langle s | - \left[ 1 + G(N-1) |\alpha|^2 \right] | w \rangle \langle w | - G |\beta|^2 \sum_{x \ne w} | x \rangle \langle x |.
\end{align*}
Recall $\gamma = \gamma_{\rm c}$ is chosen according to Eq.~\ref{eq::criticalgamma}:
\[ \gamma_{\rm c} N = 1 + G(N-1)|\alpha|^2 - G|\beta|^2, \]
which we arrange to get
\[ 1 + G(N-1)|\alpha|^2 = \gamma_{\rm c} N + G|\beta|^2. \]
Then the Hamiltonian becomes
\begin{align*}
	H
		&= - \gamma_{\rm c} N | s \rangle \langle s | - \left[ \gamma_{\rm c} N + G |\beta|^2 \right] | w \rangle \langle w | - G |\beta|^2 \sum_{x \ne w} | x \rangle \langle x | \\
		&= - \gamma_{\rm c} N \left( | s \rangle \langle s | + | w \rangle \langle w | \right) - G |\beta|^2 \mathbb{I} .
\end{align*}
The last term continuously redefines the ``zero'' of energy, so we can drop it. That is, it only changes the overall phase of the system, which has no measurable effect. Then the Hamiltonian is
\[ H = - \gamma N \left( | s \rangle \langle s | + | w \rangle \langle w | \right). \] 
Importantly, $H_{\rm FG} = -| s \rangle \langle s | - | w \rangle \langle w |$ is the Hamiltonian from Farhi and Gutmann's ``analog analogue'' of Grover's algorithm \cite{FG1998}, and it is optimal. Our nonlinear algorithm has a factor of $\gamma N$, so it effectively follows their optimal algorithm, but with a continuously rescaled time. That is, the system evolves according to
\[ i \frac{d\psi}{\gamma N dt} = H_{\rm FG} \psi. \]
Let's call the rescaled time $\tau(t)$ so that $d\tau = \gamma N dt$. Then
\[ \tau = \int \! \gamma N dt, \]
and the equation of motion becomes
\[ i \frac{d\psi}{d\tau} = H_{\rm FG} \psi. \]
This has success probability given by (11) of \cite{FG1998}:
\[ x(\tau) = \sin^2 \left( \frac{\tau}{\sqrt{N}} \right) + \frac{1}{N} \cos^2 \left( \frac{\tau}{\sqrt{N}} \right). \]
Plugging in for $\tau$,
\[ x(t) = \sin^2 \left( \frac{\int \! \gamma N dt}{\sqrt{N}} \right) + \frac{1}{N} \cos^2 \left( \frac{\int \! \gamma N dt}{\sqrt{N}} \right). \]
Since $\gamma_{\rm c} N = 1 + G\delta = 1 - G + GNx$, we get
\[ x(t) = \sin^2 \left( \frac{(1-G)t + GN \int \! x(t) dt}{\sqrt{N}} \right) + \frac{1}{N} \cos^2 \left( \frac{(1-G)t + GN \int \! x(t) dt}{\sqrt{N}} \right). \]
This integral transcendental equation gives $x(t)$. While the difficulty of solving this equation makes it less useful in practice, it does reveal our nonlinear algorithm's relationship with the linear, optimal algorithm. In particular, a different control policy for $\gamma$ will cause the system to evolve along a different, slower path. While not a proof, this is an argument for the optimality of our algorithm.

\section{Multiple Marked States}

Our analysis naturally extends to the case of $k$ marked states. Let $M$ be the set of marked basis states. As before, the system evolves in a two-dimensional subspace:
\[ | \psi(t) \rangle = \alpha(t) \frac{1}{\sqrt{k}} \sum_{x \in M} | x \rangle + \beta(t) \frac{1}{\sqrt{N-k}} \sum_{x \notin M} | x \rangle. \]
The system evolves according to
\[ \frac{d}{dt} | \psi(t) \rangle = i A | \psi \rangle, \]
where
\[ A = \gamma N | s \rangle \langle s | + \left( 1+g\frac{|\alpha|^2}{k} \right) \sum_{x \in M} | x \rangle \langle x | + g \frac{|\beta|^2}{N-k} \sum_{x \notin M} | x \rangle \langle x | \]
includes both the linear Hamiltonian and the nonlinear ``self-potential''. The eigenstates of $A$ have the form $\pm | w \rangle + | s \rangle$ when $\gamma$ is
\[ \gamma_{\rm c} = \frac{1 + G \delta}{N}, \]
where $G = g/(k(N-k))$ and $\delta = (N-k) |\alpha|^2 - k |\beta|^2$.
At $\gamma = \gamma_{\rm c}$, we can decouple these equations in the same manner as the $k = 1$ case and integrate from $t = 0$ to $t$ and $x = k/N$ to $x$ to get
\[ t = -\sqrt{\frac{N}{k (1+G (N-k)) }} \left\{ \tan^{-1}\left[\frac{\sqrt{N} \sqrt{1-x}}{\sqrt{1+G(N-k)} \sqrt{N x-k}}\right] - \frac{\pi}{2} \right\}, \]
which can be solved for a success probability of
\[ x(t) = \frac{N + k \left[ 1+G(N-k) \right] \tan^2 \left[ \frac{\pi}{2} - \sqrt{\frac{k(1+G(N-k))}{N}} t \right]}{N + N  \left[ 1+G(N-k) \right] \tan^2 \left[ \frac{\pi}{2} - \sqrt{\frac{k(1+G(N-k))}{N}} t \right]}. \]
Then the runtime is
\[ t_* = \frac{1}{\sqrt{k(1+G (N-k))}} \frac{\pi \sqrt{N}}{2}, \]
and the success probability is still periodic with period $2t_*$. At this runtime, the peak in success probability has a width of
\[ \Delta t = 2 \sqrt{\frac{N}{k (1+G (N-k)) }} \tan^{-1}\left[\frac{\sqrt{N} \sqrt{\epsilon}}{\sqrt{1+G(N-k)} \sqrt{N(1-\epsilon)-k}}\right], \]
but Taylor's theorem can be used to show that it suffices to keep the first term in the Taylor series:
\[ \Delta t = \frac{2N}{1+G(N-k)} \sqrt{\frac{\epsilon}{k(N-k)}} + O(\epsilon^{3/2}). \]

As in the case of a single marked state, we can find the scaling of $G = N^\kappa$ that optimizes the product of ``space'' and time, where ``space'' includes both the number of ions needed in an atomic clock that utilizes entanglement to achieve sufficiently high time-measurement precision, and the $\log N$ qubits needed to encode the $N$-dimensional Hilbert space. Say the number of marked sites scales as $k = N^\lambda$, with $0 \le \lambda \le 1$. When $\kappa = -\lambda/2 - 1/2$, the product of ``space'' and time takes a minimum value of $ST = N^{-\lambda/4+1/4} \log N$ (so that the runtime is $N^{-\lambda/4+1/4}$ and the time-measurement precision is constant). Note this is a square root speedup over the linear ($G = 0$) algorithm, whose product of ``space'' and time is $N^{-\lambda/2+1/2} \log N$. Thus our nonlinear method, by varying $\gamma$ and choosing an optimal nonlinear coefficient $G$, provides a significant, but not unreasonable, improvement over the continuous-time analogue of Grover's algorithm, even with multiple marked items.

Chapter 2, nearly in full, is a reprint of the material as it appears in ``Nonlinear Quantum Search Using the Gross-Pitaevskii Equation'' in New Journal of Physics 15, 063014 (2013). D.~A.~Meyer and T.~G.~Wong both contributed significantly to the work.


\chapter{Quantum Search with General Nonlinearities}

\section{Introduction}

So far in this thesis, we've only considered Schr\"odinger evolution with a cubic nonlinearity, \textit{i.e.}, evolution by the Gross-Pitaevskii equation \cite{G1961, P1961}:
\[ i \hbar \frac{\partial}{\partial t} \psi(\mathbf{r},t) = \left[ H_0 + \frac{4\pi\hbar^2a}{m} N_0 |\psi(\mathbf{r},t)|^2 \right] \psi(\mathbf{r},t), \]
where $H_0$ includes the kinetic energy and trapping potential, $m$ is the mass of the condensate atom, and $N_0$ is the number of condensate atoms. In Chapter 2, we quantified the computational advantage that this cubic nonlinear Schr\"odinger equation has in solving the unstructured search problem. To summarize, we search for one of $k$ ``marked'' basis states among $N$ orthonormal basis states $\{|0\rangle, |1\rangle, \dots, |N-1\rangle\}$. Without the nonlinearity, the optimal solution is the continuous-time analogue of Grover's algorithm \cite{Grover1996, Zalka1999, FG1998, Cleve2009}, which runs in time $O(\sqrt{N/k})$. With the nonlinearity, we can search in constant time with appropriate choice of parameters, as shown in figure \ref{fig:prob_time_critical}. This figure also reveals that the success probability spikes suddenly, so increasingly precise time measurement is necessary to catch the spike. This requires a certain number of atoms in an atomic clock that utilizes entanglement \cite{GLM2004, BIWH1996}. Jointly optimizing the runtime and number of clock ions, we achieve a resource requirement of $O((N/k)^{1/4})$---a square-root speedup over the linear quantum algorithm.

As explained in Chapter 2, Grover's algorithm is optimal \cite{Zalka1999}, so there must be additional resources such that the product of the space requirements and the square of the time requirements is lower bounded by $N$. In the case of the Gross-Pitaevskii equation, the additional resource is the condensate atoms, and the bound on the number of them is strongest at $\Omega(N/\log N)$ for the constant-runtime algorithm. Thus we've found a quantum information-theoretic lower bound on the number of condensate atoms needed for the Gross-Pitaevskii equation to be a good asymptotic description of the many-body, linear dynamics.

These two results---a significant, but not unreasonable, square-root speedup in solving the unstructured search problem, and the lower bound on the resources necessary for the Gross-Pitaevskii equation to be valid---suggest it is valuable to quantify the computational advantage that other effective nonlinearities have in solving the unstructured search problem. In particular, we consider nonlinear Schr\"odinger equations of the form
\begin{equation}
	\label{eq:NLSE}
	i \frac{\partial \psi}{\partial t} = \left[ H_0 - g f(|\psi|^2) \right] \psi,
\end{equation}
where $f$ is some real-valued function. The cubic nonlinear Schr\"odinger equation is the case when $f(p) = p$.

A reasonable way to adjust the cubic nonlinearity is to include higher-order terms, such as the quintic term that appears when three-body interactions are included in the description of a BEC \cite{Gammal2000}. Another example of including higher-order terms is the propagation of light in Kerr media \cite{Kerr1877, Kerr1878, Weinberger2008}, whose quantum origins are worked out in \cite{Takatsuji1967}. When a material is subjected to an electric field $E$, its index of refraction $n$ changes:
\[ n(E) = n + \frac{{\rm d}n}{{\rm d}E} E + \frac{1}{2} \frac{{\rm d}^2n}{{\rm d}E^2} E^2 + \dots \]
But from symmetry, many materials require that the index of refraction be an even function. Then the first-order term is zero, leaving
\[ n(E) = n + \frac{1}{2} \frac{d^2n}{dE^2} E^2 + \dots \]
The electric field needn't come from an external source---it can be from the incident light itself. For certain incident light beams, this second-order correction is self-focusing, and it appears in the equation of motion as a cubic term\footnote{Since the intensity is proportional to the square of the electric field, the index of refraction is frequently written as $n(I) = n_0 + n_2I$.}. The cubic self-focusing term, however, is sometimes insufficient to describe the propagation, and a quintic defocusing correction must be included \cite{Smektala2000, Boudebs2003, Zhan2002}. This results in the cubic-quintic nonlinear Schr\"odinger equation
\[ i \frac{{\rm d} \psi_n}{{\rm d}t} = -\gamma N \Delta \psi_n - g \left( |\psi_n|^2 - |\psi_n|^4 \right) \psi_n, \]
which naturally describes a periodic array of $N$ waveguides, where $\gamma$ is a parameter, $\psi_n$ is the amplitude of the electromagnetic wave in each waveguide, and $\Delta$ is the discrete second derivative \cite{CarreteroGonzalez2006}. This equation is of the form of \eqref{eq:NLSE} with $f(p) = p - p^2$.

The above nonlinearities, and indeed general nonlinearities of the form \eqref{eq:NLSE}, do not retain the separability of noninteracting subsystems. That is, in (linear) quantum mechanics, if a physical system consists of two noninteracting subsystems, then its state can be written as the product of the states of the subsystems (\textit{i.e.}, as a product state). Nonlinearities, however, generally cause initially uncorrelated subsystems to become correlated. The one exception \cite{BB1976} is the special case when $f(p) = \log(p)$. Then separability is retained, and the nonlinear Schr\"odinger equation \eqref{eq:NLSE} contains a loglinear term:
\[ i \frac{\partial \psi}{\partial t} = \left[ H_0 - g \log(|\psi|^2) \right] \psi. \]
Note that the limit of $\sqrt{x}\log(x)$ as $x$ goes to $0$ is $0$, so the evolution doesn't cause the wavefunction to diverge.\footnote{This concern was also addressed in \cite{Avdeenkov2011} by examining the generalized Lagrangian density and effective potential density in \cite{BB1976}.} Not only is the logarithmic\footnote{Although the nonlinearity is loglinear, the equation is typically referred to as logarithmic. This is different from the cubic and cubic-quintic nonlinearites where the equations are also referred to as cubic and cubic-quintic, respectively.} nonlinear Schr\"odinger equation important for its uniqueness in retaining separability, but it may be suitable for describing Bose liquids, which have higher densities than BECs \cite{Avdeenkov2011}.

In the following section, we write the generalized nonlinear search problem with multiple marked vertices in its two-dimensional subspace. Then we solve it, referencing the solution to the cubic nonlinear Schr\"odinger equation from Chapter 2 as we go. Finally, we end with two comprehensive examples of searching with the cubic-quintic and loglinear nonlinearities that were introduced above and give lower bounds for the physical resources needed for them hold.

\section{Setup}

On the complete graph of $N$ vertices, we have $k$ marked vertices, of which we are looking for any one. Let's call the set of marked vertices $M$. Then the nonlinear Schr\"odinger equation \eqref{eq:NLSE} has
\[ H_0 = -\gamma N | s \rangle \langle s | - \sum_{x \in M} | x \rangle \langle x |, \]
from which we subtract a nonlinear ``self-potential''
\[ V(t) = g \sum_{i=0}^{N-1} f\!\left(\left| \langle i | \psi \rangle \right|^2\right) | i \rangle \langle i |. \]
As with the cubic nonlinearity in Chapter 2, $g$ must be positive for the nonlinear algorithm to perform better since, heuristically, it causes the self-potential to act as an additional potential well, therefore attracting more probability and speeding up the search.

As the system evolves, it remains in the two-dimensional subspace spanned by orthonormal vectors
\[ \frac{1}{\sqrt{k}} \sum_{i \in M} | i \rangle \quad \text{and} \quad \frac{1}{\sqrt{N-k}} \sum_{i \notin M} | i \rangle, \]
so we can write $| \psi(t) \rangle$ as a linear combination of them:
\[ | \psi(t) \rangle = \alpha(t) \frac{1}{\sqrt{k}} \sum_{x \in M} | x \rangle + \beta(t) \frac{1}{\sqrt{N-k}} \sum_{x \notin M} | x \rangle. \]
Then the probability of measuring the system in basis state $| i \rangle$ is
\[ | \langle i | \psi \rangle |^2 = \begin{cases}
	\frac{|\alpha|^2}{k}, & i \in M \\
	\frac{|\beta|^2}{N-k}, & i \not\in M \\
\end{cases}. \]
Let's define
\[ f_\alpha = f\!\left( \frac{|\alpha|^2}{k} \right), \quad \text{and} \quad f_\beta = f\!\left( \frac{|\beta|^2}{N-k} \right). \]
Then the nonlinear Schr\"odinger equation \eqref{eq:NLSE} is written in the two-dimensional subspace as
\begin{equation}
	\label{eq:NLSEmatrix}
	\frac{{\rm d}}{{\rm d}t} \begin{pmatrix} \alpha \\ \beta \end{pmatrix} = i \begin{pmatrix}
	\gamma k + 1 + g f_\alpha & \gamma \sqrt{k} \sqrt{N-k} \\
	\gamma \sqrt{k} \sqrt{N-k} & \gamma(N-k) + g f_\beta \end{pmatrix} \begin{pmatrix} \alpha \\ \beta \end{pmatrix}.
\end{equation}

\section{Critical Gamma}

\begin{figure}
	\includegraphics[width=2.75in]{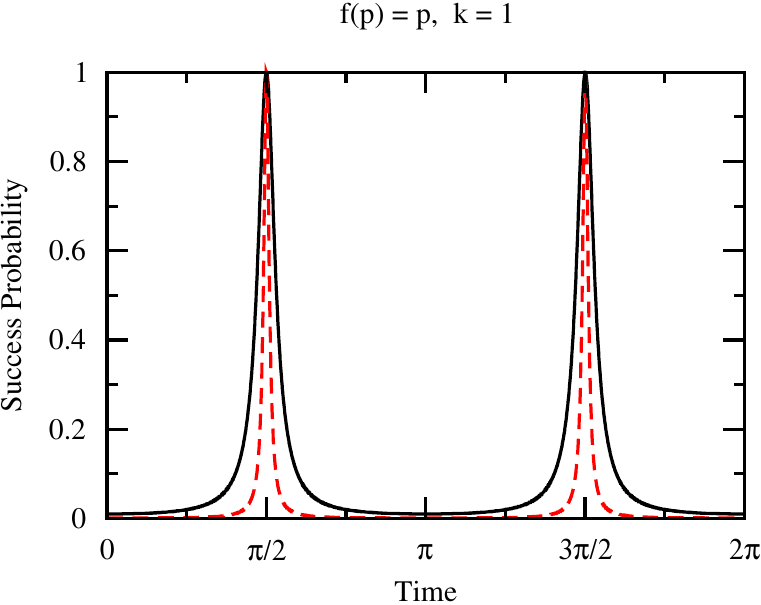}
	\quad
	\includegraphics[width=2.75in]{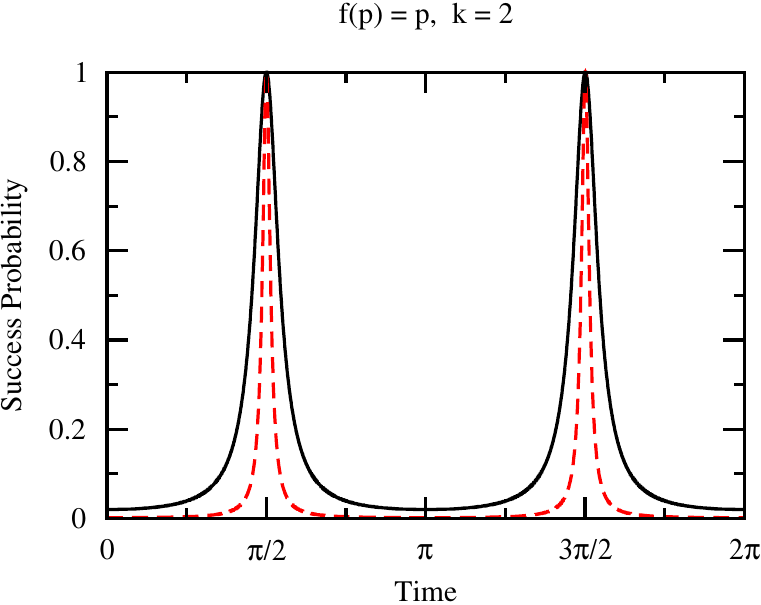}
	\vspace{0.2in}

	\includegraphics[width=2.75in]{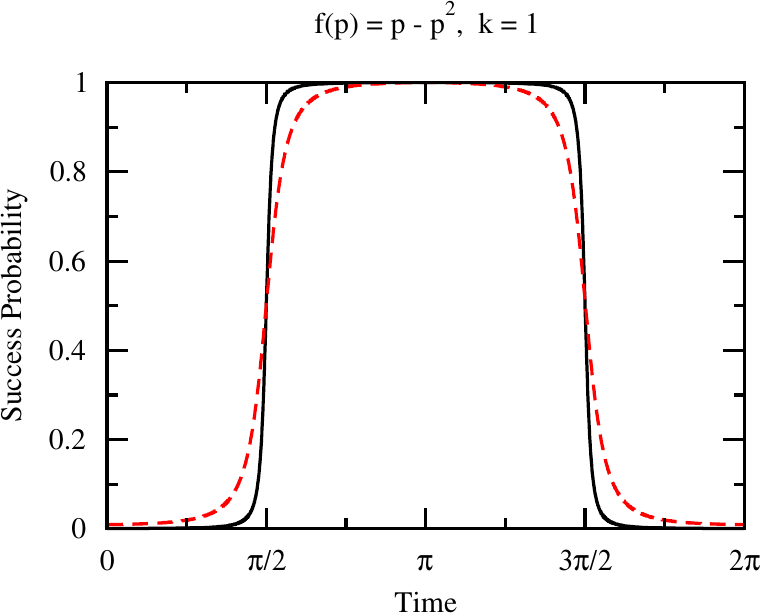}
	\quad
	\includegraphics[width=2.75in]{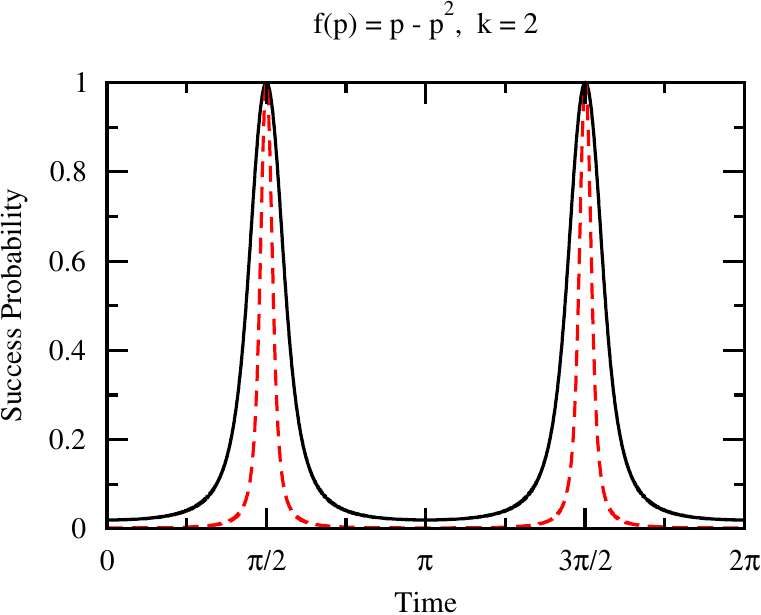}
	\vspace{0.2in}

	\includegraphics[width=2.75in]{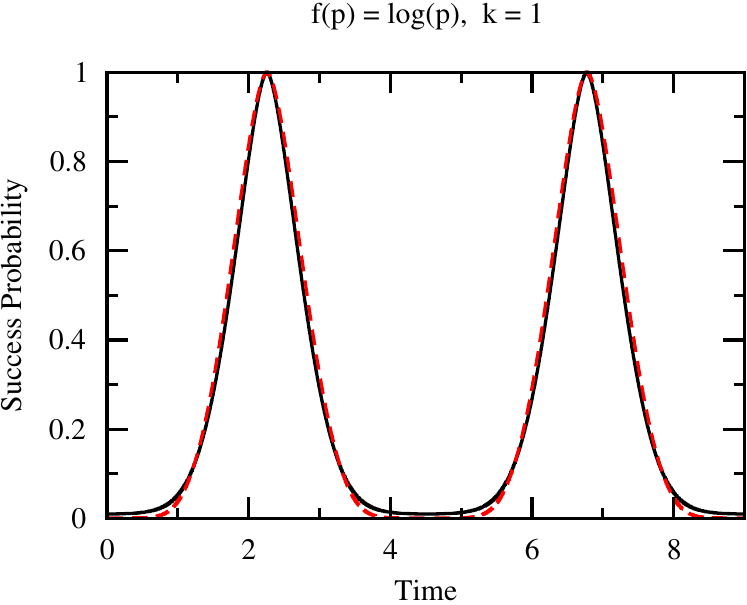}
	\quad
	\includegraphics[width=2.75in]{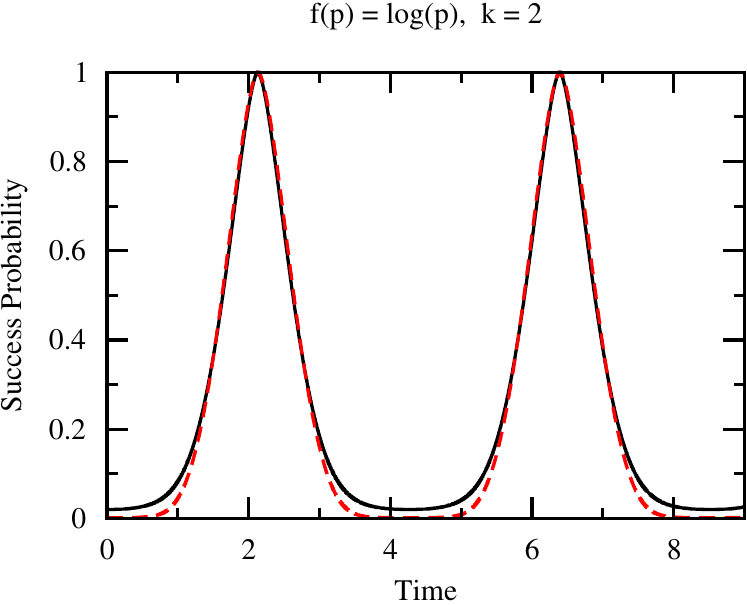}
	\caption[Success probability as a function of time for search using the cubic, cubic-quintic, and loglinear Schr\"odinger equation with $k = 1$ and $k = 2$ marked sites and $\gamma$ at its critical value given by \eqref{eq:gammac}.]{\label{fig:prob_time_mashup}Success probability as a function of time for search using the cubic, cubic-quintic, and loglinear Schr\"odinger equation with $k = 1$ and $k = 2$ marked sites and $\gamma$ at its critical value given by \eqref{eq:gammac}. The black solid curve is $N = 100$ and the red dashed curve is $N = 1000$. The nonlinearity coefficient $g$ scales as $O(N)$, $O(N)$, and $O(\sqrt{N}/\log N)$ for the respective nonlinearities so as to make the runtime constant.}
\end{figure}

We can also write $H(t) = H_0 - V(t)$ in terms of $f_\alpha$ and $f_\beta$:
\begin{align*}
	H &= -\gamma N | s \rangle \langle s | - \sum_{i \in M} | i \rangle \langle i | - g f_\alpha \sum_{i \in M} | i \rangle \langle i | - g f_\beta \sum_{i \notin M} | i \rangle \langle i | \\
	  &= -\gamma N | s \rangle \langle s | - \left( 1 + g f_\alpha - g f_\beta \right) \sum_{i \in M} | i \rangle \langle i | - g f_\beta \sum_{i = 0}^{N-1} | i \rangle \langle i |.
\end{align*}
The last term is a multiple of the identity matrix, which simply redefines the zero of energy (or contributes an overall, non-observable phase), so we can drop it. From our previous work on the cubic nonlinear Schr\"odinger equation in Chapter 2, the critical $\gamma$ causes the nonlinear system to follow the same evolution as the linear, optimal algorithm, but with rescaled time. That is, we choose
\begin{equation}
	\label{eq:gammac}
	\gamma_c = \gamma_L \left[ 1 + g \left( f_\alpha - f_\beta \right) \right] = \frac{1}{N} \left[ 1 + g \left( f_\alpha - f_\beta \right) \right],
\end{equation}
which is time-dependent, so that
\[ H = \left( 1 + g f_\alpha - g f_\beta \right) \left( -\gamma_L N | s \rangle \langle s | - \sum_{i \in M} | i \rangle \langle i | \right) = \left( 1 + g f_\alpha - g f_\beta \right) H_\text{FG}. \]
So the system evolves according to Farhi and Gutmann's Hamiltonian, but with continuously rescaled time. Thus we have the critical $\gamma$ \eqref{eq:gammac} for general nonlinearities of the form \eqref{eq:NLSE}. Note that for the cubic nonlinearity, $f(p) = p$, so if we define $G = g/[k(N-k)]$ and $\delta = (N-k)|\alpha|^2 - k|\beta|^2$, then we get the familiar result $(1+G\delta)/N$ from Chapter 2. Additionally, the critical $\gamma$ \eqref{eq:gammac} causes the eigenvectors of $H$ to be proportional to $|s\rangle \pm |w\rangle$. As explained in Chapter 1, this causes the success probability to reach a value of $1$. This is shown in figure \ref{fig:prob_time_mashup} for the cubic, cubic-quintic, and loglinear nonlinearities. A couple of observations are noteworthy. First, the cubic-quintic nonlinearity with one marked site has a wide peak in success probability, but with multiple marked sites, it has a narrow spike. Catching a narrow spike is more difficult than the wide peak, so searching with one marked site is ``easier'' than searching with multiple marked sites. This is counterintuitive, and it will be explicitly proven later. Second, for the loglinear nonlinearity, the success probability has a constant width. For the rest of the chapter, we choose $\gamma = \gamma_c$ as defined in \eqref{eq:gammac}.

\section{Runtime}

To derive the runtime of the algorithm, we follow the same procedure given in Chapter 2, generalized for \eqref{eq:NLSE}. We begin by expliciting writing out \eqref{eq:NLSEmatrix}, which yields two coupled, first-order ordinary differential equations:
\begin{subequations}
	\begin{equation} \label{eq:dadt} \frac{{\rm d}\alpha}{{\rm d}t} = i \left\{ \left[ \gamma k + 1 + g f_\alpha \right] \alpha + \gamma \sqrt{k} \sqrt{N-k} \beta \right\} \end{equation}
	\begin{equation} \label{eq:dbdt} \frac{{\rm d}\beta}{{\rm d}t} = i \left\{ \gamma \sqrt{k} \sqrt{N-k} \alpha + \left[ \gamma (N-k) + g f_\beta \right] \beta \right\}. \end{equation}
\end{subequations}
We define three real variables $x(t)$, $y(t)$, and $z(t)$ such that
\begin{subequations}
	\begin{equation} \label{eq:x} x = |\alpha|^2 \end{equation}
	\begin{equation} \label{eq:yz} y + iz = \alpha \beta^*. \end{equation}
\end{subequations}
Note that $x(t)$ is the success probability, which we want to find. To do this, we want to decouple \eqref{eq:dadt} and \eqref{eq:dbdt} for a single differential equation in terms of $x(t)$ alone and then solve it. We begin by differentiating \eqref{eq:x} by utilizing \eqref{eq:dadt}:
\[ \frac{{\rm d}x}{{\rm d}t} = \frac{{\rm d}|\alpha|^2}{{\rm d}t} = \alpha \frac{{\rm d}\alpha^*}{{\rm d}t} + \frac{{\rm d}\alpha}{{\rm d}t} \alpha^* = 2 \gamma \sqrt{k} \sqrt{N-k} z. \]
Solving for $z$,
\begin{equation}
	\label{eq:z}
	z = \frac{1}{2 \gamma \sqrt{k} \sqrt{N-k}} \frac{{\rm d}x}{{\rm d}t}.
\end{equation}
Note that the critical $\gamma$ depends on $x$:
\[ \gamma_c = \frac{1}{N} \left\{ 1 + g \left[ f\!\left( \frac{x}{k} \right) - f\!\left( \frac{1-x}{N-k} \right) \right] \right\}, \]
so we can use \eqref{eq:z} to eliminate $z$ in favor of $x$ and ${\rm d}x/{\rm d}t$. We can also find an expression for eliminating ${\rm d}z/{\rm d}t$ by differentiating this, but note that $\gamma = \gamma_c$ depends on time. Its derivative is
\[ \frac{d\gamma_c}{{\rm d}t} = \frac{g}{N} \left[ \frac{1}{k} f'_\alpha + \frac{1}{N-k} f'_\beta \right] \frac{{\rm d}x}{{\rm d}t}, \]
where we've defined in analogy to $f_\alpha$ and $f_\beta$,
\[ f'_\alpha = \left. \frac{{\rm d}f(p)}{{\rm d}p} \right|_{p = \frac{|\alpha|^2}{k} = \frac{x}{k}} \quad \text{and} \quad f'_\beta = \left. \frac{{\rm d}f(p)}{{\rm d}p} \right|_{p = \frac{|\beta|^2}{N-k} = \frac{1-x}{N-k}}. \]
Then the derivative of \eqref{eq:z} is
\begin{equation}
	\label{eq:dzdt}
	\frac{{\rm d}z}{{\rm d}t} = \frac{1}{2 \sqrt{k} \sqrt{N-k}} \left\{ \frac{-1}{\gamma^2} \frac{g}{N} \left[ \frac{1}{k} f'\left( \frac{x}{k} \right) + \frac{1}{N-k} f'\left( \frac{1-x}{N-k} \right) \right] \left( \frac{{\rm d}x}{{\rm d}t} \right)^2 + \frac{1}{\gamma} \frac{{\rm d}^2x}{{\rm d}t^2} \right\}.
\end{equation}
So now we can eliminate ${\rm d}z/{\rm d}t$ in favor of $x$, ${\rm d}x/{\rm d}t$, and ${\rm d}^2x/{\rm d}t^2$.

Now let's differentiate \eqref{eq:yz} by utilizing \eqref{eq:dadt} and \eqref{eq:dbdt}, which yields
\[ \frac{{\rm d}}{{\rm d}t} (y + iz) = \frac{{\rm d}(\alpha\beta^*)}{{\rm d}t} = \frac{{\rm d}\alpha}{{\rm d}t} \beta^* + \alpha \frac{{\rm d}\beta^*}{{\rm d}t} = -2 \gamma k z + i \left\{ 2 \gamma k y + \gamma\sqrt{k}\sqrt{N-k}(1-2x) \right\}, \]
where we've used $\gamma = \gamma_c$ to calculate the $2 \gamma k$ coefficients. Matching the real and imaginary parts, we get:
\[ \frac{{\rm d}y}{{\rm d}t} = -2 \gamma k z \]
\[ \frac{{\rm d}z}{{\rm d}t} = 2 \gamma k y + \gamma \sqrt{k} \sqrt{N-k} (1-2x). \]
In the first equation, we can eliminate $z$ using \eqref{eq:z}, which yields
\[ \frac{{\rm d}y}{{\rm d}z} = -\sqrt{\frac{k}{N-k}} \frac{{\rm d}x}{{\rm d}t}. \]
This integrates to
\[ y = -\sqrt{\frac{k}{N-k}} x + \sqrt{\frac{k}{N-k}} = -\sqrt{\frac{k}{N-k}} (x - 1) , \]
where the constant of integration was found using $y(0) = \sqrt{k(N-k)}/N$ and $x(0) = k/N$. Using this to eliminate $y$ in the second equation and simplifying,
\[ \frac{{\rm d}z}{{\rm d}t} = \gamma \sqrt{\frac{k}{N-k}} \left( N - 2Nx + k \right). \]
Eliminating ${\rm d}z/{\rm d}t$ using \eqref{eq:dzdt} and simplifying, we get
\[ \frac{{\rm d}^2x}{{\rm d}t^2} = \frac{1}{\gamma} \frac{g}{N} \left[ \frac{1}{k} f'_\alpha + \frac{1}{N-k} f'_\beta \right] \left( \frac{{\rm d}x}{{\rm d}t} \right)^2 + 2 \gamma^2 k \left( N - 2Nx + k \right), \]
which is entirely in terms of $x$ and its derivatives. Plugging in for $\gamma = \gamma_c$,
\begin{align}
	\label{eq:d2xdt2}
	\frac{{\rm d}^2x}{{\rm d}t^2} &= \frac{N}{1 + g(f_\alpha- f_\beta)} \frac{g}{N} \left[ \frac{1}{k} f'_\alpha + \frac{1}{N-k} f'_\beta \right] \left( \frac{{\rm d}x}{{\rm d}t} \right)^2 \notag \\
		&\quad+ 2 \left( \frac{1+g(f_\alpha-f_\beta)}{N} \right)^2 k \left( N - 2Nx + k \right).
\end{align}
So we've decoupled \eqref{eq:dadt} and \eqref{eq:dbdt}, yielding a second-order ordinary differential equation for $x$. To solve it, let $h(x) = ({\rm d}x/{\rm d}t)^2$ so that ${\rm d}h/{\rm d}x = 2 {\rm d}^2x/{\rm d}t^2$. Then we get a first-order ordinary differential equation for $h(x)$:
\begin{align*}
	\frac{1}{2} \frac{{\rm d}h}{{\rm d}t} &= \frac{N}{1 + g(f_\alpha- f_\beta)} \frac{g}{N} \left[ \frac{1}{k} f'_\alpha + \frac{1}{N-k} f'_\beta \right] h \\
		&\quad+ 2 \left( \frac{1+g(f_\alpha-f_\beta)}{N} \right)^2 k \left( N - 2Nx + k \right).
\end{align*}
Solving this with the initial condition $h(x=k/N) = 0$, we get
\[ h(x) = \frac{4 k (x-1) (k-Nx) \left[1 + g \left( f_\alpha - f_\beta \right) \right]^2}{N^2}. \]
Taking the square root and noting that ${\rm d}x/{\rm d}t = \pm \sqrt{h(x)}$, 
\begin{equation}
	\label{eq:dxdt}
	\frac{{\rm d}x}{{\rm d}t} = \pm \sqrt{ \frac{4 k (x-1) (k-Nx) \left[1 + g \left( f_\alpha - f_\beta \right) \right]^2}{N^2} } .
\end{equation}
We can solve this using separation of variables and integrating from $t = 0$ to $t$ and $x = k/N$ to $x$, which yields
\begin{equation}
	\label{eq:time}
	t = \frac{N}{2\sqrt{k}} \int_{x_0 = k/N}^{x} \frac{1}{1 + g(f_\alpha - f_\beta)} \sqrt{\frac{1}{(1-x)(Nx-k)}} {\rm d}x.
\end{equation}
This integral depends on the form of $f(p)$. If it is analytically integrable, we get an expression for $t(x)$, which we invert for $x(t)$. For example, for the cubic nonlinearity, $f(p) = p$. Then $f_\alpha - f_\beta = (Nx-k)/(k(N-k))$, and \eqref{eq:time} can be integrated to yield
\begin{equation}
	\label{eq:cubictime}
	t = -\sqrt{\frac{Nk}{k+g}} \left\{ \tan^{-1}\left[\frac{\sqrt{Nk} \sqrt{1-x}}{\sqrt{k+g} \sqrt{N x-k}}\right] - \frac{\pi}{2} \right\},
\end{equation}
which can be solved for a success probability of
\[ x(t) = \frac{N + (k+g) \tan^2 \left[ \frac{\pi}{2} - \sqrt{\frac{k+g}{N}} t \right]}{N + \frac{N}{k} (k+g) \tan^2 \left[ \frac{\pi}{2} - \sqrt{\frac{k+g}{N}} t \right]} . \]
This reaches a value of $1$ at a runtime of
\begin{equation}
	\label{eq:cubicruntime}
	t_* = \frac{1}{\sqrt{k+g}} \frac{\pi \sqrt{N}}{2},
\end{equation}
and the success probability is periodic with period $2t_*$. These results agree with Chapter 2.

Returning to the general nonlinearity, if we are only interested in the runtime $t_*$ and not the entire evolution of the success probability, then we can instead integrate from $x = k/N$ to $1$:
\begin{equation}
	\label{eq:runtime}
	t_* = \frac{N}{2\sqrt{k}} \int_{x_0 = k/N}^{x_* = 1} \frac{1}{1 + g(f_\alpha - f_\beta)} \sqrt{\frac{1}{(1-x)(Nx-k)}} {\rm d}x.
\end{equation}
Evaluating this for the cubic nonlinearity yields \eqref{eq:cubicruntime}, as expected.

\section{Time-Measurement Precision}

As shown in FIGs.~\ref{fig:prob_time_critical} and \ref{fig:prob_time_mashup}, the spike in success probability may be narrow. To quantify it, let's find the width of the peak at height $1 - \epsilon$.

If we can explicitly integrate \eqref{eq:time}, then we can use the result to find the width in success probability. For example, the cubic nonlinearity yielded \eqref{eq:cubictime}, which we use to find the time at which the success probability reaches a height of $1 - \epsilon$. Then the width of the peak at this height is
\[ \Delta t = 2 \sqrt{\frac{N}{k+g}} \tan^{-1}\left[\frac{\sqrt{Nk} \sqrt{\epsilon}}{\sqrt{k+g} \sqrt{N(1-\epsilon)-k}}\right]. \]
We are interested in how this time-measurement precision scales with $N$, but the inverse tangent makes it difficult to see. Instead, Taylor's theorem can be used to show that it suffices to keep the first term in the Taylor series:
\[ \Delta t^{(0)} = \frac{2Nk}{k+g} \sqrt{\frac{\epsilon}{k(N-k)}} + O(\epsilon^{3/2}). \]
If we define $G = g/[k(N-k)]$, this agrees with our result from Chapter 2.

For a general nonlinearity, we can find the leading-order formula for the time-measurement precision $\Delta t^{(0)}$ by Taylor expanding the success probability around $x = 1$, which is a maximum so the first derivative there is zero, and using \eqref{eq:d2xdt2} for the second derivative:
\begin{align*}
	x(t) 
		&= x(t_*) + x'(t_*) (t-t_*) + \frac{x''(t_*)}{2} (t-t_*)^2 + ... \\
		&\approx 1 + 0 - \left( \frac{1+g(f_\alpha|_{x=1} - f_\beta|_{x=1})}{N} \right)^2 k(N-k) (t-t_*)^2.
\end{align*}
This reaches a height of $1 - \epsilon$ at times
\[ t \approx t_* \pm \sqrt{\left( \frac{N}{1+g(f_\alpha|_{x=1} - f_\beta|_{x=1})} \right)^2 \frac{\epsilon}{k(N-k)}}. \]
So, the leading-order width of the peak is
\begin{equation}
	\label{eq:width}
	\Delta t^{(0)} = \frac{2N}{1+g(f_\alpha|_{x=1} - f_\beta|_{x=1})} \sqrt{\frac{\epsilon}{k(N-k)}}.
\end{equation}
For the cubic nonlinearity, $f_\alpha|_{x=1} - f_\beta|_{x=1} = 1/k$, so we get
\begin{equation}
	\label{eq:cubicwidth}
	\Delta t^{(0)} = \frac{2N}{1+g/k} \sqrt{\frac{\epsilon}{k(N-k)}},
\end{equation}
which agrees with our previous result.

To attain this level of time-measurement precision, say we use an atomic clock with $N_\text{clock}$ entangled ions. Then the time-measurement precision goes as $O(1/N_\text{clock})$ \cite{GLM2004, BIWH1996}. So the number of atomic clock ions we need is inversely proportional to the required time-measurement precision. This, plus the $\log N$ qubits needed to encode the $N$-dimensional Hilbert space, gives the ``space'' requirement of our algorithm. The product of ``space'' and time, which preserves the time-space tradeoff inherent in na\"ive parallelization, gives the total resource requirement.

Now that we have formulas for the runtime \eqref{eq:runtime} and time-measurement precision \eqref{eq:width} for a general nonlinearity of the form \eqref{eq:NLSE}, let's calculate them for the specific examples of the cubic-quintic and loglinear nonlinearities. But for comparison's sake, let's first review the results for the cubic nonlinearity.

\section{Cubic Nonlinearity}

The cubic nonlinear Schr\"odinger equation has the form \eqref{eq:NLSE} with $f(p) = p$. From Eqs.~\eqref{eq:cubicruntime} and \eqref{eq:cubicwidth}, we found
\[ t_* = \frac{1}{\sqrt{k+g}} \frac{\pi \sqrt{N}}{2} \]
and
\[ \Delta t^{(0)} = \frac{2N}{1+g/k} \sqrt{\frac{\epsilon}{k(N-k)}}, \]
both of which agree with Chapter 2. If $g = O(N^\kappa)$ and $k = O(N^\lambda)$ (with $0 \le \lambda \le 1$), then these become
\[ t_* = \begin{cases}
	O\left( N^{-\kappa/2 + 1/2} \right), & \kappa \ge \lambda \\
	O\left( N^{-\lambda/2 + 1/2} \right), & \kappa < \lambda
\end{cases} \]
and
\[ \Delta t^{(0)} = \begin{cases}
	O\left( N^{-\kappa + \lambda/2 + 1/2} \right), & \kappa \ge \lambda \\
	O\left( N^{-\lambda/2 + 1/2} \right), & \kappa < \lambda
\end{cases}. \]
To achieve this level of time-measurement precision, the number of ions in an atomic clock that utilizes entanglement must scale as the reciprocal of the precision \cite{GLM2004, BIWH1996}. Including the $\log N$ qubits to encode the $N$-dimensional Hilbert space, the total ``space'' requirement $S$ scales as
\[ S = \begin{cases}
	O\left( N^{\kappa - \lambda/2 - 1/2} \right), & \kappa \ge \lambda/2 + 1/2 \\
	O\left( \log N \right), & \kappa < \lambda/2 + 1/2 \\
\end{cases}. \]
Then the total resource requirement is
\[ ST = \begin{cases} 
	O\left( N^{\kappa/2 - \lambda/2}  \right), & \kappa \ge \lambda/2 + 1/2 \\
	O\left( N^{-\kappa/2 + 1/2} \log N \right), & \kappa \ge \lambda, \kappa < \lambda/2 + 1/2 \\
	O\left( N^{-\lambda/2 + 1/2} \log N  \right), & \kappa < \lambda
\end{cases} \]
This takes a minimum value of $ST = N^{-\lambda/4+1/4} \log N = (N/k)^{1/4} \log N$ when $\kappa = \lambda/2 + 1/2$, and it makes the width $\Delta t^{(0)}$ constant.

Of course, the cubic nonlinear Schr\"odinger equation, or Gross-Pitaevskii equation, is an effective nonlinear theory that only approximates the linear evolution of the multiparticle Schr\"odinger equation describing Bose-Einstein condensates. As worked out in Chapter 2 for the case of a single marked vertex, and generalized here to multiple marked vertices, since Grover's algorithm is optimal \cite{Zalka1999} for (linear) quantum computation, the number of condensate atoms $N_0$ must be included in the resource accounting such that the product of the space requirements and the square of the time requirements is lower bounded by $N$. That is, since there are $N_0$ oracles, each responding to a $\log N$ bit query,
\[ ST^2 = \begin{cases} 
	O\left( N^{-\lambda/2 + 1/2} + N^{-\kappa + 1} N_0 \log N  \right), & \kappa \ge \lambda/2 + 1/2 \\
	O\left( N^{-\kappa + 1} N_0 \log N \right), & \kappa \ge \lambda, \kappa < \lambda/2 + 1/2 \\
	O\left( N^{-\lambda + 1} N_0 \log N  \right), & \kappa < \lambda
\end{cases} = \Omega(N). \]
Then
\[ N_0 = \begin{cases}
	\Omega\left( \frac{N^\kappa}{\log N} \right), & \kappa \ge \lambda \\
	\Omega\left( \frac{N^\lambda}{\log N} \right), & \kappa < \lambda \\
\end{cases}. \]
In the first region, this bound is maximized when $\kappa = 1$, corresponding to the constant-runtime solution and beyond which it doesn't make sense to increase $\kappa$. In the second region, it is maximized when $\lambda = 1$, \textit{i.e.}, the number of marked vertices scales with $N$. In both of these cases, the bound takes its strongest value:
\[ N_0 = \Omega\left( \frac{N}{\log N} \right). \]
As expressed in Chapter 2, to the best of our knowledge, this is the first bound on the number of condensate atoms needed for the Gross-Pitaevskii equation to be a good approximation of the linear, multiparticle dynamics.

\section{Cubic-Quintic Nonlinearity}

The cubic-quintic nonlinear Schr\"odinger equation has the form \eqref{eq:NLSE} with $f(p) = p - p^2$. Then
\[ f_\alpha - f_\beta = \frac{-N(N-2k)x^2 + k(N^2-kN-2k)x - k^2(N-k-1)}{k^2(N-k)^2}. \]
Plugging this into \eqref{eq:runtime}, the runtime is given by an integral of the form
\[ t_* = \frac{Nk^2(N-k)^2}{2\sqrt{k}} \int_{x_0 = k/N}^{x_* = 1} \frac{1}{ax^2 + bx + c} \sqrt{\frac{1}{(1-x)(Nx-k)}} {\rm d}x, \]
where
\[ a = -g N (N-2k) \]
\[ b = gk(N^2-kN-2k) \]
\[ c = -gk^2(N-k-1) + k^2(N-k)^2. \]
This is analytically integrable, and the solution is
\[ t_* = \frac{\pi}{2} \frac{N k^2(N-k)^2}{2\sqrt{k}} \frac{\sqrt{2}}{\sqrt{\Sigma } \sqrt{\Delta }} \left[ \frac{2 a+b+\sqrt{\Delta}}{\sqrt{\xi +\sqrt{\Delta} (k-N)} } + \frac{-2 a-b+\sqrt{\Delta}}{\sqrt{\xi -\sqrt{\Delta}(k-N)}} \right], \]
where
\[ \Delta = b^2-4a c \]
\[ \Sigma = a+b+c \]
\[ \xi = 2ak + 2cN + b(k+N). \]
Let's find the scaling of this runtime when $g = O(N^\kappa)$ and $k = O(N^\lambda)$ (with $0 \le \lambda \le 1$) by finding the scaling of the individual terms and putting them together. We have
\[ a = O\left( N^{\kappa + 2} \right) \]
\[ b = O\left( N^{\kappa + \lambda + 2} \right) \]
\[ c = \begin{cases}
	O\left( N^{\kappa + 2\lambda + 1} \right), & \kappa \ge 1 \\
	O\left( N^{2\lambda + 2} \right), & \kappa < 1
\end{cases}. \]
Then
\[ \Delta = \begin{cases}
	O\left( N^{2\kappa + 2\lambda + 4} \right), & \kappa \ge 0 \\
	O\left( N^{\kappa + 2\lambda + 4} \right), & \kappa < 0
\end{cases} \]
\[ \Sigma = \begin{cases}
	O\left( N^{\kappa + \lambda + 2} \right), & \kappa \ge 1 \\
	O\left( N^{\kappa + \lambda + 2} \right), & \kappa < 1, \lambda \le \kappa \\
	O\left( N^{2\lambda + 2} \right), & \kappa < 1, \lambda > \kappa 
\end{cases} \]
\[ \xi = \begin{cases}
	O\left( N^{\kappa + \lambda + 3} \right), & \kappa \ge 1 \\
	O\left( N^{\kappa + \lambda + 3} \right), & \kappa < 1, \lambda \le \kappa \\
	O\left( N^{2\lambda + 3} \right), & \kappa < 1, \lambda > \kappa 
\end{cases} \]
We also have
\[ 2a + b + \sqrt{\Delta} = \begin{cases}
	O\left( N^{\kappa + \lambda + 2} \right), & \kappa \ge 0 \\
	O\left( N^{\kappa/2 + \lambda + 2} \right) & \kappa < 0
\end{cases} \]
This is different, however, from
\[ -2a - b + \sqrt{\Delta} = \begin{cases} 
	O\left( N^{\kappa + 2} \right), & \kappa \ge 1 \\ 
	O\left( N^{\kappa + 2} \right), & 0 \le \kappa < 1, \lambda \le \kappa \\ 
	O\left( N^{\lambda + 2} \right), & 0 \le \kappa < 1, \lambda > \kappa \\
	O\left( N^{\kappa/2 + \lambda + 2} \right) & \kappa < 0
\end{cases} \]
because when $\kappa \ge 0$, the dominant term in $\sqrt{\Delta}$ is $b$, which cancels with $-b$.
The expression
\[ \xi + \sqrt{\Delta} (k-N) \]
is a little tricky. The dominant terms of $\xi$ and $\sqrt{\Delta}(k-N)$ cancel in certain cases. That is, when $\kappa \ge 1$ or $\kappa < 1$ and $\lambda \le \kappa$, then $\xi = 2ak + 2cN + b(k+N)$ is dominated by the $bN$ term. When $\kappa \ge 0$, $\Delta = b^2 - 4ac$ is dominated by the $b^2$ term, so $\sqrt{\Delta}(k-N)$ is dominated by $-bN$. So in these regions, the $bN$'s cancel out, and we should ignore it when computing $\xi + \sqrt{\Delta}(k-N)$, thereby making $\xi = O(2ak + 2cN + bk)$ and $\sqrt{\Delta}(k-N) = O(bk - \frac{2ac}{b}(k-N))$. If we add them together, we get
\[ \xi + \sqrt{\Delta} (k-N) = 2ak + 2cN + 2bk - \frac{2ac}{b}(k-N). \]
Note that $2ak + 2cN + 2bk$ is dominated by $-2gkN^2 + 2k^2N^3$, and $-\frac{2ac}{b}(k-N)$ is dominated by $2gkN^2 - 2kN^3$. Adding these, the $2gkN^2$ factors cancel, leaving $\xi + \sqrt{\Delta} (k-N)$ dominated by $2k^2N^3$. So
\[ \xi + \sqrt{\Delta} (k-N) = O\left( N^{2\lambda + 3} \right). \]
It's easy to see (\textit{i.e.}, we don't have to worry about cancellations) that the scaling is also $N^{2\lambda + 3}$ for other values of $\kappa$ and $\lambda$.
Combining our results,
\[ \frac{2a + b + \sqrt{\Delta}}{\sqrt{\xi + \sqrt{\Delta} (k-N)}} = \begin{cases}
	O\left( N^{\kappa + 1/2} \right), & \kappa \ge 0 \\
	O\left( N^{\kappa/2 + 1/2} \right), & \kappa < 0 \\
\end{cases}. \]
The expression $\xi - \sqrt{\Delta} (k-N)$ is different (easier) because the dominant term no longer cancels. So we have
\[ \xi - \sqrt{\Delta} (k-N) = \begin{cases}
	O\left( N^{\kappa + \lambda + 3} \right), & \kappa \ge 1 \\
	O\left( N^{\kappa + \lambda + 3} \right), & 0 \le \kappa < 1, \lambda \le \kappa \\
	O\left( N^{2\lambda + 3} \right), & 0 \le \kappa < 1, \lambda > \kappa \\
	O\left( N^{2\lambda + 3} \right), & \kappa < 0 \\
\end{cases}. \]
Then
\[ \frac{-2a - b + \sqrt{\Delta}}{\sqrt{\xi - \sqrt{\Delta} (k-N)}} = \begin{cases}
	O\left( N^{\kappa/2 - \lambda/2 + 1/2} \right), & \kappa \ge 1 \\
	O\left( N^{\kappa/2 - \lambda/2 + 1/2} \right), & 0 \le \kappa < 1, \lambda \le \kappa \\
	O\left( N^{1/2} \right), & 0 \le \kappa < 1, \lambda > \kappa \\
	O\left( N^{\kappa/2 + 1/2} \right), & \kappa < 0 \\
\end{cases}. \]
Then
\[ \frac{2a + b + \sqrt{\Delta}}{\sqrt{\xi + \sqrt{\Delta} (k-N)}} + \frac{-2a - b + \sqrt{\Delta}}{\sqrt{\xi - \sqrt{\Delta} (k-N)}} = \begin{cases}
	O\left( N^{\kappa + 1/2} \right), & \kappa \ge 0 \\
	O\left( N^{\kappa/2 + 1/2} \right), & \kappa < 0 \\
\end{cases}. \]
We also have
\[ \sqrt{\Sigma} \sqrt{\Delta} = \begin{cases}
	O\left( N^{3\kappa/2 + 3\lambda/2 + 3} \right), & \kappa \ge 1 \\
	O\left( N^{3\kappa/2 + 3\lambda/2 + 3} \right), & 0 \le \kappa < 1, \lambda \le \kappa \\
	O\left( N^{\kappa + 2\lambda + 3} \right), & 0 \le \kappa < 1, \lambda > \kappa \\
	O\left( N^{\kappa/2 + 2\lambda + 3} \right), & \kappa < 0 \\
\end{cases} \]
Putting all this together,
\[ t_* = \begin{cases}
	O\left( N^{-\kappa/2 + 1/2} \right), & \kappa \ge 1 \\
	O\left( N^{-\kappa/2 + 1/2} \right), & 0 \le \kappa < 1, \lambda \le \kappa \\
	O\left( N^{-\lambda/2 + 1/2} \right), & 0 \le \kappa < 1, \lambda > \kappa \\
	O\left( N^{-\lambda/2 + 1/2} \right), & \kappa < 0
\end{cases} \]
Note that our formula can be reduced to two cases. When $\kappa \ge 1$, then $\lambda \le \kappa$ since $0 \le \lambda \le 1$. Similarly, when $\kappa < 0$, then $\lambda > \kappa$. So we have
\[ t_* = \begin{cases}
	O\left( N^{-\kappa/2 + 1/2} \right), & \lambda \le \kappa \\
	O\left( N^{-\lambda/2 + 1/2} \right), & \lambda > \kappa \\
\end{cases}, \]
which is the same runtime order as search with the cubic nonlinearity.

For the time-measurement precision, note that $f_\alpha|_{x=1} - f_\beta|_{x=1} = (k-1)/k^2$. Plugging this into \eqref{eq:width}, the width of the spike in success probability at height $1 - \epsilon$ is
\[ \Delta t^{(0)} = \frac{2N}{1+g(k-1)/k^2} \sqrt{\frac{1}{k(N-k)} \epsilon}. \]
When $k = 1$, the $g$ term disappears. So varying $g$, while changing the runtime, doesn't affect the width. When $k \ne 1$, it is
\[ \Delta t^{(0)} = \frac{2N}{1+g/k} \sqrt{\frac{1}{k(N-k)} \epsilon}, \]
which is the same as the cubic nonlinearity's formula. Putting these together and letting $g = O(N^\kappa)$ and $k = O(N^\lambda)$ (with $0 \le \lambda \le 1$), we get
\[ \Delta t^{(0)} = \begin{cases}
	O\left( N^{1/2} \right), & \lambda = 0 \\
	O\left( N^{-\kappa + \lambda/2 + 1/2} \right), & \lambda \ne 0, \lambda \le \kappa \\
	O\left( N^{-\lambda/2 + 1/2} \right), & \lambda \ne 0, \lambda > \kappa
\end{cases}. \]

So the runtime of search with the cubic-quintic nonlinearity scales identically to search with the cubic nonlinearity. Furthermore, when there are multiple marked sites, the time-measurement precision also scales the same. But when there is a single marked site, the time-measurement precision scales as $N^{1/2}$, which is the same as Farhi and Gutmann's linear algorithm \cite{FG1998}. This distinction between single and multiple marked sites is evident in figure \ref{fig:prob_time_mashup}. So for a single marked site, all the speedup that comes from the nonlinearity can be utilized without the expense of increasing the time-measurement precision. Thus search with the cubic-quintic nonlinearity is able to achieve a jointly-optimized runtime and time-measurement precision of $O(1)$.

As explained in Chapter 2, Grover's algorithm is optimal \cite{Zalka1999}, so there must be additional resources such that the product of the space requirements and the square of the time requirements is lower bounded by $N$. For the cubic-quintic nonlinearity, say the physical system is a periodic array of $N$ waveguides, each long enough that the electromagnetic wave propagating through it performs the calculation. So the length of the waveguide would scale with the runtime $t_*$. Keeping the cross sectional area of the waveguide constant, the number of atoms in a waveguide would also go as $t_*$. Since we have $N$ waveguides, the number of atoms would go as $Nt_*$. If the runtime is constant, then the number of atoms goes as $\Omega(N)$, satisfying the optimality proof's lower bound.

The amount of energy, or number of photons, can also be included in the resource accounting. Say a waveguide needs $P$ photons in the incident beam for it to behave like Kerr media with quintic corrections. Then we would need $PN$ photons for the whole array. But it's reasonable to say $P$ is constant, so this would scale as $N$, again satisfying the optimality proof's lower bound.

We would also need charge to create an electric field at the marked wave\-guides. Say this takes a constant number of resources. There are $k$ marked wave\-guides, so the resources for this would scale as $k$. While this may scale less than $N$, the other physical resources already satisfy the optimality proof's lower bound.

These resources may seem excessive, but if they scale linearly with $N$, then it is no different than any other database that requires the $N$ items in the database to be physically written somewhere.

For other physical systems that are effectively described by the cubic-quintic nonlinear Schr\"odinger equation, such as Bose-Einstein condensates with higher-order corrections \cite{Gammal2000}, the additional resource to the runtime and time-measurement precision is some number of particles $N_0$, each of which responds to a $\log N$ bit query. As proved above, when there are multiple marked vertices, the cubic-quintic nonlinearity solves the unstructured search problem in the same way as the cubic nonlinearity. Then the lower bound $N_0 = \Omega(N/\log N)$ from the cubic nonlinearity carries over. With a single marked vertex, the cubic-quintic nonlinearity requires a constant number of atoms in an atomic clock to achieve the necessary time-measurement precision, and this yields the same bound. Thus  the strongest lower bound on the number of particles is the same as for the cubic nonlinearity:
\[ N_0 = \Omega\left( \frac{N}{\log N} \right). \]
To the best of our knowledge, this is the first bound on the number of particles needed for the cubic-quintic Schr\"odinger equation to be a good approximation of the linear, many-body Schr\"odinger equation.

\section{Loglinear Nonlinearity}

\begin{figure}
\begin{center}
	\includegraphics{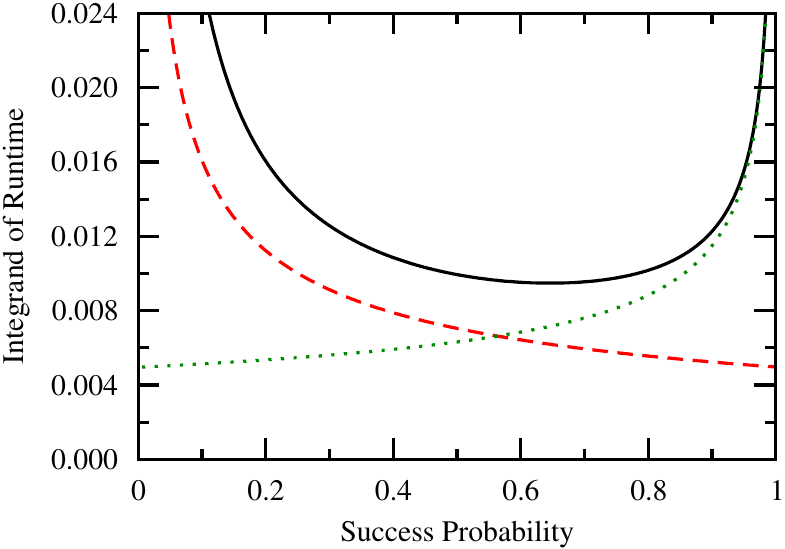}
	\caption{\label{fig:loglowerbound}The integrand of the runtime integral \eqref{eq:logtime} is the solid black curve, and the red dashed and green dotted curves are the integrands of the lower bound integrals \eqref{eq:logtimelowerbound}, all for $N = 1024$, $k = 5$, and $g = 1$.}
\end{center}
\end{figure}

The logarithmic nonlinear Schr\"odinger equation has the form \eqref{eq:NLSE} with $f(p) = \log p$. Then
\[ f_\alpha - f_\beta = \log \left( \frac{N-k}{k} \frac{x}{1-x} \right). \]
Plugging this into \eqref{eq:runtime}, the runtime is given by the integral
\begin{equation}
	\label{eq:logtime}
	t_* = \frac{N}{2\sqrt{k}} \int_{x_0 = k/N}^{1} \frac{1}{1 + g\log \left( \frac{N-k}{k} \frac{x}{1-x} \right)} \sqrt{\frac{1}{(1-x)(Nx-k)}} {\rm d}x.
\end{equation}
Although it's unclear how to directly integrate this, it is possible to bound it. Let's begin with the lower bound. Splitting the region of integration into two parts, the runtime is bounded below by
\begin{align} 
	\label{eq:logtimelowerbound}
	t_* \ge \frac{N}{2\sqrt{k}} \Bigg[ &\int_{k/N}^{1/2} \frac{1}{1 + g\log \left( \frac{N-k}{k} \frac{1/2}{1-1/2} \right)} \sqrt{\frac{1}{(1-k/N)(Nx-k)}} {\rm d}x \\
	&+ \int_{1/2}^{1} \frac{1}{1 + g\log \left( \frac{N-k}{k} \frac{1}{1-x} \right)} \sqrt{\frac{1}{(1-x)(N\cdot1-k)}} {\rm d}x \Bigg]. \notag
\end{align}
These integrands are shown in figure \ref{fig:loglowerbound} along with the original integrand, illustrating that they are indeed lower bounds. These integrate to
\begin{align*} 
	t_* \ge \frac{N}{2\sqrt{k}} \frac{1}{\sqrt{N-k}} \Bigg[ &\sqrt{\frac{2(N-2k)}{N}} \frac{1}{1 + g \log \left( \frac{N-k}{k} \right)} \\
	&- e^{\frac{1}{2g}} \frac{1}{g} \sqrt{\frac{N-k}{k}} \text{E}_1\!\left( \frac{1 + g \log \left( \frac{2(N-k)}{k} \right)}{2g} \right) \Bigg],
\end{align*}
where $\text{E}_1$ is the exponential integral
\[ \text{E}_1(x) = \int_{x}^\infty \frac{e^{-t}}{t} {\rm d}t, \]
which is bounded by
\[ \frac{1}{2} e^{-x} \log \left( 1 + \frac{2}{x} \right) < \text{E}_1(x) < e^{-x} \log \left( 1 + \frac{1}{x} \right). \]
Then the runtime is lower bounded by
\begin{align*} 
	t_* \ge \frac{N}{2\sqrt{k}} \frac{1}{\sqrt{N-k}} \Bigg[ &\sqrt{\frac{2(N-2k)}{N}} \frac{1}{1 + g \log \left( \frac{N-k}{k} \right)} \\
	&- \frac{1}{\sqrt{2}g} \log \left( 1 + \frac{2g}{1 + g \log \left( \frac{2(N-k)}{k} \right)} \right) \Bigg].
\end{align*}
Now assume that $g = O(N^\kappa)$ with $\kappa > 0$. Then for large $N$, this becomes
\[ t_* = \Omega \left( \sqrt{\frac{N}{k}} \frac{1}{g \log \left( \frac{N}{k} \right)} \right). \]

\begin{figure}
\begin{center}
	\includegraphics{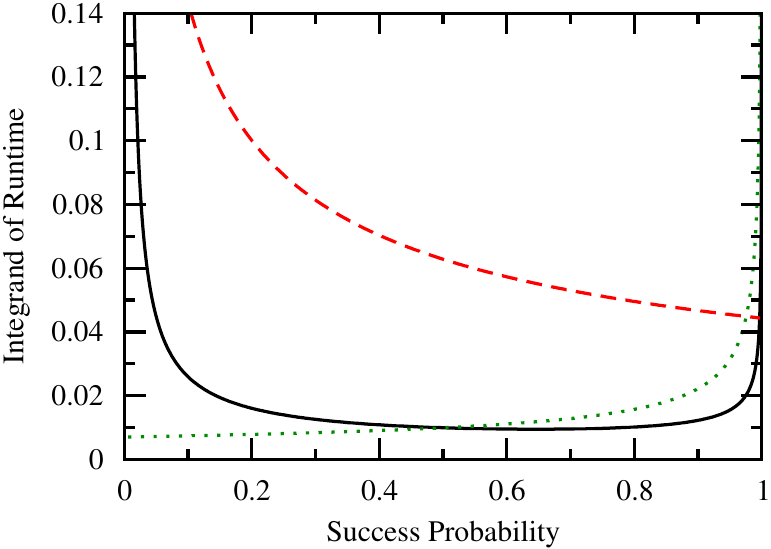}
	\caption{\label{fig:logupperbound}The integrand of the runtime integral \eqref{eq:logtime} is the solid black curve, and the red dashed and green dotted curves are the integrands of the upper bound integrals \eqref{eq:logtimeupperbound}, all for $N = 1024$, $k = 5$, and $g = 1$.}
\end{center}
\end{figure}

Now for the upper bound, we can again split the region of integration into two parts:
\begin{align} 
	\label{eq:logtimeupperbound}
	t_* \le \frac{N}{2\sqrt{k}} \Bigg[ &\int_{k/N}^{1/2} \frac{1}{1 + g\log \left( \frac{N-k}{k} \frac{k/N}{1-k/N} \right)} \sqrt{\frac{1}{(1-1/2)(Nx-k)}} {\rm d}x \\
	&+ \int_{1/2}^{1} \frac{1}{1 + g\log \left( \frac{N-k}{k} \frac{1/2}{1-1/2} \right)} \sqrt{\frac{1}{(1-x)(N/2-k)}} {\rm d}x \Bigg]. \notag
\end{align}
These integrands are shown in figure \ref{fig:logupperbound} along with the original integrand, illustrating that they are indeed upper bounds. The first region, however, is a poor bound, so we expect our result to not be tight. These integrate to
\[ t_* \le \frac{N}{2\sqrt{k}} \Bigg[ \frac{2\sqrt{N-2k}}{N} + \frac{2}{\sqrt{N-2k} \left[ 1 + g \log \left(\frac{N-k}{k}\right) \right]} \Bigg]. \]
Again assuming that $g = O(N^\kappa)$ with $\kappa > 0$ and large $N$,
\[ t_* = O\left( \sqrt{\frac{N}{k}} \right). \]
But this does not provide much insight. It simply says that the nonlinear algorithm is no worse than the linear algorithm. This is expected because our upper bound is not very tight.

\begin{figure}
\begin{center}
	\includegraphics{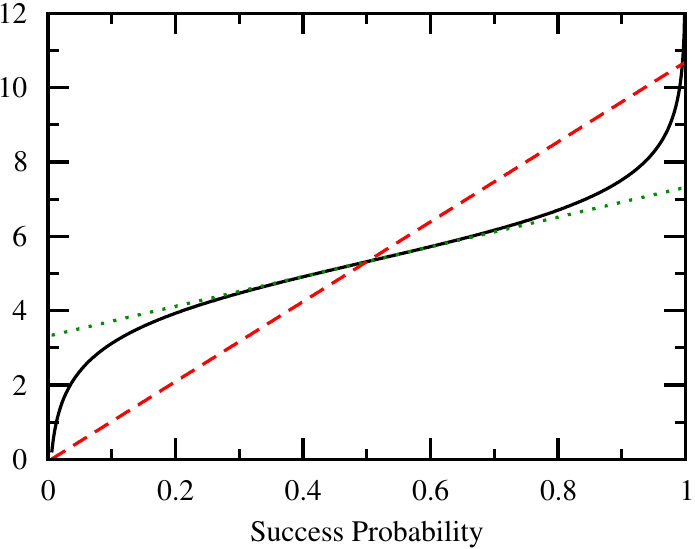}
	\caption{\label{fig:logbetterupperbound}Plot of \eqref{eq:logbetterupperbound} for $N = 1024$ and $k = 5$. The black solid curve is original logarithm, the red dashed curve is the bound from $k/N < x < 1$, and the green dotted curve is the bound from $1/2 < x < 1$.}
\end{center}
\end{figure}

\begin{figure}
\begin{center}
	\includegraphics{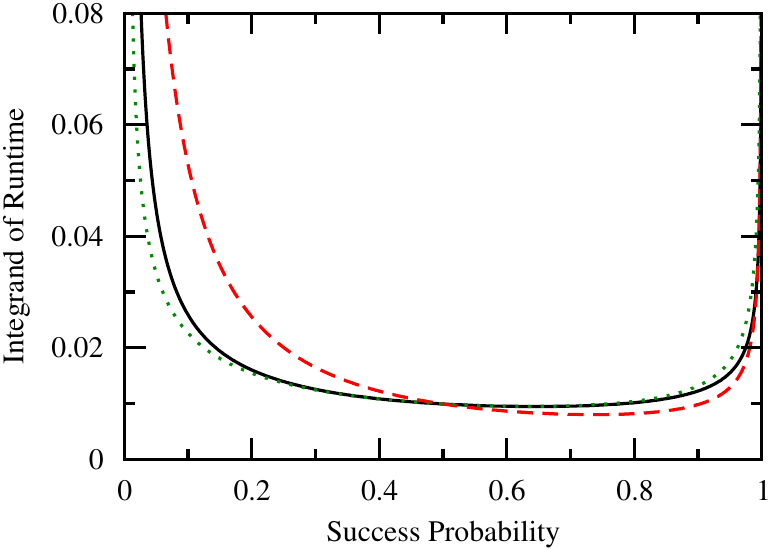}
	\caption{\label{fig:logbetterupperboundintegrand}The integrand of the runtime integral \eqref{eq:logtime} is the solid black curve, and the red dashed and green dotted curves are the integrands of the upper bound integrals \eqref{eq:logtimebetterupperbound}, all for $N = 1024$, $k = 5$, and $g = 1$.}
\end{center}
\end{figure}

\begin{figure}
\begin{center}
	\includegraphics{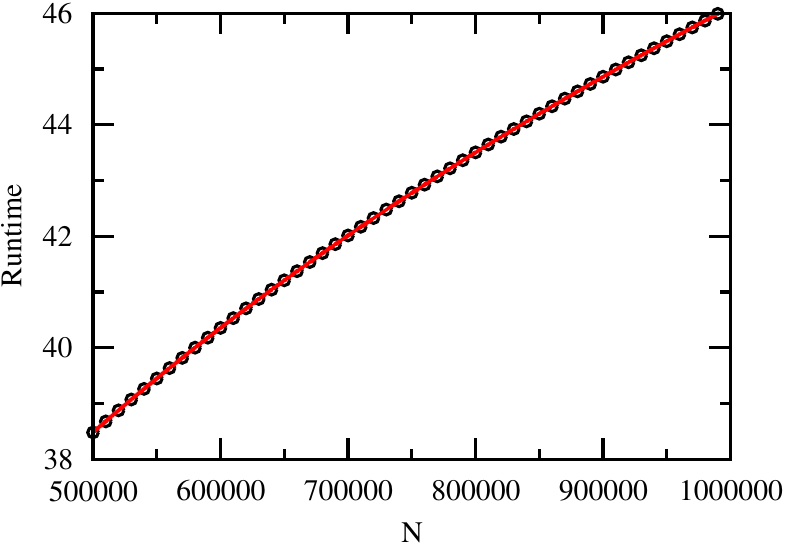}
	\caption[The runtime of search with the loglinear nonlinearity for $k = N^{1/4}$ and $g = N^{1/8}/\log(N/k)$. The black circles were numerically calculated from \eqref{eq:logtime} for $N = 500\,000$ to $N = 1\,000\,000$ with intervals of $10\,000$, and the red solid line is the best-fit.]{\label{fig:log_runtime_trend}The runtime of search with the loglinear nonlinearity for $k = N^{1/4}$ and $g = N^{1/8}/\log(N/k)$. The black circles were numerically calculated from \eqref{eq:logtime} for $N = 500\,000$ to $N = 1\,000\,000$ with intervals of $10\,000$, and the red solid line is the best-fit curve $t_* = 1.226\,N^{0.261}$.}
\end{center}
\end{figure}

To find a tighter bound for the runtime, we instead replace the logarithmic term in the denominator of the runtime integral \eqref{eq:logtime} with a smaller function. In the region $k/N < x < 1/2$, we can use the line connecting those points, and in the region $1/2 < x < 1$, we use the first-order Taylor approximation at $x = 1/2$:
\begin{equation}
	\label{eq:logbetterupperbound}
	\log \left( \frac{N-k}{k} \frac{x}{1-x} \right) \le
	\begin{cases}
		\frac{2}{N-2k} \log \left( \frac{N-k}{k} \right) (Nx-k) & k/N < x < 1/2 \\
		\log \left( \frac{N-k}{k} \right) + 4 \left( x - \frac{1}{2} \right) & 1/2 < x < 1
	\end{cases}.
\end{equation}
The bounds for this logarithm are shown in figure \ref{fig:logbetterupperbound}. Then the runtime is bounded by
\begin{align}
	\label{eq:logtimebetterupperbound}
	t_*
		&\le \frac{N}{2\sqrt{k}} \int_{k/N}^{1/2} \frac{1}{1 + g\frac{2}{N-2k} \log \left( \frac{N-k}{k} \right) (Nx-k) } \sqrt{\frac{1}{(1-x)(Nx-k)}} {\rm d}x \notag \\
		&\quad+ \frac{N}{2\sqrt{k}} \int_{1/2}^{1} \frac{1}{1 + g \left( \log \left( \frac{N-k}{k} \right) + 4 \left( x - \frac{1}{2} \right) \right)} \sqrt{\frac{1}{(1-x)(Nx-k)}} {\rm d}x.
\end{align}
These integrands are shown in figure \ref{fig:logbetterupperboundintegrand} along with the original integrand, illustrating that they are indeed upper bounds. They are also tighter than our previous attempt illustrated in figure \ref{fig:logupperbound}. The runtime integrates to
\begin{align*}
	t_* \le \frac{N}{2\sqrt{k}} \Bigg\{
		&\frac{-2 \sqrt{N-2k}}{\sqrt{N} \sqrt{N-2k+2g(N-k)\log\left(\frac{N-k}{k}\right)}} \\
		&\times \tan^{-1} \left( \frac{\sqrt{N}}{\sqrt{N-2k+2g(N-k)\log\left(\frac{N-k}{k}\right)}} \right) \\
		&+ \frac{\pi}{\sqrt{N-k}\sqrt{N}} \sqrt{\frac{N^2-3kN+2k^2}{N-2k+2g(N-k)\log\left(\frac{N-k}{k}\right)}} \\
		&+ \frac{2 \tan^{-1} \left( \frac{\sqrt{4gk+N-2gN+gN\log\left(\frac{N-k}{k}\right)}}{\sqrt{N-2k}\sqrt{1+2g+g\log\left(\frac{N-k}{k}\right)}} \right)}{\sqrt{1+2g+g\log\left(\frac{N-k}{k}\right)}\sqrt{4gk+N-2gN+gN\log\left(\frac{N-k}{k}\right)}} \Bigg\}.
\end{align*}
For $g = O(N^\kappa)$ with $\kappa > 0$ and large $N$, this is dominated by the second term and becomes
\[ t_* = O\left( \sqrt{\frac{N}{k}} \frac{1}{\sqrt{g \log \left( \frac{N}{k} \right)}} \right). \]
Combining this with the lower bound, the runtime is bounded between
\[ \sqrt{\frac{N}{k}} \frac{1}{g \log \left( \frac{N}{k} \right)} \lesssim t_* \lesssim \sqrt{\frac{N}{k}} \frac{1}{\sqrt{g \log \left( \frac{N}{k} \right)}}, \]
where the notation $f_1(N) \lesssim f_2(N)$ denotes $f_1(N) = O(f_2(N))$, which implies that $f_1(N) \gtrsim f_2(N)$ denotes $f_1(N) = \Omega(f_2(N))$. Numerically, the actual runtime seems to be closer to the lower bound. For example, when $k = N^{1/4}$ and $g = N^{1/8}/\log N$, the regression shown in figure \ref{fig:log_runtime_trend} yields a runtime scaling of $O(N^{0.261})$, whereas the lower bound is $O(N^{1/4})$ and the upper bound is $O(N^{5/16}) = O(N^{0.3125})$. Given the frequent appearance of the ratio $N/k$, we define $R = N/k$. Also, the nonlinearity coefficient $g$ appears with a factor of $\log R$, so we now let $g = O(R^\sigma / \log R)$ rather than $O(N^\kappa)$ from before. Then the bounds are
\[ R^{1/2-\sigma} \lesssim t_* \lesssim R^{1/2-\sigma/2}. \]

For the time-measurement precision, note that $f_\alpha|_{x=1} - f_\beta|_{x=1} = \infty$, so \eqref{eq:width} says the width of the success probability is zero. But from figure \ref{fig:prob_time_mashup}, that can't be right. This incorrect results arises because \eqref{eq:width} was derived by Taylor expanding the success probability about its peak, but for the loglinear nonlinearity, the second derivative at the peak is negative infinity.

To get around this, we instead Taylor expand about a nearby point $1 - \epsilon$. So we need the first and second derivative, which from \eqref{eq:dxdt} and \eqref{eq:d2xdt2} are
\[ \frac{{\rm d}x}{{\rm d}t} = \pm \sqrt{ \frac{4 k (1-x) (Nx-k) \left[1 + g \left( f_\alpha - f_\beta \right) \right]^2}{N^2} } \]
and 
\begin{align*}
	\frac{{\rm d}^2x}{{\rm d}t^2} &= \frac{N}{1 + g(f_\alpha- f_\beta)} \frac{g}{N} \left[ \frac{1}{k} f'_\alpha + \frac{1}{N-k} f'_\beta \right] \left( \frac{{\rm d}x}{{\rm d}t} \right)^2 \\
	&\quad+ 2 \left( \frac{1+g(f_\alpha-f_\beta)}{N} \right)^2 k \left( N - 2Nx + k \right).
\end{align*}
For the loglinear nonlinearity
\[ f_\alpha - f_\beta = \log \left( \frac{N-k}{k} \frac{x}{1-x} \right) \]
and
\[ \frac{1}{k} f'_\alpha + \frac{1}{N-k} f'_\beta = \frac{1}{x} - \frac{1}{1-x} = \frac{1}{x(1-x)}. \]
So near $x = 1-\epsilon$ for small $\epsilon$ and large $N$,
\[ \left. f_\alpha - f_\beta \right|_{x=1-\epsilon} \approx \log \left( \frac{N}{k} \frac{1}{\epsilon} \right). \]
Still for large $N$, the first derivative is
\[ \left. \frac{{\rm d}x}{{\rm d}t} \right|_{x=1-\epsilon} \approx \pm \sqrt{\frac{k\epsilon}{N}} g \log \left( \frac{N}{k} \frac{1}{\epsilon} \right) \]
and the second derivative is
\[ \left. \frac{{\rm d}^2x}{{\rm d}t^2} \right|_{x=1-\epsilon} \approx - \frac{k g^2 \log^2 \left( \frac{N}{k} \frac{1}{\epsilon} \right)}{N} \]
Then the Taylor expansion is
\[ x(t) \approx (1-\epsilon) + \sqrt{\frac{k\epsilon}{N}} g \log \left( \frac{N}{k} \frac{1}{\epsilon} \right) (t - t_{1-\epsilon}) - \frac{1}{2} \frac{k g^2 \log^2 \left( \frac{N}{k} \frac{1}{\epsilon} \right)}{N} (t - t_{1-\epsilon})^2, \]
where $t_{1-\epsilon}$ is the time in which the success probability is $1 - \epsilon$. Now let's consider the time in which the success probability reaches a height of $1 - \epsilon/2$, which is closer to the peak of $1$. For small $\epsilon$, the first derivative of $x(t)$ in this region is decreasing towards $0$ because the success probability is approaching the peak (where its derivative is zero). That is, for small $\epsilon$, we're considering the region after the success probability's inflection point. Then the width $\delta t = t_{1-\epsilon/2} - t_{1-\epsilon}$ is a lower bound for the width $\Delta t = t_* - t_{1-\epsilon/2}$, where $t_*$ is the time when the success probability is $1$ (\textit{i.e.}, the runtime). Then the Taylor expansion becomes
\[ 1 - \frac{\epsilon}{2} \approx (1-\epsilon) + \sqrt{\frac{k\epsilon}{N}} g \log \left( \frac{N}{k} \frac{1}{\epsilon} \right) (\delta t) - \frac{1}{2} \frac{k g^2 \log^2 \left( \frac{N}{k} \frac{1}{\epsilon} \right)}{N} (\delta t)^2. \]
This is a quadratic for $\delta t$. Solving it and keeping the highest order terms,
\[ \delta t \approx \frac{\sqrt{\frac{k\epsilon}{N}} g \log \left( \frac{N}{k} \frac{1}{\epsilon} \right)}{\frac{k}{N} g^2 \log^2 \left( \frac{N}{k} \frac{1}{\epsilon} \right)} = \sqrt{\frac{N}{k}} \frac{1}{g \log \left( \frac{N}{k} \frac{1}{\epsilon} \right)}. \]
So the width of the success probability at height $1 - \epsilon$ is bounded by
\[ \Delta t = \Omega\left( \sqrt{\frac{N}{k}} \frac{1}{g \log \left( \frac{N}{k} \frac{1}{\epsilon} \right)} \right). \]
Note that in figure \ref{fig:prob_time_mashup}, we chose $g = O(\sqrt{N}/\log N)$ since $k$ was constant, and it resulted in constant runtimes and widths. So this bound seems tight, and it is further evidence that the runtime $t_*$ is closer to its lower bound.

To achieve this level of time-measurement precision in an atomic clock that utilizes entanglement, we need the number of clock ions to scale inversely with $\Delta t$. Also including the $\log N$ qubits to encode the $N$-dimensional Hilbert space, the total ``space'' requirement $S$ is
\[ S = O \left( R^{\sigma - 1/2} + \log N \right). \]
Then the total resource requirement when $\sigma \ge 1/2$ is
\[ \left( 1 + R^{1/2-\sigma} \log N \right) \lesssim ST \lesssim \left( R^{\sigma/2} + R^{1/2-\sigma/2} \log N \right). \]
This is minimized when $\sigma = 1/2$, yielding
\[ \log N \lesssim ST \lesssim R^{1/4} \log N. \]
The upper bound equals the cubic nonlinearity's total resource requirement. So the loglinear nonlinearity is at least as good as the cubic nonlinearity in reducing the time-space resources. Given the numerical results from figure \ref{fig:log_runtime_trend}, the actual total resources seem closer to the lower bound.

Of course, there must be additional resources such that the product of the space requirements and the square of the time requirements is lower bounded by $N$ \cite{Zalka1999}. If the physical system (\textit{e.g.}, a Bose liquid) has $N_0$ particles, then each particle can be at any of the $N$ vertices of the graph, which requires $\log N$ qubits for each particle. Then the ``space'' requirement $S$ is $N_0 \log N$ plus the number of clock ions to achieve the necessary time-measurement precision. That is,
\[ S = O\left( N_0 \log N + R^{\sigma - 1/2} \right) \]
for large $N$. Then
\[ R^{1-2\sigma} (N_0 \log N + R^{\sigma - 1/2}) \lesssim ST^2 \lesssim R^{1-\sigma} (N_0 \log N + R^{\sigma-1/2}). \]
Since this must be lower bounded by $N$,
\[ R^{1-2\sigma} (N_0 \log N + R^{\sigma - 1/2}) = \Omega(N). \]
When $\sigma \le 1/2$, this bound is satisfied regardless of $N_0$. When $\sigma > 1/2$, then
\[ N_0 = \Omega\left( \frac{N R^{2\sigma - 1}}{\log N} \right). \]
As $\sigma$ increases, this bound also increases. But there is no reason to increase $\sigma$ beyond $1/2$, at which $N_0 = \Omega( N \log N )$, because that gives the optimal product of space and time when ignoring $N_0$, and numerically gives constant runtime. So we've given a quantum information-theoretic bound for the number of particles needed for the logarithmic nonlinear Schr\"odinger equation to describe the physical system (\textit{e.g.}, the number of atoms in a Bose liquid), and to the best of our knowledge, it is the first such result.

\section{Conclusion}

Our results indicate that a host of physically realistic nonlinear quantum systems of the form \eqref{eq:NLSE} can be used to perform continuous-time computation faster than (linear) quantum computation. In particular, we've quantified this speedup by analyzing the quantum search problem, and the particular choice of nonlinearity gives rise to different runtimes, requires different levels of time-measurement precision, and necessitates a different number of particles for the nonlinearity to be an asymptotic description of the many-body quantum dynamics. 

Chapter 3, nearly in full, is a reprint of the material as it appears in ``Quantum Search with General Nonlinearities'' in Physical Review A 89, 012312 (2014). D.~A.~Meyer and T.~G.~Wong both contributed significantly to the work.


\chapter{Quantum Search on Strongly Regular Graphs}


\section{Introduction}

\begin{figure}
\begin{center}
	\includegraphics[width=1.9in]{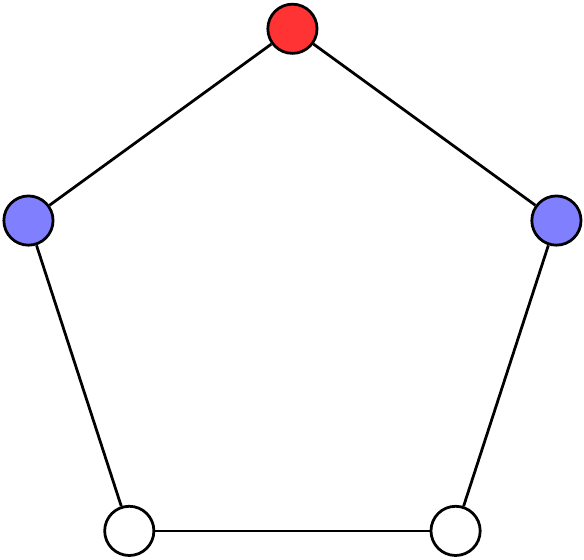}
	\includegraphics[width=1.9in]{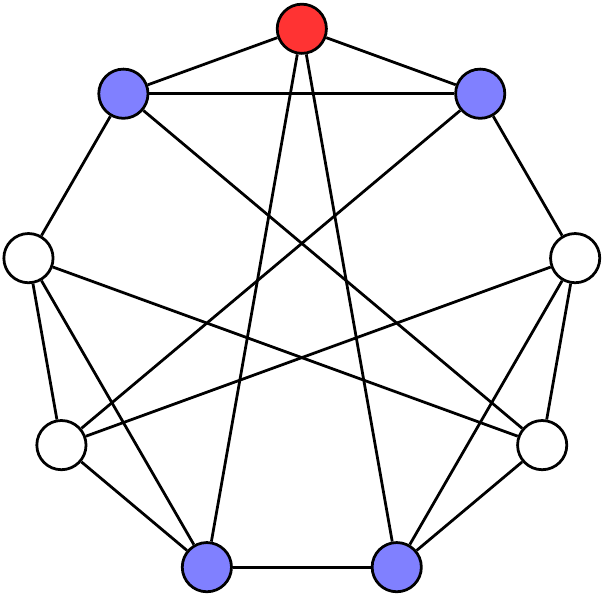}
	\includegraphics[width=1.9in]{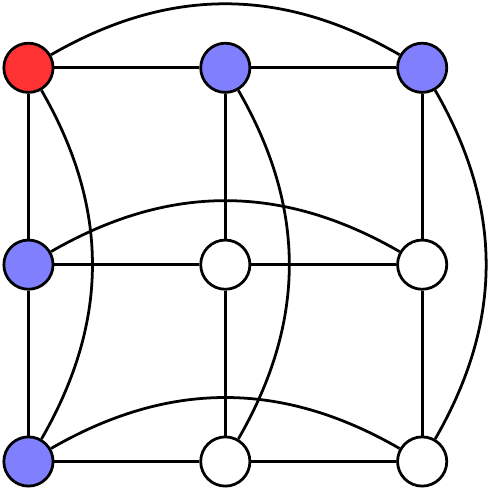}
	\caption[Some examples of strongly regular graphs. From left to right: the Paley graph with parameters (5,2,0,1), the Paley graph with parameters (9,4,1,2), and the square graph $L_2(3)$ with parameters (9,4,1,2).]{\label{fig:srgsmarked}Some examples of strongly regular graphs. From left to right: the Paley graph with parameters (5,2,0,1), the Paley graph with parameters (9,4,1,2), and the square graph $L_2(3)$ with parameters (9,4,1,2). Without loss of generality, a ``marked'' vertex is colored red, vertices adjacent to it are colored blue, and vertices not adjacent to it are colored white.}
\end{center}
\end{figure}

As explained in Chapter 1, the quantum search problem can be formulated as a quantum random walk on the complete graph of $N$ vertices, as shown in figure \ref{fig:complete}, where the randomly walking quantum particle is initially in an equal superposition over all $N$ vertices. Then the system evolves in a two-dimensional subspace spanned by the marked vertex and superposition of non-marked vertices. The next step in difficulty would be search on a graph where the system evolves in a three-dimensional subspace, namely that spanned by the marked vertex, the superposition of vertices adjacent to (or ``one away'' from) the marked vertex, and the superposition of vertices non-adjacent to (or ``two away'' from) the marked vertex. But this is precisely a strongly regular graph, examples of which are shown in figure~\ref{fig:srgsmarked} with the vertices of each respective subspace colored red, blue, and white. A strongly regular graph with parameters ($N$, $k$, $\lambda$, $\mu$) is a graph with $N$ vertices, each with $k$ neighbors, where adjacent vertices have $\lambda$ common neighbors and non-adjacent vertices have $\mu$ common neighbors.

As one might expect, only certain choices of parameters ($N$, $k$, $\lambda$, $\mu$) give rise to strongly regular graphs. One constraint is that the parameters satisfy
\begin{equation}
	k(k - \lambda - 1) = (N - k - 1)\mu,
	\label{eq:params}
\end{equation}
which is proved by counting the number of pairs of adjacent blue and white vertices. On the left hand side of Eq.~\ref{eq:params}, the marked red vertex has $k$ neighbors, so there are $k$ blue vertices. Each blue vertex has $k$ neighbors, one of which is the red marked vertex, and $\lambda$ of which are other blue vertices. So it is adjacent to $k - \lambda - 1$ white vertices. So the number of pairs of blue and white vertices that are adjacent to each other is $k(k - \lambda - 1)$. On the right hand side of Eq.~\ref{eq:params}, we count the number of pairs another way, beginning with the number of white vertices. There are $N$ total vertices in the graph, one of which is red and $k$ of which are blue. So there are $N - k - 1$ white vertices. Each of these white vertices has is adjacent to $\mu$ blue vertices. So there are $(N - k - 1)\mu$ pairs of blue and white vertices that are connected to each other. Equating these expressions gives Eq.~\ref{eq:params}. Note this is a necessary, but not sufficient, condition for a strongly regular graph to exist.

Equation \ref{eq:params} also implies that that $k$, the degree of the vertices, must be lower bounded by $\sqrt{N}$. That is,
\[ k^2 > k(k-\lambda-1) = (N-k-1)\mu. \]
Then
\begin{equation}
	k = \Omega(\sqrt{N}).
	\label{eq:kN}
\end{equation}

Additional constraints on the parameters ($N$, $k$, $\lambda$, $\mu$) divide strongly regular graphs into two types \cite{Janmark2013, Cameron1991}:
\begin{enumerate}
	\item	\emph{Type I graphs}, also called conference graphs, satisfy
		\[ (N-1)(\lambda - \mu) + 2k = 0, \]
		which means ($N$, $k$, $\lambda$, $\mu$) can be parameterized by
		\[ N = 4t + 1, \quad k = 2t, \quad \lambda = t - 1, \quad \text{and} \quad \mu = t. \]
		This parameterization reveals that
		\begin{equation}
			k = \Theta(N),
			\label{eq:type1k}
		\end{equation}
		which will be useful later. Furthermore, Type I graphs exist if and only if $N$ is the sum of two squares (one of the squares can be zero, so $N = 9 = 3^2 + 0^2$ is acceptable).
		
		A large number of Type I graphs are Paley graphs, where $N$ is congruent to $1\text{ mod }4$. The two smallest Paley graphs, where $N = 5$ and $N = 9$, are shown in figure \ref{fig:srgsmarked}.

	\item	\emph{Type II graphs} satisfy
		\[ \frac{(N-1)(\mu - \lambda) - 2k}{d} = N - 1 \mod 2 \]
		This condition is not sufficient for the existence of a strongly regular graph. That is, just because a set of parameters ($N$, $k$, $\lambda$, $\mu$) satisfies this equation does \emph{not} mean a strongly regular graph exists with such parameters. There are, however, some parameter families that do exist, three examples of which we now discuss.

		\emph{A. Square lattice graphs}, an example of which is shown in figure \ref{fig:srgsmarked}, can be pictured as a square lattice of $t^2$ vertices, where vertices are connected if and only if they are in the same row or column. They are denoted $L_2(t)$ according to the parameterization
		\[ N = t^2, \quad k = 2(t-1), \quad \lambda = t - 2, \quad \text{and} \quad \mu = 2. \]

		\emph{B. Latin square graphs} are similar to square lattice graphs, except each vertex is given a symbol that only appears once in each row and column. Then vertices with the same symbol are additionally connected. An example is given in figure \ref{fig:latinsquare}. They are denoted $L_3(t)$ according to the parameterization
		\[ N = t^2, \quad k = 3(t-1), \quad \lambda = t, \quad \text{and} \quad \mu = 6, \]
		with $t \ge 3$.

		\begin{figure}
		\begin{center}
			\includegraphics[width=2in]{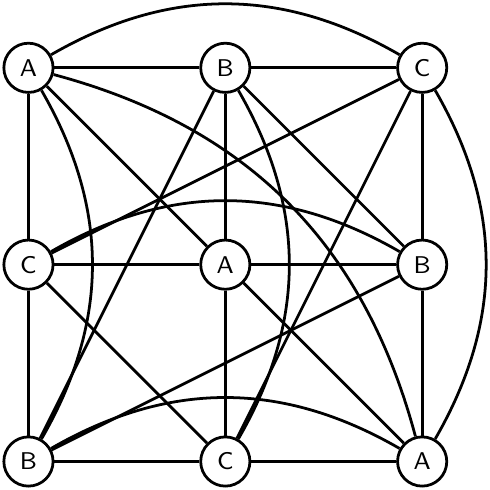}
			\caption{\label{fig:latinsquare}The Latin square graph $L_3(3)$ with parameters (9,6,3,6).}
		\end{center}
		\end{figure}

		\emph{C. Triangular graphs}, an example of which is shown in figure \ref{fig:triangular}, are denoted $T(t)$ and are parameterized by
		\[ N = \frac{1}{2} t (t-1), \quad k = 2(t-2), \quad \lambda = t - 2, \quad \text{and}\quad \mu = 4, \]
		with $t \ge 4$. They are constructed by labeling each vertex with a different unordered pair of different numbers in the set $\{1,2,\dots,t\}$. Two vertices are connected if they have a common number.

		\begin{figure}
		\begin{center}
			\includegraphics[width=2in]{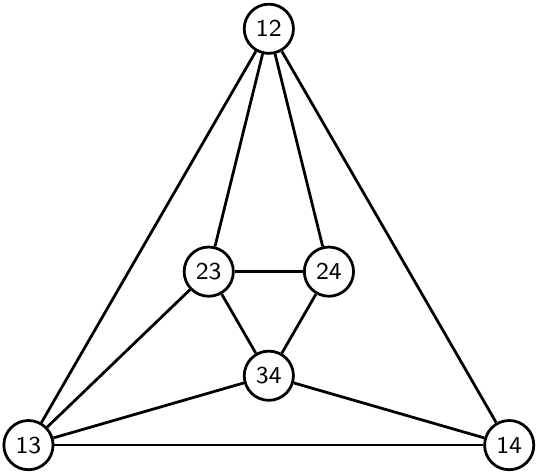}
			\caption{\label{fig:triangular}The triangular graph $T(4)$ with parameters (6,4,2,4).}
		\end{center}
		\end{figure}

		Note that each of these examples of existing parameter families have
		\begin{equation}
			k = \Theta(\sqrt{N}),
			\label{eq:type2k}
		\end{equation}
		meaning they reach the lower bound given by Eq.~\ref{eq:kN}.
\end{enumerate}

It seems likely that almost all strongly regular graphs are asymmetric, meaning their automorphism groups are trivial. While this has not been proved in general, it has been proved for Latin square graphs, which were introduced above and shown in figure \ref{fig:latinsquare} \cite{Babai1995}. Thus, they lack global symmetry, which is intuitively believed to be necessary for fast quantum search \cite{CG2004}. In this chapter, we show this intuition to be false, \textit{i.e.}, that a randomly walking quantum particle on strongly regular graphs optimally \cite{Zalka1999} solves the quantum search problem in $O(\sqrt{N})$ time for large $N$.


\section{Setup}

The vertices serve as a $N$-dimensional computational basis, which we label $\{\ket{0}, \ket{1}, \dots, \ket{N-1}\}$ and reduce to a three-dimensional subspace because there are three types of vertices: the red marked vertex, the $k$ blue vertices that are adjacent to the red marked vertex, and the $N - k - 1$ white vertices that are not adjacent to the red marked vertex. Let's call the basis states corresponding to these $\ket{w}$, $\ket{a}$ (for adjacent), and $\ket{b}$ (since we called the other basis state $\ket{a}$), respectively. That is,
\[ \ket{w} = \begin{pmatrix} 1 \\ 0 \\ 0 \end{pmatrix}\!\!, \quad \ket{a} = \frac{1}{\sqrt{k}} \sum_{(x,w) \in \mathcal{E}} \ket{x} = \begin{pmatrix} 0 \\ 1 \\ 0 \end{pmatrix}\!\!, \quad \ket{b} = \frac{1}{\sqrt{N-k-1}} \sum_{(x,w) \not\in \mathcal{E}} \ket{x} = \begin{pmatrix} 0 \\ 0 \\ 1 \end{pmatrix}\!\!. \]
The system begins in the equal superposition of all vertices, which we can write in the three-dimensional $\{\ket{w}, \ket{a}, \ket{b}\}$ basis:
\begin{align*}
	\ket{s} &= \frac{1}{\sqrt{N}} \sum_x \ket{x} \\
		&= \frac{1}{\sqrt{N}} \left( \ket{w} + \sum_{x \sim w} \ket{x} + \sum_{x \not\sim w} \ket{x} \right) \\
		&= \frac{1}{\sqrt{N}} \left( \ket{w} + \sqrt{k} \ket{a} + \sqrt{N-k-1} \ket{b} \right) \\
		&= \frac{1}{\sqrt{N}} \begin{pmatrix} 1 \\ \sqrt{k} \\ \sqrt{N-k-1} \end{pmatrix}.
\end{align*}
The system evolves by Schr\"odinger's equation with Hamiltonian
\[ H = -\gamma L - \ketbra{w}{w}, \]
where $\gamma$ is the amplitude per unit time of the randomly walking quantum particle transitioning from one vertex to another, and $L$ is the graph Laplacian which effects a quantum random walk on the graph \cite{CG2004}. More specifically, $L = A - D$, where $A_{ij} = 1$ if $(i,j) \in \mathcal{E}$ (and $0$ otherwise) is the adjacency matrix indicating which vertices are connected to one another, and $D_{ii} = \text{deg}(i)$ (and $0$ otherwise) is the degree matrix indicating how many neighbors each vertex has. In the case of strongly regular graphs, each vertex has degree $k$, so the degree matrix is a multiple of the identity matrix: $D = kI$. This is simply a rescaling of energy, so we can drop it without observable effects. Then the Hamiltonian is
\[ H = -\gamma A - \ketbra{w}{w}. \]

\begin{figure}
\begin{center}
	\includegraphics[width=2in]{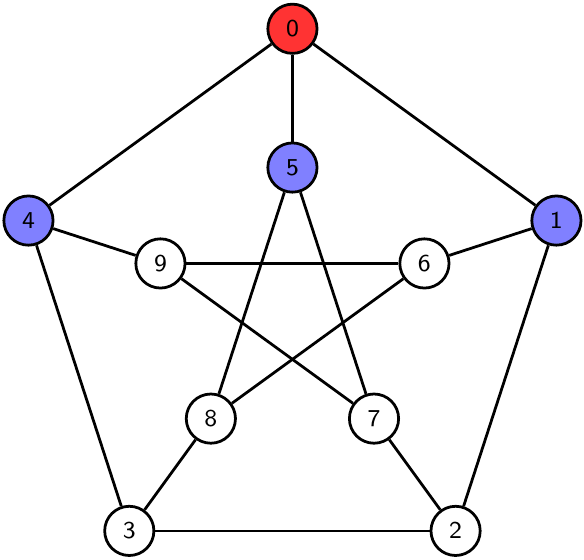}
	\caption{\label{fig:petersen}The Petersen graph, which is a strongly regular graph with parameters (10,3,0,1). Without loss of generality, the vertices have been numbered and the marked vertex was chosen to be $0$.}
\end{center}
\end{figure}

Now let's determine $H$ in the three-dimensional $\{\ket{w}, \ket{a}, \ket{b}\}$ basis. The $\ketbra{w}{w}$ term is simply a 3x3 matrix with a $1$ in the top-left corner and $0$'s everywhere else:
\[ \ketbra{w}{w} = \begin{pmatrix}
	1 & 0 & 0 \\
	0 & 0 & 0 \\
	0 & 0 & 0 \\
\end{pmatrix}. \]
The adjacency matrix $A$ can be a little tricky, so let's work it out explicitly for the Petersen graph, as shown in figure \ref{fig:petersen}, in its $(N = 10)$-dimensional computational basis. With the labeling in figure \ref{fig:petersen}, the adjacency matrix is
\[ A = \begin{pmatrix}
	0 & 1 & 0 & 0 & 1 & 1 & 0 & 0 & 0 & 0 \\ 
	1 & 0 & 1 & 0 & 0 & 0 & 1 & 0 & 0 & 0 \\ 
	0 & 1 & 0 & 1 & 0 & 0 & 0 & 1 & 0 & 0 \\ 
	0 & 0 & 1 & 0 & 1 & 0 & 0 & 0 & 1 & 0 \\ 
	1 & 0 & 0 & 1 & 0 & 0 & 0 & 0 & 0 & 1 \\ 
	1 & 0 & 0 & 0 & 0 & 0 & 0 & 1 & 1 & 0 \\ 
	0 & 1 & 0 & 0 & 0 & 0 & 0 & 0 & 1 & 1 \\ 
	0 & 0 & 1 & 0 & 0 & 1 & 0 & 0 & 0 & 1 \\ 
	0 & 0 & 0 & 1 & 0 & 1 & 1 & 0 & 0 & 0 \\ 
	0 & 0 & 0 & 0 & 1 & 0 & 1 & 1 & 0 & 0 \\
\end{pmatrix}. \]
The basis states $\{\ket{w}, \ket{a}, \ket{b}\}$ in the computational basis are
\[
	\ket{w} = \begin{pmatrix} 1 \\ 0 \\ 0 \\ 0 \\ 0 \\ 0 \\ 0 \\ 0 \\ 0 \\ 0 \end{pmatrix}, \quad
	\ket{a} = \frac{1}{\sqrt{3}} \begin{pmatrix} 0 \\ 1 \\ 0 \\ 0 \\ 1 \\ 1 \\ 0 \\ 0 \\ 0 \\ 0 \end{pmatrix}, \quad
	\ket{b} = \frac{1}{\sqrt{6}} \begin{pmatrix} 0 \\ 0 \\ 1 \\ 1 \\ 0 \\ 0 \\ 1 \\ 1 \\ 1 \\ 1 \end{pmatrix}.
\]
Then the adjacency matrix acting on each basis state is
\[ A \ket{w} = \begin{pmatrix} 0 \\ 1 \\ 0 \\ 0 \\ 1 \\ 1 \\ 0 \\ 0 \\ 0 \\ 0 \end{pmatrix} = \sqrt{3}\ket{a} \]
\[ A \ket{a} = \frac{1}{\sqrt{3}} \begin{pmatrix} 3 \\ 0 \\ 1 \\ 1 \\ 0 \\ 0 \\ 1 \\ 1 \\ 1 \\ 1 \end{pmatrix} = \frac{1}{\sqrt{3}}3\ket{w} + \frac{\sqrt{6}}{\sqrt{3}} \ket{b} \]
\[ A \ket{b} = \frac{1}{\sqrt{6}} \begin{pmatrix} 0 \\ 2 \\ 2 \\ 2 \\ 2 \\ 2 \\ 2 \\ 2 \\ 2 \\ 2 \end{pmatrix} = \frac{\sqrt{3}}{\sqrt{6}}2\ket{a} + 2\ket{b}. \]
So the adjacency matrix in the three-dimensional $\{\ket{w}, \ket{a}, \ket{b}\}$ basis is
\[ A = \begin{pmatrix}
	0 & \frac{1}{\sqrt{3}}3 & 0 \\
	\sqrt{3} & 0 & \frac{\sqrt{3}}{\sqrt{6}}2 \\
	0 & \frac{\sqrt{6}}{\sqrt{3}} & 2 \\
\end{pmatrix}. \]
This is symmetric, as expected.

Let's examine where these terms come from. The normalization factors of $\ket{w}$, $\ket{a}$, and $\ket{b}$ are $1$, $1/\sqrt{3}$, and $1/\sqrt{6}$, respectively. By the definition of the adjacency matrix, we can write it as a matrix of normalization conversions multiplied component-by-component with another matrix that describes the connection between vertices in $\ket{w}$, $\ket{a}$, and $\ket{b}$:
\[ A = \begin{pmatrix}
	1 & \frac{1}{\sqrt{3}} & \frac{1}{\sqrt{6}} \\
	\sqrt{3} & 1 & \frac{\sqrt{3}}{\sqrt{6}} \\
	\sqrt{6} & \frac{\sqrt{6}}{\sqrt{3}} & 1 \\
\end{pmatrix} *\hat{} \begin{pmatrix}
	0 & 3 & 0 \\
	1 & 0 & 2 \\
	0 & 1 & 2 \\
\end{pmatrix}, \]
where $*\hat{}$ indicates component-by-component multiplication (as in Matlab syntax). That is, the $(i,j)$-th entry of the first matrix is the normalization factor of the $j$th subspace basis vector divided by the $i$th (\textit{e.g.}, the entry at $(2,3)$ is the normalization factor of $\ket{b}$ divided by the normalization factor of $\ket{a}$, or $1/\sqrt{6}$ divided by $1/\sqrt{3}$), and the $(i,j)$-th entry of the second matrix is the number of vertices in the $j$th subspace basis vector that are adjacent to a single vertex in the $i$th subspace basis vector (\textit{e.g.}, the entry at $(2,3)$ is 2 because there are two white vertices connected to a single blue vertex). 

This example of the Petersen graph reveals what $A$ should be for a general strongly regular graph with parameters $(N,k,\lambda,\mu)$. First, recall that the normalization factors for $\ket{w}$, $\ket{a}$, and $\ket{b}$ are $1$, $1/\sqrt{k}$, and $1/\sqrt{N-k-1}$, respectively. Then the adjacency matrix is 
\[ A = \begin{pmatrix}
	1 & \frac{1}{\sqrt{k}} & \frac{1}{\sqrt{N-k-1}} \\
	\sqrt{k} & 1 & \frac{\sqrt{k}}{\sqrt{N-k-1}} \\
	\sqrt{N-k-1} & \frac{\sqrt{N-k-1}}{\sqrt{k}} & 1 \\
\end{pmatrix} *\hat{} \begin{pmatrix}
	\cdot & \cdot & \cdot \\
	\cdot & \cdot & \cdot \\
	\cdot & \cdot & \cdot \\
\end{pmatrix}. \]
Now let's determine the second matrix. First, there are $k$ blue vertices that go into the red vertex. So we have
\[ A = \begin{pmatrix}
	1 & \frac{1}{\sqrt{k}} & \frac{1}{\sqrt{N-k-1}} \\
	\sqrt{k} & 1 & \frac{\sqrt{k}}{\sqrt{N-k-1}} \\
	\sqrt{N-k-1} & \frac{\sqrt{N-k-1}}{\sqrt{k}} & 1 \\
\end{pmatrix} *\hat{} \begin{pmatrix}
	0 & k & 0 \\
	\cdot & \cdot & \cdot \\
	\cdot & \cdot & \cdot \\
\end{pmatrix}. \]
Next, there is one red vertex, $\lambda$ blue vertices, and $(k-\lambda-1)$ white vertices that go into a blue vertex (for a total of $k$, the degree of a vertex). So we have
\[ A = \begin{pmatrix}
	1 & \frac{1}{\sqrt{k}} & \frac{1}{\sqrt{N-k-1}} \\
	\sqrt{k} & 1 & \frac{\sqrt{k}}{\sqrt{N-k-1}} \\
	\sqrt{N-k-1} & \frac{\sqrt{N-k-1}}{\sqrt{k}} & 1 \\
\end{pmatrix} *\hat{} \begin{pmatrix}
	0 & k & 0 \\
	1 & \lambda & k-\lambda-1 \\
	\cdot & \cdot & \cdot \\
\end{pmatrix}. \]
Finally, there are $\mu$ blue vertices and $(k-\mu)$ white vertices that go into a white vertex, so
\[ A = \begin{pmatrix}
	1 & \frac{1}{\sqrt{k}} & \frac{1}{\sqrt{N-k-1}} \\
	\sqrt{k} & 1 & \frac{\sqrt{k}}{\sqrt{N-k-1}} \\
	\sqrt{N-k-1} & \frac{\sqrt{N-k-1}}{\sqrt{k}} & 1 \\
\end{pmatrix} *\hat{} \begin{pmatrix}
	0 & k & 0 \\
	1 & \lambda & k-\lambda-1 \\
	0 & \mu & k-\mu \\
\end{pmatrix}. \]
Multiplying these component-by-component,
\[ A = \begin{pmatrix}
	0 & \sqrt{k} & 0 \\
	\sqrt{k} & \lambda & \frac{\sqrt{k}}{\sqrt{N-k-1}}(k-\lambda-1) \\
	0 & \frac{\sqrt{N-k-1}}{\sqrt{k}}\mu & k-\mu \\
\end{pmatrix}. \]
The adjacency matrix must be symmetric. Let's check the two non-obvious terms:
\begin{align*}
	\frac{\sqrt{k}}{\sqrt{N-k-1}}(k-\lambda-1) &\stackrel{?}{=} \frac{\sqrt{N-k-1}}{\sqrt{k}}\mu \\
	k(k-\lambda-1) &\stackrel{?}{=} (N-k-1)\mu.
\end{align*}
But this is precisely Eq.~\ref{eq:params}, so they are equal. Furthermore, it says that both of these are equal to $\sqrt{\mu}\sqrt{k-\lambda-1}$. So we have
\[ A = \begin{pmatrix}
	0 & \sqrt{k} & 0 \\
	\sqrt{k} & \lambda & \sqrt{\mu}\sqrt{k-\lambda-1} \\
	0 & \sqrt{\mu}\sqrt{k-\lambda-1} & k-\mu \\
\end{pmatrix}. \]
Thus the Hamiltonian $H = -\gamma A - \ketbra{w}{w}$ in the three-dimensional $\{\ket{w}, \ket{a}, \ket{b}\}$ basis is
\begin{equation}
	H = -\gamma \begin{pmatrix}
		\frac{1}{\gamma} & \sqrt{k} & 0 \\
		\sqrt{k} & \lambda & \sqrt{\mu}\sqrt{k-\lambda-1} \\
		0 & \sqrt{\mu}\sqrt{k-\lambda-1} & k-\mu \\
	\end{pmatrix}.
	\label{eq:Hamiltonian}
\end{equation}

		
\section{Critical Gamma from Eigenstate Overlaps}

\begin{figure}
\begin{center}
	\includegraphics[width=3in]{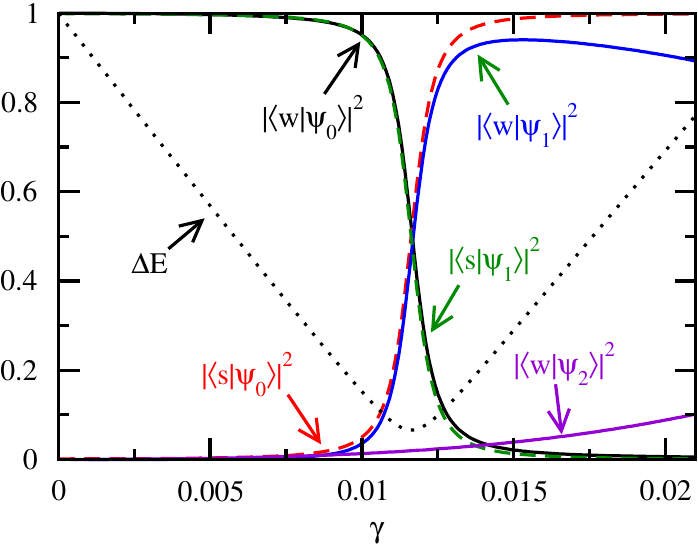}
	\caption[For the Latin square graph with parameters (900,87,30,6), overlaps of the eigenstates of the Hamiltonian in Eq.~\ref{eq:Hamiltonian} with $\ket{w}$ and $\ket{s}$ and the gap between the two lowest energy eigenvalues for various values of $\gamma$.]{\label{fig:overlaps}For the Latin square graph with parameters (900,87,30,6), overlaps of the eigenstates of the Hamiltonian in Eq.~\ref{eq:Hamiltonian} with $\ket{w}$ and $\ket{s}$ and the gap between the two lowest energy eigenvalues for various values of $\gamma$. Note that $|\braket{s}{\psi_2}|^2$ is near zero and not visible.}
\end{center}
\end{figure}

For the complete graph, the critical $\gamma$ caused the eigenstates of the Hamiltonian to be proportional to $\ket{s} \pm \ket{w}$, which caused the system to evolve from $\ket{s}$ to $\ket{w}$ in time $\pi\sqrt{N}/2$. We want the energy eigenstates to have the same form, but there are three of them $\ket{\psi_0}$, $\ket{\psi_1}$, and $\ket{\psi_2}$. As shown in figure \ref{fig:overlaps}, this isn't a problem---the projection of the highest energy eigenstate $\ket{\psi_2}$ onto $\ket{w}$ and $\ket{s}$ is small, so it doesn't contribute significantly to the evolution of the system. So only the ground and first excited states matter, and there is a point near the middle of the plot where they are roughly proportional to $\ket{s} \pm \ket{w}$. This corresponds to the critical value of $\gamma$.

So we expect to find a value of $\gamma$ that causes
\[ \ket{s} \pm \ket{w} = \left( \frac{1}{\sqrt{N}} \pm 1 \right) \ket{w} + \frac{\sqrt{k}}{\sqrt{N}} \ket{a} + \frac{\sqrt{N-k-1}}{\sqrt{N}} \ket{b} = \begin{pmatrix}
		\frac{1}{\sqrt{N}} \pm 1 \\
		\frac{\sqrt{k}}{\sqrt{N}} \\
		\frac{\sqrt{N-k-1}}{\sqrt{N}} \\
\end{pmatrix} \]
to be an energy eigenvector. So let's find the eigenvectors of $H$ and choose $\gamma$ so that they have the desired form. The characteristic equation for the eigenvalues $\chi$ (since $\lambda$ is already used as a parameter of the strongly regular graph) of $-H/\gamma$ is
\begin{align*}
      0 &= \det\left( \frac{-1}{\gamma}H - \chi I \right) \\
	&= \begin{vmatrix}
		\frac{1}{\gamma}-\chi & \sqrt{k} & 0 \\
		\sqrt{k} & \lambda-\chi & \sqrt{\mu}\sqrt{k-\lambda-1} \\
		0 & \sqrt{\mu}\sqrt{k-\lambda-1} & k-\mu-\chi \\
           \end{vmatrix} \\
	&= \left( \frac{1}{\gamma} - \chi \right) \left[ (\lambda-\chi)(k-\mu-\chi) - \mu (k-\lambda-1) \right] - \sqrt{k} \left[ \sqrt{k} (k-\mu-\chi) \right] \\
	&= -\chi^3 + \left(\frac{1}{\gamma}+k-\lambda-\mu\right)\chi^2 \\
	&\quad\quad + \left[ \frac{-1}{\gamma}(k+\lambda-\mu) - \lambda(k-\mu) + \mu(k-\lambda-1) + k \right] \chi \\
	&\quad\quad + \frac{1}{\gamma}\lambda(k-\mu) - \frac{1}{\gamma}\mu(k-\lambda-1) - k(k-\mu)
\end{align*}
This is a cubic equation of the form $a\chi^3 + b\chi^2 + c\chi + d = 0$. Solving it seems really messy, as does finding the eigenvectors. So we'll need another approach.


\section{Perturbation Theory in the $\{\ket{w}, \ket{a}, \ket{b}\}$ Basis}

Recall that the critical $\gamma$ for the complete graph can be found using degenerate perturbation theory. Let's try the same approach for strongly regular graphs. The leading order terms of the Hamiltonian in Eq.~\ref{eq:Hamiltonian} depend on the relative sizes of the parameters, so we break the problem into two cases: when $k$ scales as $N$, and when $k$ scales less than $N$ (but still bounded by $\sqrt{N}$ as in Eq.~\ref{eq:kN}).

\subsection{Case 1: $k = \Theta(N)$}
	
\begin{figure}
	\begin{center}
		\includegraphics[width=3in]{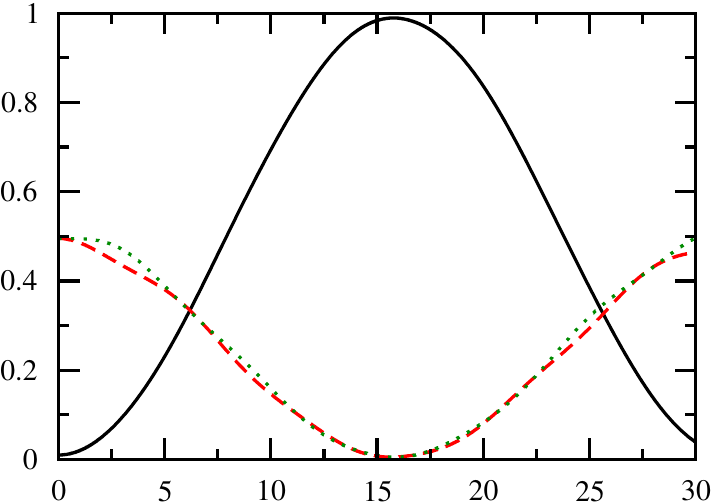}
		\caption[Evolution in each subspace for search on the Paley graph with parameters (101,50,24,25) with $\gamma_{c1}$ in Eq.~\ref{eq:gammac1}. The black solid curve is $\left| \braket{w}{\psi} \right|^2$, the red dashed curve is $\left| \braket{a}{\psi} \right|^2$, and the green dotted curve is $\left| \braket{b}{\psi} \right|^2$.]{\label{fig:EvoPaley}Evolution in each subspace for search on the Paley graph with parameters (101,50,24,25) with $\gamma_{c1}$ in Eq.~\ref{eq:gammac1}. The black solid curve is $\left| \braket{w}{\psi} \right|^2$, the red dashed curve is $\left| \braket{a}{\psi} \right|^2$, and the green dotted curve is $\left| \braket{b}{\psi} \right|^2$. As expected, the success probability nears $1$.}
	\end{center}
	\end{figure}

\begin{figure}
	\begin{center}
		\includegraphics[width=3in]{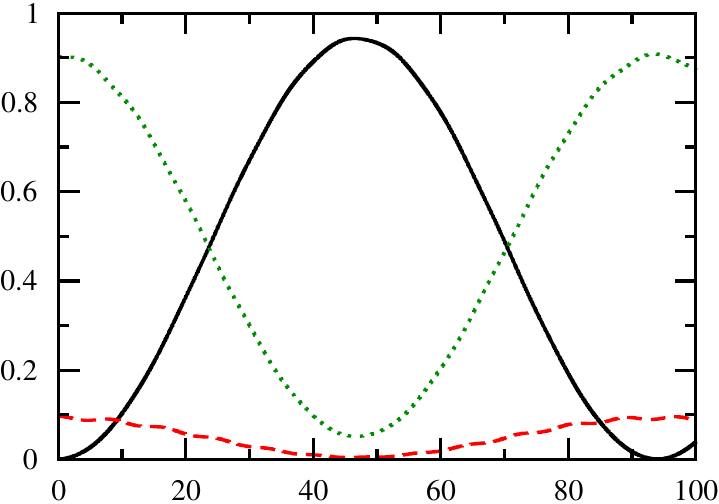}
		\caption[Evolution in each subspace for search on the Latin square graph with $m = 30$ (so parameters (900,87,30,60) with $\gamma_{c1}$ in Eq.~\ref{eq:gammac1}. The black solid curve is $\left| \braket{w}{\psi} \right|^2$, the red dashed curve is $\left| \braket{a}{\psi} \right|^2$, and the green dotted curve is $\left| \braket{b}{\psi} \right|^2$.]{\label{fig:EvoLatin}Evolution in each subspace for search on the Latin square graph with $m = 30$ (so parameters (900,87,30,60) with $\gamma_{c1}$ in Eq.~\ref{eq:gammac1}. The black solid curve is $\left| \braket{w}{\psi} \right|^2$, the red dashed curve is $\left| \braket{a}{\psi} \right|^2$, and the green dotted curve is $\left| \braket{b}{\psi} \right|^2$. As expected, the success probability doesn't reach 1.}
	\end{center}
	\end{figure}

	When $k$ scales the same as $N$, the leading order terms in the Hamiltonian from Eq.~\ref{eq:Hamiltonian} are
	\[ H^{(0)} = -\gamma \begin{pmatrix}
		\frac{1}{\gamma} & 0 & 0 \\
		0 & \lambda & \sqrt{\mu}\sqrt{k-\lambda-1} \\
		0 & \sqrt{\mu}\sqrt{k-\lambda-1} & k-\mu \\
	\end{pmatrix}. \]
	Clearly, $\ket{w}$ is an eigenstate of this with eigenvalue $-1$. It's straightforward to show that
	\[ \ket{r} = \frac{1}{\sqrt{N-1}} \sum_{x \not\sim w} \ket{x} = \frac{1}{\sqrt{N-1}} \left( \sqrt{k}\ket{a} + \sqrt{N-k-1}\ket{b} \right), \]
	which is approximately $\ket{s}$, is also an eigenvector of $H^{(0)}$, but with eigenvalue $-\gamma k$:
	\begin{align*}
		H^{(0)} \ket{r} 
			&= -\gamma \begin{pmatrix}
				\frac{1}{\gamma} & 0 & 0 \\
				0 & \lambda & \sqrt{\mu}\sqrt{k-\lambda-1} \\
				0 & \sqrt{\mu}\sqrt{k-\lambda-1} & k-\mu \\
			\end{pmatrix} \frac{1}{\sqrt{N-1}} \begin{pmatrix}
				0 \\
				\sqrt{k} \\
				\sqrt{N-k-1}
			\end{pmatrix} \\
			&= \frac{-\gamma}{\sqrt{N-1}} \begin{pmatrix}
				0 \\
				\lambda \sqrt{k} + \sqrt{\mu}\sqrt{k-\lambda-1}\sqrt{N-k-1} \\
				\sqrt{k}\sqrt{\mu}\sqrt{k-\lambda-1} + (k-\mu)\sqrt{N-k-1}
			\end{pmatrix} \\
			&= \frac{-\gamma}{\sqrt{N-1}} \begin{pmatrix}
				0 \\
				\lambda \sqrt{k} + \sqrt{k}(k-\lambda-1) \\
				\mu\sqrt{N-k-1} + (k-\mu)\sqrt{N-k-1}
			\end{pmatrix} \\
			&= \frac{-\gamma}{\sqrt{N-1}} \begin{pmatrix}
				0 \\
				\sqrt{k}(k-1) \\
				k\sqrt{N-k-1}
			\end{pmatrix} \\
			&\approx -\gamma k \frac{1}{\sqrt{N-1}} \begin{pmatrix}
				0 \\
				\sqrt{k} \\
				\sqrt{N-k-1}
			\end{pmatrix} \\
			&= -\gamma k \ket{r},
	\end{align*}
	where we assumed that $N$ is sufficiently large such that $k \approx k-1$. We want $\ket{w}$ and $\ket{r}$ to have the same eigenvalue so that the perturbation will cause the eigenstates of the full Hamiltonian to be a linear combination of $\ket{w}$ and $\ket{r}$. So we want $-\gamma k = -1$, or
	\begin{equation}
		\gamma_{c1} = \frac{1}{k}.
		\label{eq:gammac1}
	\end{equation}
	So with this value of $\gamma$, the success probability should go to $1$ for large $N$ when $k$ scales larger than $\sqrt{N}$. This is verified in figure \ref{fig:EvoPaley} for a Type I graph. When $k = \Theta(\sqrt{N})$, as in our examples of Type II graphs, we expect the success probability to \emph{not} reach 1, and this is verified in figure \ref{fig:EvoLatin}.
	
	Finally, the third eigenvector of $H^{(0)}$ is
	\[ \ket{e_3} = \frac{1}{\sqrt{N-1}} \left( \sqrt{N-k-1} \ket{a} - \sqrt{k} \ket{b} \right), \]
	which is necessarily orthogonal to both $\ket{w}$ and $\ket{r}$ and has eigenvalue $-(\lambda-\mu)/k$:
	\begin{align*}
		H^{(0)} \ket{e_3} 
			&= -\gamma \begin{pmatrix}
				\frac{1}{\gamma} & 0 & 0 \\
				0 & \lambda & \sqrt{\mu}\sqrt{k-\lambda-1} \\
				0 & \sqrt{\mu}\sqrt{k-\lambda-1} & k-\mu \\
			\end{pmatrix} \frac{1}{\sqrt{N-1}} \begin{pmatrix}
				0 \\
				\sqrt{N-k-1} \\
				-\sqrt{k}
			\end{pmatrix} \\
			&= \frac{-\gamma}{\sqrt{N-1}} \begin{pmatrix}
				0 \\
				\lambda \sqrt{N-k-1} - \sqrt{k}\sqrt{\mu}\sqrt{k-\lambda-1} \\
				\sqrt{\mu}\sqrt{k-\lambda-1}\sqrt{N-k-1} - \sqrt{k}(k-\mu)
			\end{pmatrix} \\
			&= \frac{-\gamma}{\sqrt{N-1}} \begin{pmatrix}
				0 \\
				\lambda \sqrt{N-k-1} - \mu\sqrt{N-k-1} \\
				\sqrt{k}(k-\lambda-1) - \sqrt{k}(k-\mu)
			\end{pmatrix} \\
			&= \frac{-\gamma}{\sqrt{N-1}} \begin{pmatrix}
				0 \\
				(\lambda - \mu) \sqrt{N-k-1} \\
				(\lambda - \mu + 1) (-\sqrt{k})
			\end{pmatrix} \\
			&\approx -\gamma (\lambda-\mu) \frac{1}{\sqrt{N-1}} \begin{pmatrix}
				0 \\
				\sqrt{N-k-1} \\
				-\sqrt{k}
			\end{pmatrix} \\
			&= -\gamma (\lambda-\mu) \ket{e_3}.
	\end{align*}

	Since $\ket{w}$ and $\ket{r}$ are degenerate eigenvectors of $H^{(0)}$, degenerate perturbation theory says that the eigenstates of the perturbed system are linear combinations of them:
	\[ \ket{\psi_\pm} = \alpha_w \ket{w} + \alpha_r \ket{r}. \]
	The coefficients $\alpha_w$ and $\alpha_r$ can be found by solving the eigenvalue problem
	\[ \begin{pmatrix} H_{ww} & H_{wr} \\ H_{rw} & H_{rr} \end{pmatrix} \begin{pmatrix} \alpha_w \\ \alpha_r \end{pmatrix} = E_\pm \begin{pmatrix} \alpha_w \\ \alpha_r \end{pmatrix}, \]
	where $H_{wr} = \langle w | H^{(0)} + H^{(1)} | r \rangle$ and
	\[ H^{(1)} = -\gamma \begin{pmatrix}
		0 & \sqrt{k} & 0 \\
		\sqrt{k} & 0 & 0 \\
		0 & 0 & 0 \\
	\end{pmatrix}. \]
	Evaluating the matrix components, we get
	\[ \begin{pmatrix} -1 & \frac{-\gamma k}{\sqrt{N-1}} \\ \frac{-\gamma k}{\sqrt{N-1}} & -1 \end{pmatrix} \begin{pmatrix} \alpha_w \\ \alpha_r \end{pmatrix} = E_\pm \begin{pmatrix} \alpha_w \\ \alpha_r \end{pmatrix}. \]
	Since $\gamma = \gamma_{c1} = 1/k$, this is
	\[ \begin{pmatrix} -1 & \frac{-1}{\sqrt{N-1}} \\ \frac{-1}{\sqrt{N-1}} & -1 \end{pmatrix} \begin{pmatrix} \alpha_w \\ \alpha_r \end{pmatrix} = E_\pm \begin{pmatrix} \alpha_w \\ \alpha_r \end{pmatrix}. \]
	Solving this eigenvalue problem, we get eigenvectors
	\[ \frac{1}{\sqrt{2}} \begin{pmatrix} -1 \\ 1 \end{pmatrix} \text{ with eigenvalue } E_+ = -1 + \frac{1}{\sqrt{N-1}} \]
	\[ \frac{1}{\sqrt{2}} \begin{pmatrix} 1 \\ 1 \end{pmatrix} \text{ with eigenvalue } E_- = -1 - \frac{1}{\sqrt{N-1}} \]
	Then the eigenstates of $H$ are
	\[ \ket{\psi_\pm} = \frac{1}{\sqrt{2}} \left( \mp \ket{w} + \ket{r} \right) \]
	with eigenvalues
	\[ E_\pm = -1 \pm \frac{1}{\sqrt{N-1}}. \]
	Note that the energy gap is $\Delta E = \frac{2}{\sqrt{N-1}}$. Since $\ket{r} \approx \ket{s}$, the system evolves from $\ket{s}$ to nearly $\ket{w}$ in time $t_* = \pi / \Delta E = \pi \sqrt{N-1} / 2 \approx \pi \sqrt{N} / 2$.

\subsection{Case 2: $k = o(N)$}
		
	When $k$ scales less than $N$, the leading order terms in the Hamiltonian from Eq.~\ref{eq:Hamiltonian} are
	\[ H^{(0)} = -\gamma \begin{pmatrix}
		\frac{1}{\gamma} & 0 & 0 \\
		0 & \lambda & 0 \\
		0 & 0 & k-\mu \\
	\end{pmatrix}. \]
	Clearly, the eigenstates of this are $\ket{w}$, $\ket{a}$, and $\ket{b}$ with corresponding eigenvalues $-1$, $-\gamma \lambda$, and $-\gamma(k-\mu)$. Although $\ket{r} \approx \ket{s}$ isn't an eigenstate, from figure \ref{fig:EvoLatin}, the system is hardly in $\ket{a}$ for the Latin square graph, so it might be sufficient to use $\ket{b}$ instead of $\ket{r}$. Doing this, we want $\ket{w}$ and $\ket{b}$ to have degenerate eigenvalues, or $-\gamma(k-\mu) = -1$, which implies
	\begin{equation}
		\gamma_{c2}' = \frac{1}{k-\mu},
		\label{eq:gammac2prime}
	\end{equation}
	where the prime denotes that this is different from another (better) critical value of $\gamma$ for the second case that we derive later.
	
\begin{figure}
	\begin{center}
		\includegraphics{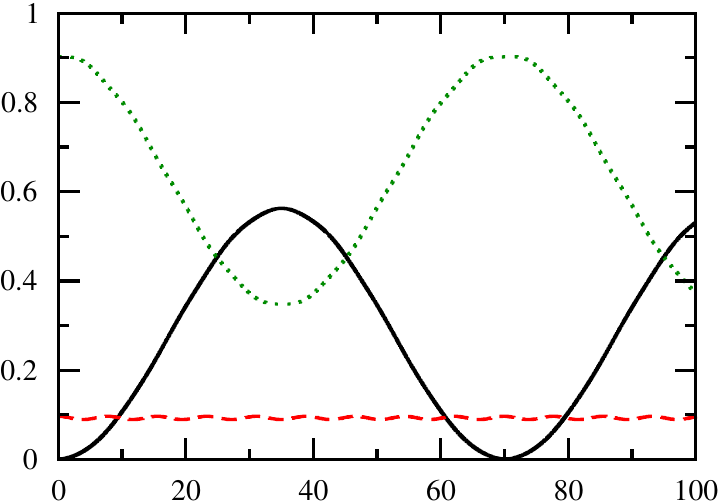}
		\caption[Evolution in each subspace for search on the Latin square graph with $m = 30$ (so parameters (900,87,30,60) with $\gamma_{c2}'$ in Eq.~\ref{eq:gammac2prime}. The black solid curve is $\left| \braket{w}{\psi} \right|^2$, the red dashed curve is $\left| \braket{a}{\psi} \right|^2$, and the green dotted curve is $\left| \braket{b}{\psi} \right|^2$.]{\label{fig:EvoLatinGammac2Prime}Evolution in each subspace for search on the Latin square graph with $m = 30$ (so parameters (900,87,30,60) with $\gamma_{c2}'$ in Eq.~\ref{eq:gammac2prime}. The black solid curve is $\left| \braket{w}{\psi} \right|^2$, the red dashed curve is $\left| \braket{a}{\psi} \right|^2$, and the green dotted curve is $\left| \braket{b}{\psi} \right|^2$. As expected, the success probability doesn't reach 1.}
	\end{center}
	\end{figure}

	Figure \ref{fig:EvoLatinGammac2Prime} shows the evolution of search on a Latin square graph. It reveals that $\gamma_{c2}'$ is worse than $\gamma_{c1} = 1/k$, even though the latter was already suboptimal. This is expected, however, since more terms were dropped in the leading-order Hamiltonian $H^{(0)}$ used to derive $\gamma_{c2}'$ than $\gamma_{c1}$. It is not apparent which terms to drop in the Hamiltonian in this $\{ \ket{w}, \ket{a}, \ket{b} \}$ basis.

\section{Perturbation Theory in the $\{\ket{w}, \ket{r}, \ket{e_3}\}$ Basis}

In the last section, we used perturbation theory in the $\{\ket{w}, \ket{a}, \ket{b}\}$ basis. When $k = \Theta(N)$ it was clear which terms were best to drop in the leading-order Hamiltonian $H^{(0)}$. When $k = o(N)$, however, too many terms were dropped. In order to drop fewer terms, we switch from the $\{\ket{w}, \ket{a}, \ket{b}\}$ basis we've been using to the basis $\{\ket{w}, \ket{r}, \ket{e_3}\}$. We can transform the Hamiltonian Eq.~\ref{eq:Hamiltonian} to this basis by multiplying $T^{-1} H T$, where
\[ T = \begin{pmatrix} \ket{w} & \ket{r} & \ket{e_3} \end{pmatrix} = \begin{pmatrix}
	1 & 0 & 0 \\
	0 & \frac{\sqrt{k}}{\sqrt{N-1}} & \frac{\sqrt{N-k-1}}{\sqrt{N-1}} \\
	0 & \frac{\sqrt{N-k-1}}{\sqrt{N-1}} & -\frac{\sqrt{k}}{\sqrt{N-1}}
\end{pmatrix}. \]
Note that $T^{-1} = T^{\top} = T$, and $T$ is a reflection in the $(\ket{a},\ket{b})$-plane about the line through the origin in the direction of the vector $\langle \sqrt{1-\xi}, \sqrt{1+\xi} \rangle$, where $\xi = \sqrt{k/(N-1)}$.
Multiplying $T^{-1} H T$, the Hamiltonian in the new basis is
\begin{equation}
	H' = - \gamma \begin{pmatrix}
		\frac{1}{\gamma} & \frac{k}{\sqrt{N-1}} & \frac{\sqrt{k}\sqrt{N-k-1}}{\sqrt{N-1}} \\
		\frac{k}{\sqrt{N-1}} & \frac{k(N-2)}{N-1} & \frac{-\sqrt{k}\sqrt{N-k-1}}{N-1} \\
		\frac{\sqrt{k}\sqrt{N-k-1}}{\sqrt{N-1}} & \frac{-\sqrt{k}\sqrt{N-k-1}}{N-1} & \frac{(\lambda-\mu)(N-1) + k}{N-1} \\
	\end{pmatrix}.
	\label{eq:HamiltonianPrime}
\end{equation}
As before we break the problem into two cases: when $k$ scales as $N$, and when $k$ scales less than $N$.

\subsection{Case 1: $k = \Theta(N)$}

	Although we've already solved this case, we can replicate the result in the $\{\ket{w}, \ket{r}, \ket{e_3}\}$ basis. The leading-order terms in the Hamiltonian in Eq.~\ref{eq:HamiltonianPrime} are
	\[ H^{(0)} = -\gamma \begin{pmatrix}
		\frac{1}{\gamma} & 0 & 0 \\
		0 & k & 0 \\
		0 & 0 & \lambda - \mu \\
	\end{pmatrix}. \]
	Clearly, the eigenvectors of this are $\ket{w}$, $\ket{r}$, and $\ket{e_3}$ with corresponding eigenvalues $-1$, $-\gamma k$, and $-\gamma(\lambda - \mu)$, which is consistent with what we found in the $\{\ket{w}, \ket{a}, \ket{b}\}$ basis. We want $\ket{w}$ and $\ket{r}$ to be degenerate so that the perturbation causes the eigenstates to be linear combinations of $\ket{w}$ and $\ket{r}$. The value of $\gamma$ that does this is
	\[ \gamma_{c1} = \frac{1}{k}, \]
	which is the same result as Eq.~\ref{eq:gammac1}.

\subsection{Case 2: $k = o(N)$}

	When $k$ scales less than $N$, which includes Latin square graphs where $k = \Theta(\sqrt{N})$, the leading-order terms in the Hamiltonian in Eq.~\ref{eq:HamiltonianPrime} are
	\[ H^{(0)} = -\gamma \begin{pmatrix}
		\frac{1}{\gamma} & 0 & \sqrt{k} \\
		0 & k & 0 \\
		\sqrt{k} & 0 & \lambda-\mu \\
	\end{pmatrix}. \]
	It's clear that $\ket{r}$ is an eigenvector of $H^{(0)}$ with eigenvalue $-\gamma k$. Then the two other eigenvectors have the form
	\[ \begin{pmatrix} c_1 \\ 0 \\ c_3 \end{pmatrix}. \]
	We want one of these to have the same eigenvalue $-\gamma k$ so that $H^{(0)}$ is degenerate:
	\[ H^{(0)} \begin{pmatrix} c_1 \\ 0 \\ c_3 \end{pmatrix} = -\gamma k \begin{pmatrix} c_1 \\ 0 \\ c_3 \end{pmatrix}. \]
	Solving this in the two-dimensional subspace:
	\[ \begin{pmatrix} \frac{1}{\gamma} & \sqrt{k} \\ \sqrt{k} & \lambda-\mu \end{pmatrix} \begin{pmatrix} c_1 \\ c_3 \end{pmatrix} = k \begin{pmatrix} c_1 \\ c_3 \end{pmatrix} \]
	\[ \det \begin{pmatrix} \frac{1}{\gamma} - k & \sqrt{k} \\ \sqrt{k} & \lambda-\mu -k \end{pmatrix} = 0 \]
	\[ \left( \frac{1}{\gamma} - k \right) \left( \lambda - \mu - k \right) - k = 0 \]
	\begin{align*}
		\gamma
			&= \frac{1}{k} \left( 1 - \frac{1}{k - \lambda + \mu} \right)^{-1} \\
			&= \frac{1}{k} \frac{k - \lambda + \mu}{k - \lambda + \mu - 1} \\
			&= \frac{1}{k} \left( 1 + \frac{1}{k - \lambda - 1 + \mu} \right) \\
			&= \frac{1}{k} \left( 1 + \frac{1}{(N-k-1)\frac{\mu}{k}+ \mu} \right) \\
			&= \frac{1}{k} \left( 1 + \frac{k}{(N-1)\mu} \right)
	\end{align*}
	This is the critical $\gamma$ when $k = o(N)$:
	\begin{equation}
		\gamma_{c2} = \frac{1}{k} + \frac{1}{(N-1)\mu}.
		\label{eq:gammac2}
	\end{equation}
	Finding the corresponding eigenvector is straightforward:
	\[ -\gamma \begin{pmatrix}
		\frac{1}{\gamma} & 0 & \sqrt{k} \\
		0 & k & 0 \\
		\sqrt{k} & 0 & \lambda-\mu \\
	\end{pmatrix} \begin{pmatrix} c_1 \\ 0 \\ c_3 \end{pmatrix} 
	= -\gamma k \begin{pmatrix} c_1 \\ 0 \\ c_3 \end{pmatrix} \]
	\[ \begin{pmatrix}
		\frac{1}{\gamma} - k & 0 & \sqrt{k} \\
		0 & 0 & 0 \\
		\sqrt{k} & 0 & \lambda-\mu-k \\
	\end{pmatrix} \begin{pmatrix} c_1 \\ 0 \\ c_3 \end{pmatrix} 
	= \begin{pmatrix} 0 \\ 0 \\ 0 \end{pmatrix} \]
	Using the third line (or equivalently the first line),
	\[ \sqrt{k} c_1 + (\lambda - \mu - k)c_3 = 0 \]
	\[ c_1 = \frac{k - \lambda + \mu}{\sqrt{k}} c_3 \]
	Normalization requires $c_1^2 + c_3^2 = 1$:
	\[ \frac{(k - \lambda + \mu)^2}{k} c_3^2 + c_3^2 = 1 \]
	\[ c_3 = \left( 1 + \frac{(k - \lambda + \mu)^2}{k} \right)^{-1/2} \]
	So the eigenvector is
	\[ \ket{c} = \begin{pmatrix} c_1 \\ 0 \\ c_3 \end{pmatrix} = \left( 1 + \frac{(k - \lambda + \mu)^2}{k} \right)^{-1/2} \begin{pmatrix} \frac{k - \lambda + \mu}{\sqrt{k}} \\ 0 \\ 1 \end{pmatrix}. \]
	Note that the third eigenvector of $H^{(0)}$ is
	\[ \left( 1 + \frac{k}{(k - \lambda + \mu)^2} \right)^{-1/2} \begin{pmatrix} \frac{-\sqrt{k}}{k - \lambda + \mu} \\ 0 \\ 1 \end{pmatrix} \]
	with corresponding eigenvalue
	\[ \frac{(\lambda-\mu)^2 + k(1-\lambda+\mu)}{k(-1+k-\lambda+\mu)}. \]

	The perturbation
	\[ H^{(1)} = -\gamma \begin{pmatrix}
		0 & \frac{k}{\sqrt{N}} & 0 \\
		\frac{k}{\sqrt{N}} & 0 & -\frac{k}{\sqrt{N}} \\
		0 & -\frac{k}{\sqrt{N}} & 0 \\
	\end{pmatrix} \]
	causes the eigenstates of $H^{(0)} + H^{(1)}$ to be a linear combination of $\ket{r}$ and $\ket{c}$:
	\[ \ket{\psi_\pm} = \alpha_r \ket{r} + \alpha_c \ket{c} \]
	To find $\alpha_r$ and $\alpha_c$, we solve the eigenvalue problem
	\[ \begin{pmatrix} H_{rr} & H_{rc} \\ H_{cr} & H_{cc} \end{pmatrix} \begin{pmatrix} \alpha_w \\ \alpha_r \end{pmatrix} = E_\pm \begin{pmatrix} \alpha_w \\ \alpha_r \end{pmatrix}, \]
	where $H_{rc} = \langle r | H^{(0)} + H^{(1)} | c \rangle$, etc. These terms are straightforward to calculate. We get
	\[ \begin{pmatrix} -\gamma k & -\gamma A \frac{N-1}{\sqrt{kN}} \mu \\ -\gamma A \frac{N-1}{\sqrt{kN}} \mu & -\gamma k \end{pmatrix} \begin{pmatrix} \alpha_w \\ \alpha_r \end{pmatrix} = E_\pm \begin{pmatrix} \alpha_w \\ \alpha_r \end{pmatrix}, \]
	where $A = \left( 1 + (k - \lambda + \mu)^2/k \right)^{-1/2}$ is the normalization constant of $\ket{c}$, and we used Eq.~\ref{eq:params} for the off-diagonal terms. For large $N$, this is
	\[ \begin{pmatrix} -\gamma k & -\gamma A \sqrt{\frac{N}{k}} \mu \\ -\gamma A \sqrt{\frac{N}{k}} \mu & -\gamma k \end{pmatrix} \begin{pmatrix} \alpha_w \\ \alpha_r \end{pmatrix} = E_\pm \begin{pmatrix} \alpha_w \\ \alpha_r \end{pmatrix}. \]
	Solving this, we get eigenvectors
	\[ \frac{1}{\sqrt{2}} \begin{pmatrix} 1 \\ -1 \end{pmatrix} \text{ with eigenvalue } E_+ = -\gamma k - \gamma A \sqrt{\frac{N}{k}} \mu \]
	\[ \frac{1}{\sqrt{2}} \begin{pmatrix} 1 \\ 1 \end{pmatrix} \text{ with eigenvalue } E_- = -\gamma k + \gamma A \sqrt{\frac{N}{k}} \mu \]
	Then the eigenstates of $H^{(0)} + H^{(1)}$ are
	\[ \ket{\psi_\pm} = \frac{1}{\sqrt{2}} \left( \ket{r} \mp \ket{c} \right) \]
	with eigenvalues
	\[ E_\pm = -\gamma k \pm \gamma A \sqrt{\frac{N}{k}} \mu . \]

\begin{figure}
	\begin{center}
		\includegraphics{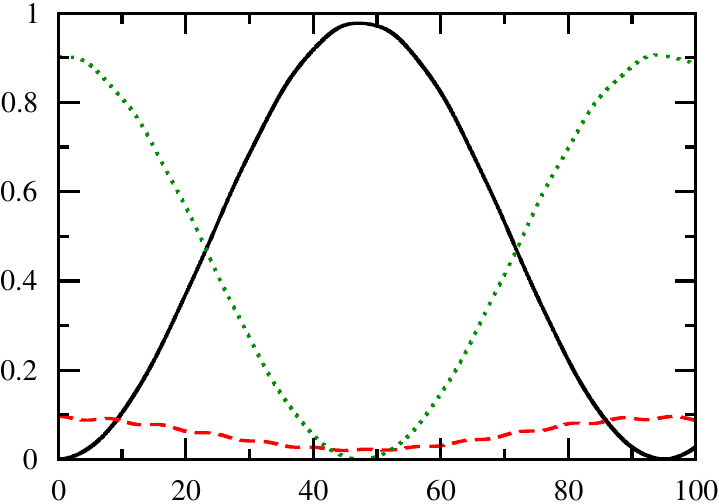}
		\caption[Evolution in each subspace for search on the Latin square graph with $m = 30$ (so parameters (900,87,30,60) with $\gamma_{c2}$ in Eq.~\ref{eq:gammac2}. The black solid curve is $\left| \braket{w}{\psi} \right|^2$, the red dashed curve is $\left| \braket{a}{\psi} \right|^2$, and the green dotted curve is $\left| \braket{b}{\psi} \right|^2$.]{\label{fig:EvoLatinGammac2}Evolution in each subspace for search on the Latin square graph with $m = 30$ (so parameters (900,87,30,60) with $\gamma_{c2}$ in Eq.~\ref{eq:gammac2}. The black solid curve is $\left| \braket{w}{\psi} \right|^2$, the red dashed curve is $\left| \braket{a}{\psi} \right|^2$, and the green dotted curve is $\left| \braket{b}{\psi} \right|^2$. As expected, the success probability nears 1.}
	\end{center}
	\end{figure}

\begin{figure}
	\begin{center}
		\includegraphics{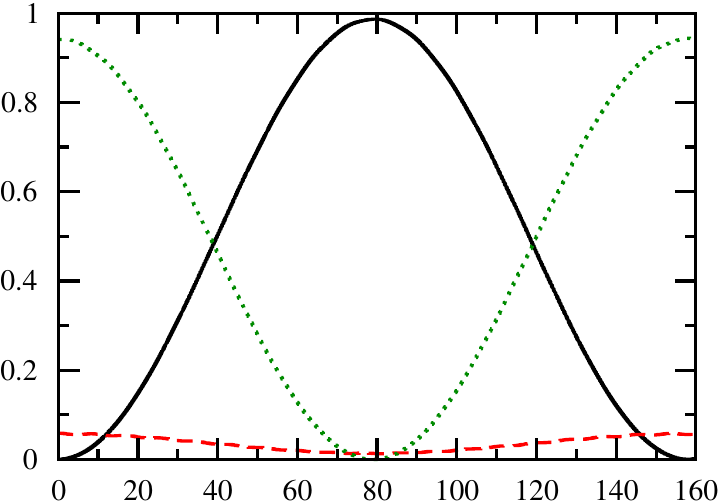}
		\caption[Evolution in each subspace for search on the Latin square graph with $m = 50$ (so parameters (2500,147,50,6) with $\gamma_{c2}$ in Eq.~\ref{eq:gammac2}. The black solid curve is $\left| \braket{w}{\psi} \right|^2$, the red dashed curve is $\left| \braket{a}{\psi} \right|^2$, and the green dotted curve is $\left| \braket{b}{\psi} \right|^2$.]{\label{fig:EvoLatinBigGammac2}Evolution in each subspace for search on the Latin square graph with $m = 50$ (so parameters (2500,147,50,6) with $\gamma_{c2}$ in Eq.~\ref{eq:gammac2}. The black solid curve is $\left| \braket{w}{\psi} \right|^2$, the red dashed curve is $\left| \braket{a}{\psi} \right|^2$, and the green dotted curve is $\left| \braket{b}{\psi} \right|^2$. As expected, the success probability nears 1.}
	\end{center}
	\end{figure}

	Now let's find the success probability as a function of time.
	The initial state of the system is $\ket{s}$, which is close to $\ket{r}$. So the evolution of the system is approximately spanned by these two eigenstates:
	\[ \ket{\psi(t)} = e^{-iH't} \ket{s}, \]
	where $H' = H^{(0)} + H^{(1)}$ is the approximate Hamiltonian. That is, we're ignoring
	\[ H^{(2)} = -\gamma \begin{pmatrix}
		0 & 0 & 0 \\
		0 & 0 & 0 \\
		0 & 0 & \frac{k}{N} \\
	\end{pmatrix}. \]
	Then the state of the system approximately evolves in the subspace spanned by $\ket{\psi_\pm}$:
	\[ \ket{\psi(t)} \approx e^{-iE_+t} \ket{\psi_+} \braket{\psi_+}{s} + e^{-iE_-t} \ket{\psi_-} \braket{\psi_-}{s}. \]
	Note that $\braket{\psi_\pm}{s} = \frac{1}{\sqrt{2}} \left( \braket{r}{s} \mp \braket{c}{s} \right) \approx \frac{1}{\sqrt{2}} (1 \mp 0) = \frac{1}{\sqrt{2}}.$
	Then the success amplitude as a function of time is
	\[ \braket{w}{\psi(t)} \approx \frac{1}{\sqrt{2}} \left( e^{-iE_+t} \braket{w}{\psi_+} + e^{-iE_-t} \braket{w}{\psi_-} \right). \]
	Now let's compute the inner products $\braket{w}{\psi_\pm}$. To evaluate them, note that
	\begin{align}
		k - \lambda + \mu 
			&= k - \lambda - 1 + \mu + 1 \notag \\
			&= (N-k-1)\frac{\mu}{k} + \mu + 1 \notag \\
			&= \frac{\mu}{k}(N-1) + 1 \notag \\
			&\approx \frac{\mu N}{k} + 1 \notag \\
			&\approx \frac{\mu N}{k}, \label{eq:klm}
	\end{align}
	where we've used Eq.~\ref{eq:params}, $k = o(N)$, and assumed large $N$. Then $\braket{w}{\psi_\pm} = \mp \frac{1}{\sqrt{2}} \braket{w}{c} \approx \mp \frac{1}{2} A \mu N / k^{3/2}$. Plugging this in,
	\[ \braket{w}{\psi(t)} \approx \frac{1}{2} A \frac{\mu N}{k^{3/2}} \left( -e^{-iE_+t} + e^{-iE_-t} \right). \]
	Plugging in for the energy eigenvalues,
	\[ \braket{w}{\psi(t)} \approx e^{-i \gamma k t} \frac{1}{2} A \frac{\mu N}{k^{3/2}} \left( -e^{-i\gamma A \sqrt{N/k} \mu t} + e^{i\gamma A \sqrt{N/k} \mu t} \right). \]
	The exponentials give us $2i\sin(\cdot)$.
	\[ \braket{w}{\psi(t)} \approx e^{-i \gamma k t} A \frac{\mu N}{k^{3/2}} i \sin \left(\gamma A \sqrt{\frac{N}{k}} \mu t \right). \]
	Then the success probability is
	\[ \left| \braket{w}{\psi(t)} \right|^2 \approx \left( A \frac{\mu N}{k^{3/2}} \right)^2 \sin^2 \left(\gamma A \sqrt{\frac{N}{k}} \mu t \right). \]
	Using $\gamma \approx 1/k$,
	\[ \left| \braket{w}{\psi(t)} \right|^2 \approx \left( A \frac{\mu N}{k^{3/2}} \right)^2 \sin^2 \left(A \frac{\sqrt{N}}{k^{3/2}} \mu t \right). \]
	Using Eq.~\ref{eq:klm}, the normalization constant of $\ket{c}$ becomes
	\[ A = \left( 1 + \frac{(k - \lambda + \mu)^2}{k} \right)^{-1/2} \approx \left( 1 + \frac{(\mu N)^2}{k^3} \right)^{-1/2} \approx \frac{k^{3/2}}{\mu N}, \]
	when $k$ scales less than or equal to $(\mu N)^{2/3}$, which is true for the known parameter families of Latin square graphs (which are proved asymmetric \cite{Babai1995}), pseudo-Latin square graphs, negative Latin square graphs, square lattice graphs, negative Latin square graphs, square lattice graphs, triangular graphs, and point graphs of partial geometries \cite{Cameron1991}. For these, the success probability is
	\[ \left| \braket{w}{\psi(t)} \right|^2 \approx \sin^2 \left( \frac{t}{\sqrt{N}} \right), \]
	which is $1$ at time $t_* = \pi \sqrt{N} / 2$, which is the same as on the complete graph. This is shown in figure \ref{fig:EvoLatinGammac2} for a Latin square graph, and it outperforms figures \ref{fig:EvoLatin} and \ref{fig:EvoLatinGammac2Prime}, as expected. We expect the success probability to approach $1$ for large $N$, which we confirm in figure \ref{fig:EvoLatinBigGammac2}

	Thus, we've shown that quantum search on known strongly regular graphs behaves like search on the complete graph for large $N$, reaching a success probability of $1$ at time $O(\sqrt{N})$. This requires choosing $\gamma = \gamma_{c1} = 1/k$ when $k = \Theta(N)$ and $\gamma_{c2} = 1/k + 1/[(N-1)\mu]$ when $k = o(N)$. Since this includes strongly regular graphs that are asymmetric, it disproves the intuition that fast quantum search requires global symmetry.

	Chapter 4 is based on a paper, ``Global Symmetry is Unnecessary for Fast Quantum Search,'' published in Physical Review Letters 112, 210502 (2014). J.~Janmark, D.~A.~Meyer and T.~G.~Wong all contributed significantly to the work.


\chapter{Nonlinear Quantum Search on Sufficiently Complete Graphs}

\section{Introduction}

In Chapter 1, we showed that a randomly walking quantum particle can locate a marked vertex on the complete graph with probability $1$ in time $\pi \sqrt{N} / 2$. It does this by evolving in a two-dimensional subspace with energy eigenstates
\begin{equation}
	\label{eq:defn}
	\ket{\psi_{0,1}} = \frac{1}{\sqrt{2}} \sqrt{\frac{\sqrt{N}}{\sqrt{N} \pm 1}} \left( | s \rangle \pm | w \rangle \right),
\end{equation}
at the critical $\gamma$, where $\ket{s}$ is the equal superposition of the vertices, and $\ket{w}$ is the marked vertex that we want to find. The corresponding energy eigenvalues of these eigenstates are
\begin{equation}
	\label{eq:defn_energies}
	E_{0,1} = -1 \mp \frac{1}{\sqrt{N}},
\end{equation}
so the system evolves from $\ket{s}$ to $\ket{w}$ in time $\pi / \Delta E = \pi \sqrt{N} / 2$.

In the previous chapter, we showed that a randomly walking quantum particle can search for a marked vertex on a strongly regular graph with the same asymptotic behavior as on the complete graph. That is, although strongly regular graphs are not complete (and even lack global symmetry), they are ``complete enough'' for the search to primarily evolve in its two lowest energy eigenstates, which take the form $(\ket{r} \pm \ket{w})/\sqrt{2}$, where $\ket{r}$ is the equal superposition of non-marked vertices, at the critical $\gamma$ and for large $N$. This is the same as \eqref{eq:defn} up to terms of order $1/\sqrt{N}$. We call such graphs that evolve according to \eqref{eq:defn} with error terms that tend to zero for large $N$ \emph{sufficiently complete}.

\begin{figure}
\begin{center}
	\includegraphics[width=2in]{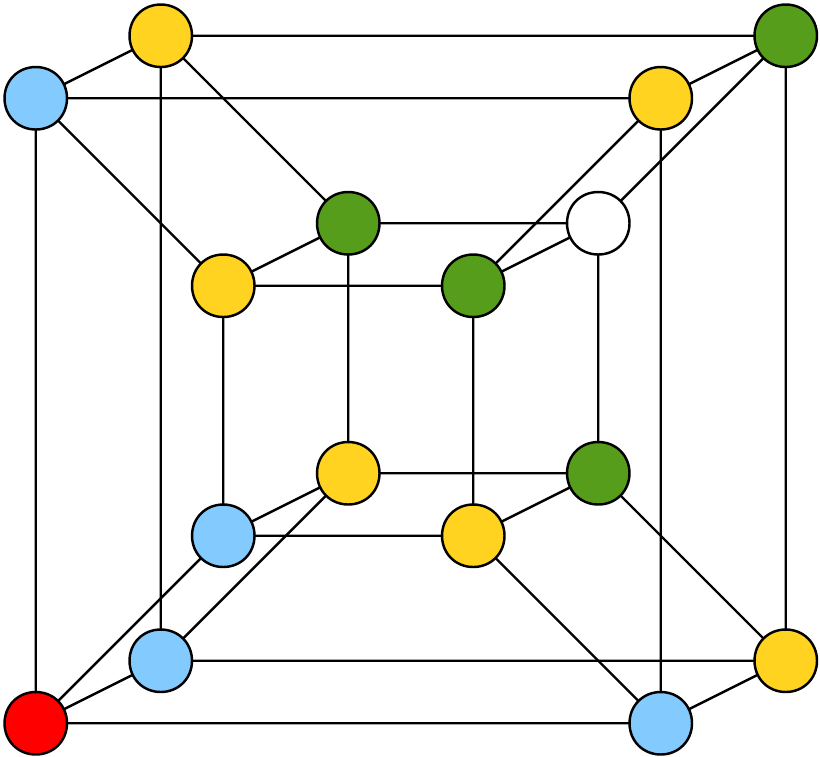}
	\caption{\label{fig:hypercube}The 4-dimensional hypercube, which has $2^4 = 16$ vertices. A single marked vertex is colored red. Vertices ``one away'' (adjacent) are colored blue, ``two away'' are yellow, ``three away'' are green, and ``four away'' is white.}
\end{center}
\end{figure}

\begin{figure}
\begin{center}
	\includegraphics{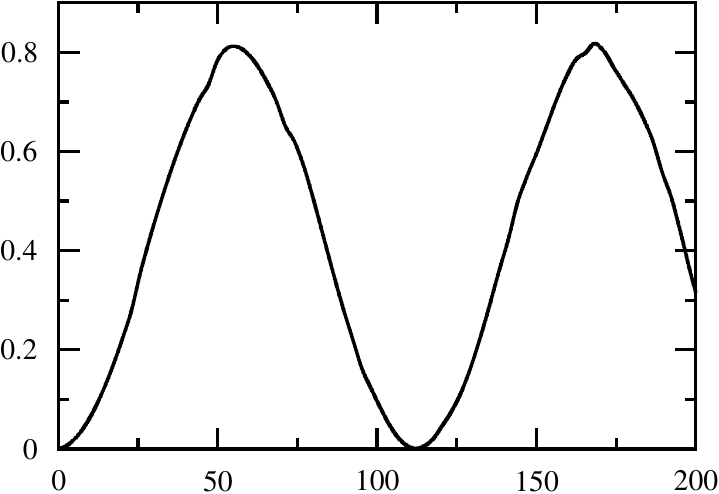}
	\caption{\label{fig:hypercube_prob_time_linear}Success probability as a function of time for search on the 10-dimensional hypercube, which has $2^{10} = 1024$ vertices, at the critical $\gamma$ derived by Childs and Goldstone \cite{CG2004}.}
\end{center}
\end{figure}

Strongly regular graphs are not the only sufficiently complete graphs---the hypercube is as well. The hypercube is even ``less complete'' than strongly regular graphs---whereas search on strongly regular graphs evolve in a three-dimensional subspace, search on the $n$-dimensional hypercube evolves in a $(n+1)$-dimensional subspace, which grows with $N = 2^n$. An example of this for four dimensions is shown in figure \ref{fig:hypercube}, where vertices that evolve identically are the same color. Nonetheless, the hypercube is complete enough for search on it to behave like search on the complete graph. That is, search on the $n$-dimensional hypercube also primarily evolves in its two lowest eigenstates, which take the form of \eqref{eq:defn}, up to terms of order $1/n$, at the critical $\gamma$ and large $N$ \cite{FGGS2000, CDFGGL2002, CG2004}. As an example, the evolution of the success probability for the $10$-dimensional hypercube is shown in figure \ref{fig:hypercube_prob_time_linear}; while $N = 2^{10} = 1024$ is large enough for the runtime to be near $\pi \sqrt{1024} / 2 \approx 50.265$, it is not large enough for the second peak to be near $3 \pi \sqrt{1024} / 2 \approx 150.80$, or for the success probability to be near $1$.

In Chapters 2 and 3, we showed that a fundamental nonlinearity provides a computational advantage in searching on the complete graph. In this chapter show that nonlinearities sometimes provide the same speedup when searching on sufficiently complete graphs, depending on the nonlinearity and the graph.

\section{Linear Search}

We begin by introducing notation to describe linear search on sufficiently complete graphs. First we assume the system exactly evolves with eigenstates and eigenvalues of the form \eqref{eq:defn} and \eqref{eq:defn_energies}, respectively, showing that it finds the marked vertex with probability $1$ in time $\pi \sqrt{N} / 2$. Then we introduce error in the eigenstates and show how it propagates to the success probability and runtime. Since search on strongly regular graphs evolves in a three-dimensional subspace, and search on the hypercube evolves in a $(n+1)$-dimensional subspace, search on a general sufficiently complete graph may evolve in an $M$-dimensional subspace. That is, the $N$ vertices of the graph can be grouped together in $M$ sets $m_i$ of size $|m_i|$, where all vertices in a set evolve identically. Then the equal superpositions of identically evolving vertices
\[ | m_i \rangle = \frac{1}{\sqrt{|m_i|}} \sum_{j \in m_i} | j \rangle \]
form an orthonormal basis $\{ | m_0 \rangle, | m_1 \rangle, \dots, | m_{M-1} \rangle \}$ for the $M$-dimensional subspace.
Without loss of generality, we pick the marked site to be $\ket{m_0} = \ket{w}$, so $|m_0| = 1$. In the case of a strongly regular graph with parameters $(N,k,\lambda,\mu)$, the basis states of the subspace are
\begin{align*}
	\ket{m_0} &= \ket{w} \\
	\ket{m_1} &= \frac{1}{\sqrt{k}} \sum_{(x,w) \in \mathcal{E}} \ket{x} \\
	\ket{m_2} &= \frac{1}{\sqrt{N-k-1}} \sum_{(x,w) \not\in \mathcal{E}} \ket{x},
\end{align*}
which correspond to the marked vertex, vertices adjacent to the marked vertex, and vertices not adjacent to the marked vertex. For the $n$-dimensional hypercube, we first label each of the $N = 2^n$ vertices with an $n$-bit string $\ket{z_1 \dots z_n}$. Without loss of generality, we choose the marked vertex to be the string of all zeros: $\ket{w} = \ket{0 \dots 0}$. Then the vertices ``one away'' are bit strings with a single one (\textit{i.e.}, with Hamming weight 1), the vertices ``two away'' are bit strings with two ones (\textit{i.e.}, with Hamming weight 2), and so forth. Taking the superposition of vertices equally far from the marked vertex, we get
\[ \ket{m_k} = \comb{n}{k}^{-1/2} \sum_{z_1 + \dots + z_n = k} \ket{z_1 \dots z_n}. \]
Then the set $\{ \ket{m_k} : k = 0, 1, \dots, n \}$ is an orthonormal basis for the $(n+1)$-dimensional subspace.

In this $\{ | m_0 \rangle, | m_1 \rangle, \dots, | m_{M-1} \rangle \}$ subspace, the state $| \psi(t) \rangle$ of the system can be written as a linear combination of the basis states:
\[ | \psi(t) \rangle = \sum_{i=0}^{M-1} c_i(t) | m_i \rangle. \]
We assume that the system evolves in its two lowest energy eigenstates, having the form \eqref{eq:defn}. Then the amplitudes are
\[ c_i(t) = \braket{m_i}{\psi(t)} = \langle m_i | e^{-iHt} | s \rangle = \braket{m_i}{\psi_0} \braket{\psi_0}{s} e^{-iE_0t} + \braket{m_i}{\psi_1} \braket{\psi_1}{s} e^{-iE_1t}. \]
Then from the definition of sufficiently complete graphs in \eqref{eq:defn},
\[ \braket{\psi_{0,1}}{s} = \frac{1}{\sqrt{2}} \sqrt{\frac{\sqrt{N}}{\sqrt{N} \pm 1}} \left( 1 \pm \frac{1}{\sqrt{N}} \right) = \frac{1}{\sqrt{2}} \sqrt{\frac{\sqrt{N}}{\sqrt{N} \pm 1}} \frac{\sqrt{N} \pm 1}{\sqrt{N}} \]
and
\[ \braket{m_i}{\psi_{0,1}} = \frac{1}{\sqrt{2}} \sqrt{\frac{\sqrt{N}}{\sqrt{N} \pm 1}} \left( \frac{\sqrt{|m_i|}}{\sqrt{N}} \pm \delta_{i0} \right), \]
where $\delta_{i0} = 1$ when $i = 0$ and $0$ otherwise is the Kronecker delta.
Then the amplitudes are
\[ c_i(t) = \begin{cases}
	e^{-it} \left[ \frac{1}{\sqrt{N}} \cos \left( \frac{t}{\sqrt{N}} \right) + i \sin \left( \frac{t}{\sqrt{N}} \right) \right], & i = 0 \\
	e^{-it} \frac{\sqrt{|m_i|}}{\sqrt{N}} \cos \left( \frac{t}{\sqrt{N}} \right), & i \ne 0 \\
\end{cases}. \]
Squaring them, the probabilities are
\begin{equation}
	\label{eq:probs}
	| c_i(t) |^2 = 
	\begin{cases}
		\frac{1}{N} \cos^2 \left( \frac{t}{\sqrt{N}} \right) + \sin^2 \left( \frac{t}{\sqrt{N}} \right), & i = 0 \\
		\frac{|m_i|}{N} \cos^2 \left( \frac{t}{\sqrt{N}} \right), & i \ne 0 \\
	\end{cases}
\end{equation}
This reveals that the success probability $|c_0|^2$ reaches $1$ at time $\pi \sqrt{N} / 2$.

Now, to introduce error, a sufficiently complete graph has eigenstates of the form \eqref{eq:defn} and corresponding eigenenergies of the form \eqref{eq:defn_energies}, both up to terms of some order $\epsilon$, where $\epsilon$ tends to zero for large $N$. For strongly regular graphs, $\epsilon = 1/\sqrt{N}$, and for the $n$-dimensional hypercube, $\epsilon = 1/n$. Propagating these errors through the previous calculations, we get probabilities
\[ \left| c_0(t) \right|^2 = \left[ \frac{1}{N} + O \left( \epsilon \right) \right] \cos^2 \left( \frac{1}{\sqrt{N}} + O \left( \epsilon \right) \right) t + \left[ 1 + O \left( \epsilon \right) \right]\sin^2 \left( \frac{1}{\sqrt{N}} + O \left( \epsilon \right) \right) t, \]
and
\[ \left| c_{i \ne 0}(t) \right|^2 = \left[ \frac{|m_i|}{N} + O \left( \epsilon \right) \right] \cos^2 \left( \frac{1}{\sqrt{N}} + O \left( \epsilon \right) \right) t, \]
where the $t$'s are inside the trigonometric functions. So if the error in the eigenstates \eqref{eq:defn} and eigenenergies \eqref{eq:defn_energies} tends to zero for large $N$, then the evolution also tends to the complete graph's evolution, as expected.

\section{Nonlinear Search}

For the nonlinear algorithm, we subtract from $H_0$ an additional nonlinear ``self-potential'' $V(t) = g f(|\psi(\mathbf{r},t)|^2)$, where $f$ is a real-valued function, so that the system evolves according to the nonlinear Schr\"odinger equation:
\[ i \frac{\partial}{\partial t} \psi(\mathbf{r},t) = \big[ H_0 - \underbrace{g f\!\left( |\psi(\mathbf{r},t)|^2 \right)}_{V(t)} \big] \psi(\mathbf{r},t). \]
Then for positive $g$ this speeds up the buildup of probability amplitude. In the computational basis, the self-potential is
\begin{equation}
	\label{eq:v}
	V(t) = g \sum_{i=0}^{N-1} f\!\left( \left| \langle i | \psi \rangle \right|^2 \right) | i \rangle \langle i |.
\end{equation}
Even with this nonlinearity, the system still evolves in the $M$-dimensional subspace spanned by $\{ | m_0 \rangle, | m_1 \rangle, \dots, | m_{M-1} \rangle \}$. In this $M$-dimensional subspace, the self-potential \eqref{eq:v} has off-diagonal elements equal to zero. Its diagonal terms are
\begin{align*}
	\left\langle m_i \middle| V(t) \middle| m_i \right\rangle 
		&= g \sum_{j = 0}^{N-1} f\!\left( \left| \langle j | \psi \rangle \right|^2 \right) \langle m_i | j \rangle \langle j | m_i \rangle \\
		&= g \frac{1}{|m_i|} \sum_{j \in m_i} f\!\left( \left| \langle j | \psi \rangle \right|^2 \right) \\
		&= g \frac{1}{|m_i|} |m_i| f\!\left( \frac{|c_i|^2}{|m_i|} \right) \\
		&= g f\!\left( \frac{|c_i|^2}{|m_i|} \right).
\end{align*}
For ease of notation, let's define
\[ f_i = f\!\left( \frac{|c_i|^2}{|m_i|} \right). \]
Then in the $M$-dimensional subspace, the nonlinearity is
\[ V(t) = g \begin{pmatrix}
	f_0 & 0 & \cdots & 0 \\
	0 & f_1 & \cdots & \vdots \\
	\vdots & \vdots & \ddots & 0 \\
	0 & \cdots & 0 & f_{M-1}
\end{pmatrix}. \]

In the $M$-dimensional subspace, the equation of motion is governed by
\begin{align*}
	H 
	&= -\gamma L - \ketbra{m_0}{m_0} - g \sum_{i = 0}^{M-1} f_i \ketbra{m_i}{m_i} \\
	&= -\gamma L - \left( 1 + g f_0 \right) \ketbra{m_0}{m_0} - g \sum_{i = 1}^{M-1} f_i \ketbra{m_i}{m_i} \\
	&= -\gamma L - \left( 1 + g f_0  - g f_1 \right) \ketbra{m_0}{m_0} - g f_1 \mathbb{I} - g \sum_{i = 2}^{M-1} \left( f_i - f_1 \right) \ketbra{m_i}{m_i} .
\end{align*}
The term proportional to the identity matrix can be dropped since it is a rescaling of energy (or an overall phase), which is unobservable. We want to show that there exists a critical $\gamma$ that causes the nonlinear evolution to follow the same path as the linear evolution for large $N$, but with rescaled time $\tau$. That is, we want to show that $f_{i \ge 2} - f_1$ can be dropped in comparison to $f_0 - f_1$ when we use the linear evolution, but with $t \to \tau$. Assume for the moment that we can do this. Then we have
\[ H = -\gamma L - \left( 1 + g f_0 - g f_1 \right) \ketbra{m_0}{m_0} \]
for large $N$. Then the critical $\gamma$ for the nonlinear algorithm is
\[ \gamma_c = \gamma_L \left( 1 + g f_0 - g f_1 \right), \]
where $\gamma_L$ is the linear algorithm's critical $\gamma$. At $\gamma_c$, we have
\[ H = \left( 1 + g f_0 - g f_1 \right) \left( -\gamma_L L - \ketbra{m_0}{m_0} \right), \]
which is the linear Hamiltonian at its critical $\gamma$ with a rescaled factor that depends on $f_0 - f_1$. For large $N$, the leading-order behavior of $f_0 - f_1$ is
\begin{align*}
	f_0 - f_1 = f\!\left( \frac{1}{N} \cos^2 \left( \frac{\tau}{\sqrt{N}} \right) + \sin^2 \left( \frac{\tau}{\sqrt{N}} \right) \right) - f\!\left( \frac{1}{N} \cos^2 \left( \frac{\tau}{\sqrt{N}} \right) \right)
\end{align*}
which is the same for all sufficiently complete graphs that can be sped up by the nonlinearity, including complete graphs, and so we expect the nonlinearity to speed them up the same way.

\begin{figure}
\begin{center}
	\includegraphics{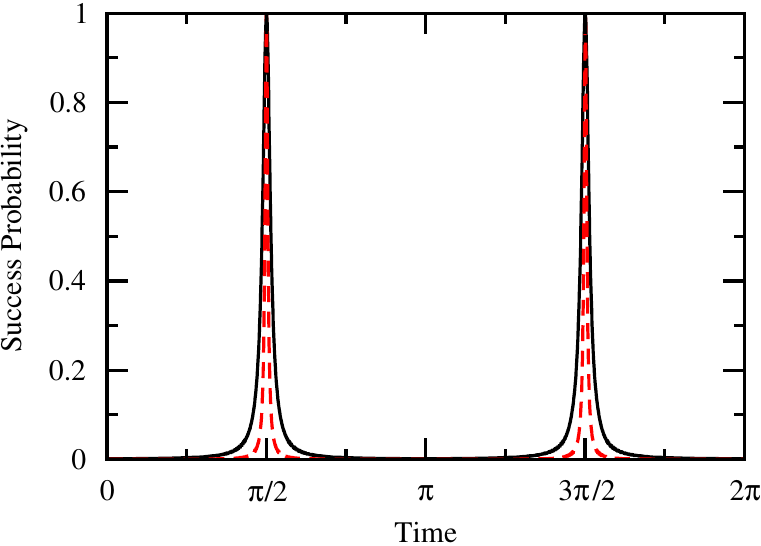}
	\caption[Success probability for search with a cubic nonlinearity and $g = N - 1$ on strongly regular graphs at the critical $\gamma$.]{\label{fig:srg_prob_time_cubic}Success probability for search with a cubic nonlinearity and $g = N - 1$ on strongly regular graphs at the critical $\gamma$. The black solid curve is with parameters $(N, k, \lambda, \mu)$ = (509,254,126,127), and the red dashed curve is (4001,2000,999,1000). With this choice of $g$, the runtime is constant for large $N$.}
\end{center}
\end{figure}

\begin{figure}
\begin{center}
	\includegraphics{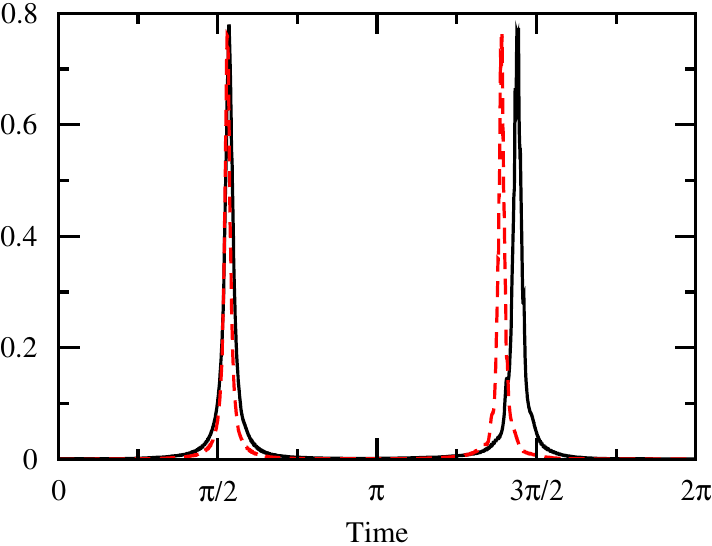}
	\caption[Success probability for search with a cubic nonlinearity and $g = N - 1$ on the hypercube.]{\label{fig:hypercube_prob_time_cubic}Success probability for search with a cubic nonlinearity and $g = N - 1$ on the hypercube. The black solid curve is the $10$-dimensional hypercube with $N = 2^{10} = 1024$ vertices, and the red dashed curve is the $11$-dimensional hypercube with $N = 2^{11} = 2048$ vertices.}
\end{center}
\end{figure}

Now let's prove that such a critical $\gamma$ exists, that we can drop $f_{i \ge 2} - f_1 $ compared to $f_0 - f_1$. This depends on the graph and the form of $f$. Assuming as before that the error in the eigenstates \eqref{eq:defn} and eigenenergies \eqref{eq:defn_energies} is of order $\epsilon$, let's first work out $f_0 - f_1$ and $f_{i \ge 2} - f_1$ for $f(p) = p$ (\textit{i.e.}, the cubic nonlinearity) so that $f_i = |c_i|^2 / |m_i|$. We get
\[ f_0 - f_1 = O \left( \epsilon \right) \cos^2 \left( \frac{1}{\sqrt{N}} + O \left( \epsilon \right) \right) \tau + \left( 1 + O \left( \epsilon \right) \right) \sin^2 \left( \frac{1}{\sqrt{N}} + O \left( \epsilon \right) \right) \tau \]
and
\[ f_{i \ge 2} - f_1 = O \left( \frac{\epsilon}{|m_i|} - \frac{\epsilon}{|m_1|} \right) \cos^2 \left( \frac{1}{\sqrt{N}} + O \left( \epsilon \right) \right) \tau, \vspace{.07in} \]
where the $\tau$'s are inside the trigonometric functions. Clearly, $f_0 - f_1$ dominates $f_{i \ge 2} - f_1$ at later time because of the sine piece, but what about at short time when cosine dominates? Recall that the set $m_0$ corresponds to the marked vertex, and the other sets $m_{i \ne 0}$ correspond to identically evolving vertices. Then for strongly regular graphs, the $m_{i \ne 0}$'s correspond to the $k = \Omega(\sqrt{N})$ vertices adjacent to the marked vertex and the $N - k - 1 = \Theta(N)$ vertices not adjacent to the marked vertex. Since both $|m_1|$ and $|m_2|$ scale as $N$ to a positive power, $f_{i \ge 2} - f_1$ scales smaller than $f_0 - f_1$, so we can drop it. Thus a critical $\gamma$ exists that causes search with a cubic nonlinearity on strongly regular graphs to behave like search on the complete graph, as shown in figure \ref{fig:srg_prob_time_cubic}. This argument doesn't work with the hypercube, however, because there is an $m_i$ that corresponds to the vertex furthest from the marked vertex, which is the white vertex in figure \ref{fig:hypercube}, so it has size $|m_i| = 1$. For this vertex, $f_{i \ge 2} - f_1$ scales the same than $f_0 - f_1$, so we can't use this argument to drop it. Nonetheless, figure \ref{fig:hypercube_prob_time_cubic} indicates that a critical $\gamma$ might still exist for the hypercube---our argument has not ruled it out.

\begin{figure}
\begin{center}
	\includegraphics{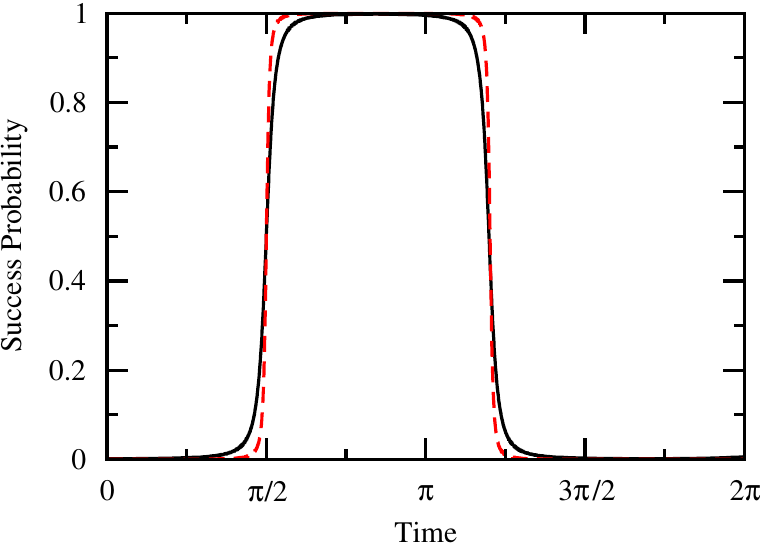}
	\caption[Success probability for search with a cubic-quintic nonlinearity and $g = N - 1$ on strongly regular graphs at the critical $\gamma$.]{\label{fig:srg_prob_time_cubicquintic}Success probability for search with a cubic-quintic nonlinearity and $g = N - 1$ on strongly regular graphs at the critical $\gamma$. The black solid curve is with parameters $(N, k, \lambda, \mu)$ = (509,254,126,127), and the red dashed curve is (4001,2000,999,1000). With this choice of $g$, the runtime is constant for large $N$.}
\end{center}
\end{figure}

\begin{figure}
\begin{center}
	\includegraphics{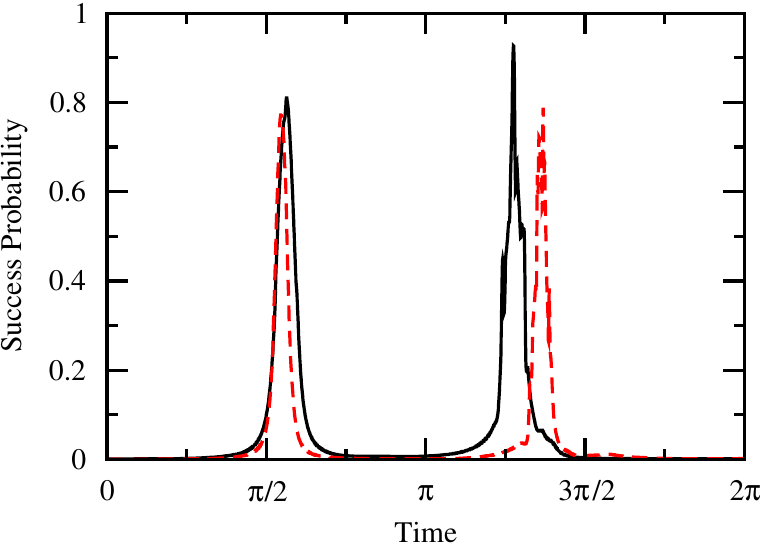}
	\caption[Success probability for search with a cubic-quintic nonlinearity and $g = N - 1$ on the hypercube.]{\label{fig:hypercube_prob_time_cubicquintic}Success probability for search with a cubic-quintic nonlinearity and $g = N - 1$ on the hypercube. The black solid curve is the $10$-dimensional hypercube with $N = 2^{10} = 1024$ vertices, and the red dashed curve is the $11$-dimensional hypercube with $N = 2^{11} = 2048$ vertices.}
\end{center}
\end{figure}

These arguments persist when we consider nonlinearities where $f(p)$ is a polynomial, which includes the cubic-quintic nonlinearity. If the polynomial is order $q$, so that $f(p) = \Theta(p^q)$, then $f_0 - f_1$ would be
\[ O \left( \epsilon^p \right) \cos^{2p} \left( \frac{1}{\sqrt{N}} + O \left( \epsilon \right) \right) \tau \]
plus terms with sine. We compare this to
\[ f_{i \ge 2} - f_1 = O \left( \frac{\epsilon^p}{|m_i|} - \frac{\epsilon^p}{|m_1|} \right) \cos^{2p} \left( \frac{1}{\sqrt{N}} + O \left( \epsilon \right) \right) \tau. \vspace{.07in} \]
Again we have the same sufficient condition for the existence of a critical $\gamma$, that if the sets $m_{i \ne 0}$ grow with $N$, then we can drop $f_{i \ge 2} - f_1 $ compared to $f_0 - f_1$. This is shown in figure \ref{fig:srg_prob_time_cubicquintic} for strongly regular graphs. Note that the peak in success probability is not as wide as for the complete graph in figure \ref{fig:prob_time_mashup}; this is expected because the width broadens as the success probability approaches 1. That is, as it approaches 1, $f_0$ goes to zero while $f_1$ remains nonzero. Then the rescaling of time slows down the evolution, causing a broad peak. For strongly regular graphs, the error with which the success probability approaches $1$ decreases as $N$ increases, so the peak is wider for large $N$. For the hypercube, figure \ref{fig:hypercube_prob_time_cubicquintic} indicates that a critical $\gamma$ may exist, although the quintic term seems to have magnified the errors compared to the cubic case in figure \ref{fig:hypercube_prob_time_cubic}. Since its success probability only reaches 0.8, we don't expect the peak to broaden at all.

\begin{figure}
\begin{center}
	\includegraphics{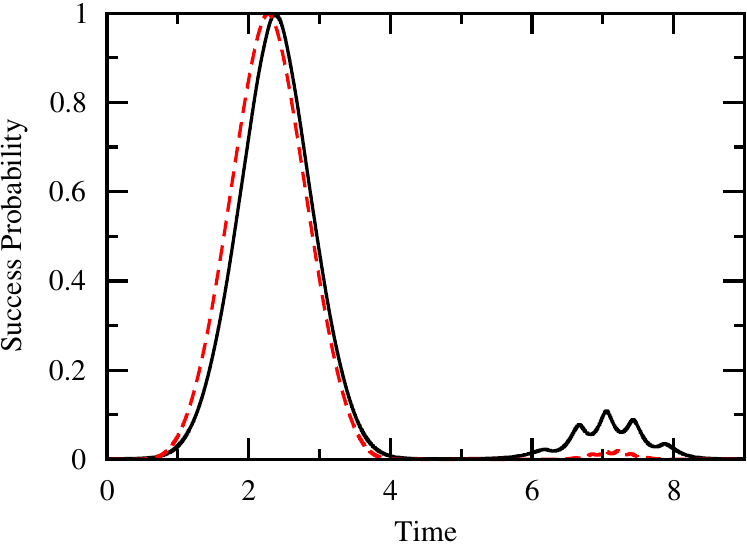}
	\caption[Success probability for search with a loglinear nonlinearity and $g = \sqrt{N}/\log N$ on strongly regular graphs at the critical $\gamma$.]{\label{fig:srg_prob_time_log}Success probability for search with a loglinear nonlinearity and $g = \sqrt{N}/\log N$ on strongly regular graphs at the critical $\gamma$. The black solid curve is with parameters $(N, k, \lambda, \mu)$ = (509,254,126,127), and the red dashed curve is (4001,2000,999,1000). With this choice of $g$, the runtime is constant for large $N$.}
\end{center}
\end{figure}

\begin{figure}
\begin{center}
	\includegraphics{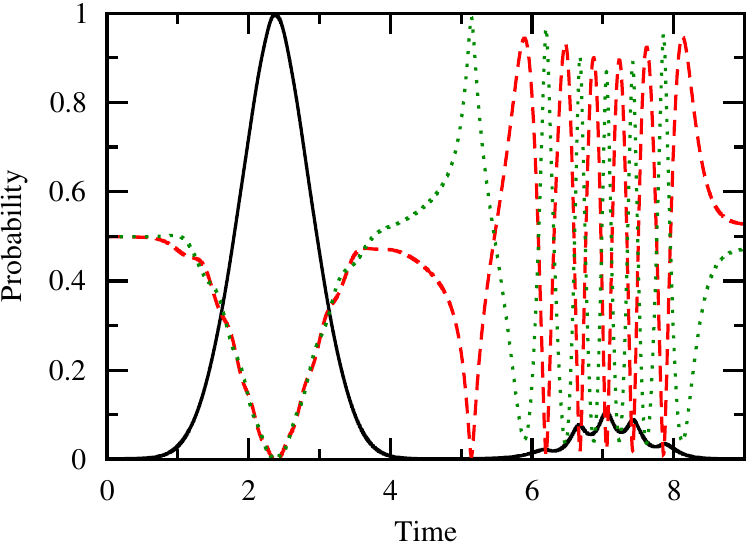}
	\caption[Evolution of search with a loglinear nonlinearity on a strongly regular graphs with parameters (509,254,126,127).]{\label{fig:srg_evolution_log}Evolution of search with a loglinear nonlinearity and $g = \sqrt{N}/\log N$ on a strongly regular graph with parameters $(N, k, \lambda, \mu)$ = (509,254,126,127). The black solid curve is $|c_0(t)|^2$, the red dashed curve is $|c_1(t)|^2$, and the green dotted curve is $|c_2(t)|^2$; they correspond to the marked vertex, vertices adjacent to the marked vertex, and vertices not adjacent to the marked vertex, respectively.}
\end{center}
\end{figure}

For the loglinear nonlinearity $f(p) = \log p$, we similarly find that whether $f_0 - f_1$ dominates $f_i - f_1$ depends on whether the sets of non-marked, identically evolving vertices have sizes that increase with $N$ or not. Since $\log(a) - \log(b) = \log(a/b)$, we have
\begin{align*}
	f_0 - f_1 
		&= \log \frac
			{\left[ \frac{1}{N} + O(\epsilon) \right] \cos^2 \left( \frac{1}{\sqrt{N}} + O(\epsilon) \right) \tau + \left[ 1 + O(\epsilon) \right]\sin^2 \left( \frac{1}{\sqrt{N}} + O(\epsilon) \right) \tau}
			{\left[ \frac{1}{N} + O\!\left( \frac{\epsilon}{|m_1|} \right) \right] \cos^2 \left( \frac{1}{\sqrt{N}} + O(\epsilon) \right) \tau} \\
		&= \log \frac
			{\left[ 1 + O(N \epsilon) \right] \cos^2 \left( \frac{1}{\sqrt{N}} + O(\epsilon) \right) \tau + \left[ N + O(N \epsilon) \right]\sin^2 \left( \frac{1}{\sqrt{N}} + O(\epsilon) \right) \tau}
			{\left[ 1 + O\!\left( \frac{N \epsilon}{|m_1|} \right) \right] \cos^2 \left( \frac{1}{\sqrt{N}} + O(\epsilon) \right) \tau}.
\end{align*}
We also have
\begin{align*}
	f_{i \ge 2} - f_1 
		&= \log \frac
	{\left[ \frac{1}{N} + O\!\left( \frac{\epsilon}{|m_i|} \right) \right] \cos^2 \left( \frac{1}{\sqrt{N}} + O(\epsilon) \right) \tau} 
			{\left[ \frac{1}{N} + O\!\left( \frac{\epsilon}{|m_1|} \right) \right] \cos^2 \left( \frac{1}{\sqrt{N}} + O(\epsilon) \right) \tau} \\
		&= \log \frac
			{1 + O\!\left( \frac{N \epsilon}{|m_i|} \right)} 
			{1 + O\!\left( \frac{N \epsilon}{|m_1|} \right)}.
\end{align*}
At later time, $f_0 - f_1$ clearly dominates $f_{i \ge 2} - f_1$ because of the sine piece. At earlier time, it is dominated by the cosine piece, so it reduces to
\[ 
	       	\log \frac
			{1 + O\!\left( N \epsilon \right)}
			{1 + O\!\left( \frac{N \epsilon}{|m_1|} \right)}.
\]
For strongly regular graphs, $\epsilon = 1/\sqrt{N}$ and $|m_1| = \Omega(\sqrt{N})$, so $N \epsilon / |m_1| = O(1)$. Then $f_0 - f_1$ for small time is dominated by $\log [O(N\epsilon)]$. Similarly, $f_{i \ge 2} - f_1$ is dominated by $\log [O(N \epsilon / |m_i|)]$, which is smaller, so we can drop it compared to $f_0 - f_1$, showing there exists a critical $\gamma$. This is shown in figure \ref{fig:srg_prob_time_log} for strongly regular graphs. The second ``peak'' is strange because of numerical error; the derivative of $\log x$ at $x = 0$ is nonzero, which makes the nonlinearity highly susceptible to noise, as shown in figure \ref{fig:srg_evolution_log}, where the evolution begins to vary wildly shortly after the first peak.

\begin{figure}
\begin{center}
	\includegraphics{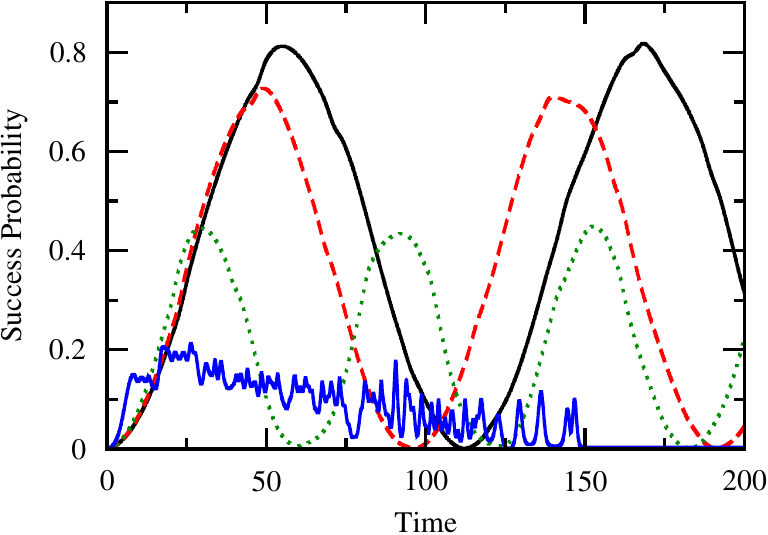}
	\caption[Success probability for search with a loglinear nonlinearity on the 10-dimensional hypercube.]{\label{fig:hypercube_prob_time_log}Success probability for search with a loglinear nonlinearity on the 10-dimensional hypercube. The black solid curve is $g = 0$ (linear), the red dashed curve is $g = 0.01$, the green dotted curve is $g = 0.05$, and the blue solid curve that wildly varies is $g = 0.5$.}
\end{center}
\end{figure}

For the hypercube, we have the same issue as before; the set corresponding to the vertex furthest from the marked vertex has size $1$, so we can't justify dropping $f_{i \ge 2} - f_1$ compared to $f_0 - f_1$. As shown in figure \ref{fig:hypercube_prob_time_log}, our formulation doesn't yield a critical $\gamma$; if it did, the peak of the success probability would stay near 0.8 with the nonlinearity.

So whether a sufficiently complete graph can be sped up by the nonlinear Schr\"odinger equation depends on the graph and the nonlinearity. Nonetheless, we've shown that even with a degree of noncompleteness, some nonlinearities speed up search on certain sufficiently complete graphs in the same way as on the complete graph for large $N$.

Chapter 5 is preliminary work for a paper to be published. D.~A.~Meyer and T.~G.~Wong both contributed significantly to the work.


\chapter{Conclusion}

\section{Summary}

To summarize our results, although quantum mechanics, which is governed by Schr\"odinger's equation, is linear, there are nevertheless many-body quantum systems whose effective evolutions are governed by nonlinear Schr\"odinger equations of the form
\begin{equation}
	\label{eq:Conclusion_NLSE}
       	i \frac{\partial \psi}{\partial t} = \left[ H_0 - g f(|\psi|^2) \right] \psi,
\end{equation}
where $f$ is a real-valued function. For example, Bose-Einstein condensates \cite{Bose1924, Einstein1924, Einstein1925} can be described by the Gross-Pitaevskii equation \cite{G1961, P1961}, which takes the form of \eqref{eq:Conclusion_NLSE} with $f(p) = p$. Nonlinear Kerr media with defocusing corrections \cite{Smektala2000, Boudebs2003, Zhan2002} can be described with $f(p) = p - p^2$. Bose liquids \cite{Avdeenkov2011} may be described with $f(p) = \log p$, which also retains the separability of noninteracting subsystems \cite{BB1976}.

We showed that the nonlinear Schr\"odinger equation \eqref{eq:Conclusion_NLSE} can be used to solve the unstructured quantum search problem faster than standard quantum computation, using the cubic, cubic-quintic, and loglinear nonlinearities as specific examples. In some instances, however, this causes the success probability to suddenly spike, which requires a certain number of atoms in an atomic clock to achieve the necessary time-measurement precision to catch the spike. This consideration of time-measurement precision as a physical resource is new. Even with it, we jointly optimized the runtime and time-measurement precision to outperform Grover's algorithm, indicating that evolution by \eqref{eq:Conclusion_NLSE} is takes fewer resources than evolution by Schr\"odinger's equation, assuming the nonlinearity is fundamental.

Of course, the nonlinearity is not fundamental, but arises as an effective description of the underlying linear dynamics. Since Grover's algorithm is optimal \cite{Zalka1999}, the speedup must be at the expense of increasing the number of particles in the physical system. Taking this into account, we arrived at bounds on the number of particles needed for the systems to be effectively described by the nonlinear equations. These are the first such bounds, and we novelly determined them by quantum information-theoretic means.

The quantum search problem can be formulated as search on the complete graph, which evolves in a two-dimensional subspace. The next level of difficulty is search in a three-dimensional subspace, which strongly regular graphs support. Although strongly regular graphs are not complete, they are ``complete enough'' such that search on them behaves like search on the complete graph for large $N$, which we novelly used degenerate perturbation theory to show. Since this includes strongly regular graphs that are asymmetric, it disproves the intuition that global symmetry is needed for fast quantum search.

The hypercube is even less complete than strongly regular graphs, yet is still ``sufficiently complete'' for search on it to behave like search on the complete graph for large $N$. The nonlinear Schr\"odinger equation \eqref{eq:Conclusion_NLSE} speeds up search on certain sufficiently complete graphs in the same way as it sped up search on the complete graph, so our results from nonlinear search on the complete graph carry over to these graphs.

Thus we have proposed a new type of quantum computer that utilizes physically realistic nonlinearities to compute in continuous time faster than (linear) quantum computation, even when some non-completeness is introduced into the underlying graph. 

\section{Future Directions}

To physically implement our nonlinear search algorithms, the underlying search graph would need to be physically encoded in our three-dimensional world. Thus the complete graph, strongly regular graphs, and the hypercube are not viable graphs to search on for large $N$. Square and cubic lattices, however, would be, and linear search on them has already been considered \cite{CG2004}. Preliminary results indicate that search by the nonlinear Schr\"odinger equation \eqref{eq:Conclusion_NLSE} also yields a speedup on arbitrary dimensional cubic lattices. But these graphs have sets of identically evolving vertices of constant size, so our analysis in Chapter 5 must be refined before we can analytically determine the precise computational advantage on cubic lattices. This would give a concrete runtime that experimentalists can seek to achieve, giving them a way to test the viability of our scheme for nonlinear, analog quantum computation.


\bibliographystyle{my-h-physrev}
\bibliography{refs}

\end{document}